\documentclass[3p,sort&compress,]{elsarticle} 

\usepackage{amsmath,amssymb,amsfonts,amsthm,color}
\usepackage[colorlinks]{hyperref}
\usepackage{graphicx} 
\usepackage{subfigure}
\usepackage{dcolumn} 
\usepackage{bm} 
\usepackage{bbm}
\usepackage{cleveref}
\usepackage{hyperref, mathtools,graphicx,soul,enumitem}
\usepackage{tcolorbox} 

\usepackage{xcolor}
\definecolor{CHIGUSA}{RGB}{58,143,183}
\definecolor{AOTAKE}{RGB}{0,137,108}
\definecolor{HATOBANEZUMI}{RGB}{114,99,110}

\newdefinition{df}{Definition}
\newdefinition{ds}{Description}
\newdefinition{crit}{Criterion}

\newproof{pf}{proof}

\newtheorem{lemma}{Lemma}

\tcbuselibrary{theorems}
\newtcbtheorem{theo}{Theorem}%
{fonttitle=\bfseries}{th}

\newcommand{\rhos}{\rho_{\rm s}}
\newcommand{\la}{\langle}
\newcommand{\ra}{\rangle}

\newcommand{\tr}{{\rm Tr}}
\newcommand{\trse}{{\rm Tr_{se}}}
\newcommand{\tre}{{\rm Tr_e}}
\newcommand{\trs}{{\rm Tr_s}}
\newcommand{\rhose}{\rho_{\rm se}}
\newcommand{\rhoe}{\rho_{\rm e}}

\newcommand{\trhoe}{\tilde{\rho}_{\rm e}}

\newcommand{\hs}{\mathbb{H}_{\rm s}}

\newcommand{\he}{\mathbb{H}_{\rm e}}
\newcommand{\hse}{\mathbb{H}_{\rm se}}
\newcommand{\fbs}{{\mathfrak B}( {\mathbb H}_{\rm s} ) }
\newcommand{\fbe}{{\mathfrak B}( {\mathbb H}_{\rm e} ) }
\newcommand{\fbse}{{\mathfrak B}( {\mathbb H}_{\rm se} ) }
\newcommand{\fus}{{\mathfrak U}( {\mathbb H}_{\rm s} ) }
\newcommand{\fue}{{\mathfrak U}( {\mathbb H}_{\rm e} ) }

\newcommand{\ha}{\hat{A}}
\newcommand{\hb}{\hat{B}}

\newcommand{\cu}{{\cal U}}
\newcommand{\ce}{{\cal E}}

\newcommand{\cq}{{\cal Q}}

\newcommand{\tce}{\tilde{\ce}}
\newcommand{\cw}{{\cal W}}

\newcommand{\cs}{{\cal S}}

\newcommand{\clu}[2]{\cu_{t_#1}^{t_#2}}
\newcommand{\cm}{{\mathfrak M}}
\newcommand{\me}[1]{\langle #1\rangle}

\newcommand{\dt}{{\rm d}t}
\newcommand{\dtau}{{\rm d}\tau}

\newcommand{\eq}[1]{{\rm Eq.~}\eqref{#1}}

\newcommand{\hep}[1]{{\mathbb H}_{\rm p_{#1}}}
\newcommand{\hef}[1]{{\mathbb H}_{\rm f_{#1}}}

\newcommand{\intfr}[1]{{#1}^\prime}
\newcommand{\hf}[1]{\mathbb{H}_{{\rm f}_{#1}}}
\newcommand{\hn}[2]{\mathbb{H}_{{\rm e}_{#1}^{#2}}}
\newcommand{\hp}[1]{\mathbb{H}_{{\rm p}_{#1}}}
\newcommand{\eeq}{\end{equation}}

\newcommand{\eqa}{\end{eqnarray}}

\newcommand{\nn}{\nonumber}

\newcommand{\dg}{^\dagger}

\newcommand{\erf}[1]{Eq.~(\ref{#1})}

\newcommand{\erfand}[2]{Eqs.~(\ref{#1}) and (\ref{#2})}

\newcommand{\smallfrac}[2]{\mbox{$\frac{#1}{#2}$}}
\newcommand{\bra}[1]{ \langle{#1} |}
\newcommand{\ket}[1]{ |{#1} \rangle}

\newcommand{\sch}{Schr\"odinger}
\newcommand{\hei}{Heisenberg}
\newcommand{\half}{\smallfrac{1}{2}}

\newcommand{\srf}[1]{Sec.~\ref{#1}}
\newcommand{\ea}{{\it et al.}}

\newcommand{\ito}{It\^o }

\newcommand{\s}[1]{\hat\sigma_{#1}}

\newcommand{\blk}{\color{black}}

\definecolor{maroon}{rgb}{0.7,0,0}

\definecolor{ngreen}{rgb}{0.3,0.7,0.3}

\definecolor{golden}{rgb}{0.8,0.6,0.1}

\definecolor{npurple}{rgb}{0.3,0,0.6}

\newcommand{\beq}{\begin{equation} }

\tcbset{parskip/.style={before={\par\pagebreak[0]\parindent=0pt},
after={\par}}}

\makeatletter
\def\appendixname{Appendix}
\appto\appendix{%
  \addtocontents{toc}{\patch@l@section}
  \appto\appendixname{ }
}
\protected\def\patch@l@section{%
  \patchcmd{\l@section}{1.5em}{\widthof{\appendixname\space}+2.5em}{}{}%
}
\makeatother

\begin{document}

\begin{frontmatter}

\title{Concepts of quantum non-Markovianity: a hierarchy }

\author{Li Li$^1$}
\ead{kenny.li@griffithuni.edu.au}

\author{Michael J. W. Hall$^{1,2}$}
\ead{michael.hall@griffith.edu.au}

\author{Howard M. Wiseman$^1$*}
\ead{h.wiseman@griffith.edu.au}

\cortext[cor]{Corresponding author}
\address{$^1$Centre for Quantum Computation and Communication Technology (Australian Research Council),\\ Centre for Quantum Dynamics, Griffith University, Brisbane, Queensland 4111, Australia.\\
$^2$Department of Theoretical Physics, Research School of Physics and Engineering, \\Australian National University, Canberra ACT 0200, Australia.}

\begin{abstract}
Markovian approximation is a widely-employed idea in descriptions of the dynamics of open quantum systems (OQSs). Although it is usually claimed to be a concept inspired by classical Markovianity, the term quantum Markovianity is used inconsistently and often unrigorously in the literature. In this report we compare the descriptions of classical stochastic processes and quantum stochastic processes (as arising in OQSs), and show that there are inherent differences that lead to the non-trivial problem of characterizing quantum non-Markovianity. Rather than proposing a single definition of quantum Markovianity, we study a host of Markov-related concepts in the quantum regime. Some of these concepts have long been used in quantum theory, such as \emph{quantum white noise}, \emph{factorization approximation}, \emph{divisibility}, \emph{GKS-Lindblad master equation}, \emph{etc.}. Others are first proposed in this report, including those we call \emph{past-future independence}, \emph{no (quantum) information backflow},  and \emph{composability}.
All of these concepts are defined under a unified framework, which allows us to rigorously build hierarchy relations among them. With various examples, we argue that the current most often used definitions of quantum Markovianity in the literature do \emph{not} fully capture the memoryless property of OQSs. In fact, quantum non-Markovianity is highly context-dependent. The results in this report, summarized as a hierarchy figure, bring clarity to the nature of quantum non-Markovianity.
\end{abstract}


\journal{Physics Reports}
\end{frontmatter}

\tableofcontents

\section{Introduction}
\label{intro}

\subsection{Markovian processes: from classical to quantum} \label{introctq}

History shapes, but does not determine, the present world.  So for any physical process. The study of a dynamical system aims to model its evolution, using the language of mathematics.  This places no specific restrictions on the modeling, other than agreement with experimental observations and consistency with relevant physical theories. However, because the `history' of the process is involved, modeling the dynamics in the generic case can be very complicated. Nevertheless, there are some physical processes which can be well approximated using assumptions that vastly simplify the description. Such processes, descriptions, and assumptions provide insights into the characterization of generic dynamics. One great simplification of this type comes from the assumption that the system evolution at any future time does not depend on its dynamical history, but only on its current state. A process that can be modeled under this assumption is usually referred to as a~\emph{Markov} process, named after the Russian mathematician Andrey Markov~\cite{BLN04}. 

For a classical stochastic process, the assumption of Markovianity can be straightforwardly translated into a rigorous mathematical definition. Given a classical  system, the system state at a single time is fully described by a vector of random variables $\bm x$. At any time $t$, $\bm x_t$ is a (vector-valued) random variable. The complete description of system dynamics from time $t_0$ to $t$ is the complete collection of random vectors at all times, which will naturally be correlated at different times. If we consider a finite set of times $t \geq t_n \geq t_{n-1} ,\ldots, t_1 \geq t_0$ (where we follow this ordering convention in the remainder of this report), we can partially characterize the dynamical process by the joint probability distribution $P(\bm x_t, \bm x_{t_n},\ldots,\bm x_{t_0})$. According to Bayes' theorem, this can be rewritten as
$
		P(\bm x_t, \bm x_{t_n},\ldots,\bm x_{t_0})
		= P(\bm x_t | \bm x_{t_n},\ldots,\bm x_{t_0}) P(\bm x_{t_n},\ldots,\bm x_{t_0})
$.
That is, the future system evolution is generally conditional on the whole of its dynamical  history. A classical stochastic process is defined to be a Markov process if, for any sequence of times $t_1, \cdots t_n$ within the interval $[ t_0, \  t]$, the conditional probability satisfies
\begin{align}
\label{def_classical_Markovianity}
	P(\bm x_t | \bm x_{t_n},\ldots,  \bm x_{t_0}) = P(\bm x_t | \bm x_{t_n}) \ .
\end{align}
This is well known as the Markov/Markovian/memoryless condition, and clearly formalizes the loose characterization stated above, that the future  state of the system depends statistically only on its current state. The condition is particularly natural for discrete-time stochastic processes, for which all times appear in Eq.~\eqref{def_classical_Markovianity}~\cite{GS02}.
There is general consensus that the simple and elegant formula  in \erf{def_classical_Markovianity} fully characterizes classical Markovianity. Markov processes include  well-studied examples, such as Wiener processes~\cite{WN23},  and  Poisson processes~\cite{FW50}, which provide basic analytical tools for understanding classical stochastic processes in general.

It is natural to attempt to employ a similar idea to define  Markovianity for an open quantum system (OQS). Unlike an ideal closed quantum system, which undergoes unitary evolution, any real quantum system is open, namely, coupled to another quantum system, typically much larger, which is usually referred to as the \emph{environment} or~\emph{bath} (we will use these terms interchangeably throughout this report)~\cite{BP02,WU99}. This coupling between the system and its environment leads to correlations between them, introducing decoherence into the dynamics of the system. Such non-unitary dynamics can be described in a variety of ways, including stochastic Heisenberg equations for system operators and evolution equations for the system density operator~\cite{CH09}.

However, the generalization of the Markovian assumption from the classical regime to the quantum regime  immediately meets some difficulties. The quantum analogues of random variables of the classical system are operators representing system observables. If one restricted to a set or vector of operators $\hat{\bf x}$ that, at any one time $t$, are mutually commuting, then one could determine a probability distribution $P(\hat{\bf x})$ via the quantum state $\rho$ at that time. But even such a restricted set of operators will not be mutually commuting at different times. Thus there is no analogue of the classical conditional distribution appearing in the Markovian condition of Eq.~(\ref{def_classical_Markovianity}).\footnote{Implicit in this statement is the assumption that the $t_j$, or, in the discrete case, the labels $j$, really do refer to different times. In the classical case this is neither here nor there; classical statistics makes no fundamental distinction between correlations of variables for the `same' system at `different' times $j$, and correlations of variables for `different' systems $j$ at the `same' time. In the quantum case, the temporal and `spatial' (for want of a better word) situations are quite distinct~\cite{RAV15}. Markov-related concepts have also been studied for the `spatial' situation, for states (``quantum Markov chains'') on a tensor-product Hilbert space~\cite{LA83,HJP04}. In this report
we are concerned only with concepts pertaining to the dynamical properties of OQSs.}
Hence, a different approach is needed to the issue of Markovian and non-Markovian evolution for OQSs. In particular, we will argue, there is no one concept that should be identified with Markovianity in the quantum case, but rather a host of related concepts.

\begin{figure}[htbp]
	\centering
	\includegraphics[width=1.0\linewidth]{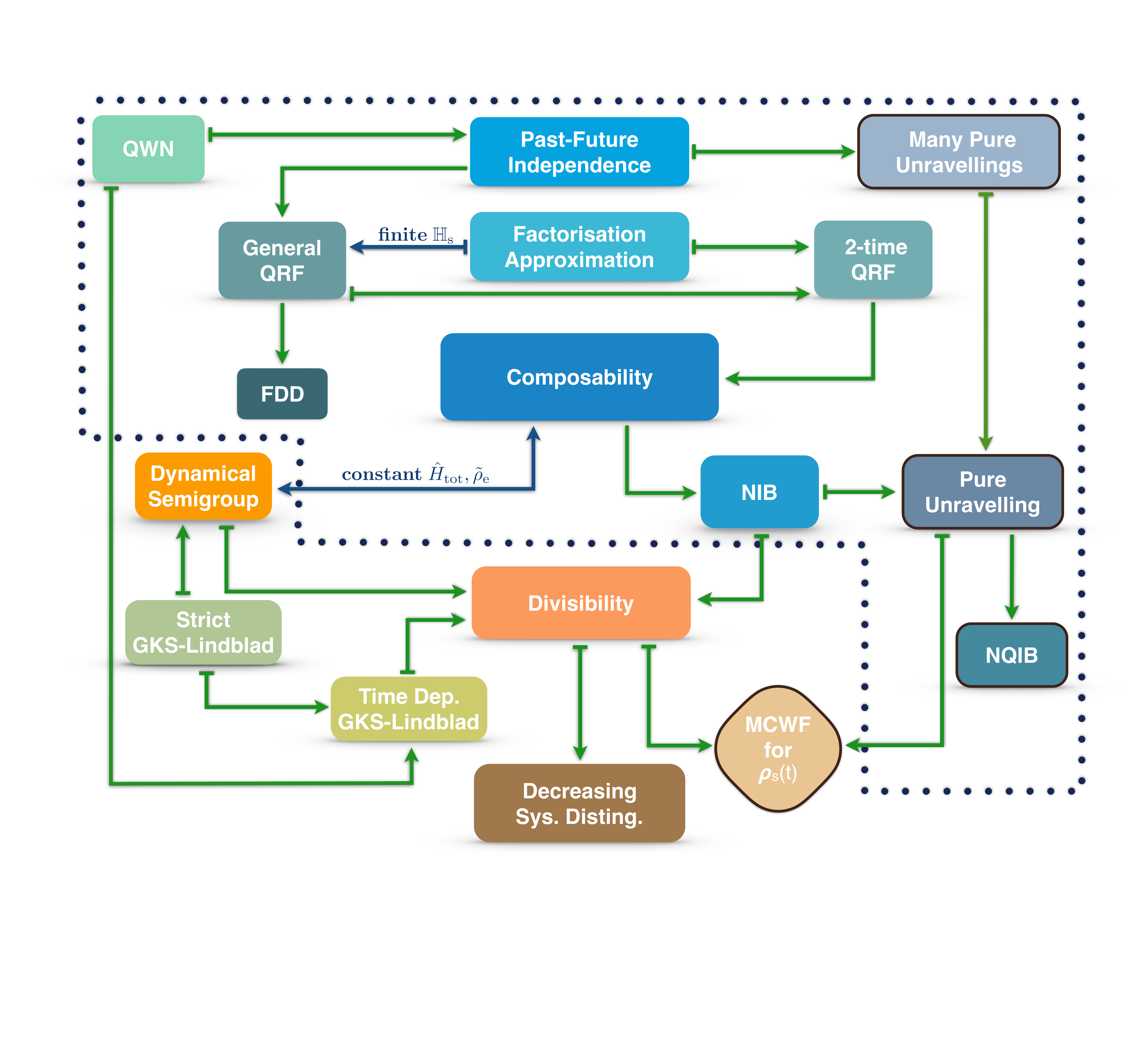}
	\caption{Hierarchy of relationships between concepts related to quantum Markovianity. We use the following abbreviations: QWN for quantum white noise; QRF for quantum regression  formula; FDD for failure of dynamical decoupling; N(Q)IB for no (quantum) information backflow; GKS-Lindblad for Gorini-Kossakowsi-Sudarshan-Lindblad type master equation; and MCWF for the Monte Carlo wave function  method. All  concepts within the  blue-dotted frame explicitly require knowledge of the system--environment interaction, while  those outside this frame are formulated solely in terms of the system dynamical map. The concepts in boxes depicted with a black border line --- MPU, PU, NQIB and MCWF ---  should not be considered as candidate definitions of quantum Markovianity, but are interesting concepts related to it. An arrow from one concept to another means that the first concept is sufficient to imply the second. Thus a double-headed arrow means two concepts are equivalent. A single-headed arrow with a bar on the tail means that the head-end concept is provably {\em not} sufficient to imply the tail-end concept. A single direction arrow without a bar means that it is unknown whether the concepts are equivalent. A blue (rather than green) arrow means the relation holds only when the additional specified constraint is satisfied.  We use a differently shaped box for MCWF because for it we give only a description, not a definition.  See detailed explanations in the text.}
	\label{fighie}
\end{figure}

In the theory of OQSs, many methods and criteria have been developed that are conceptually related to the general idea of  Markovianity. For example, the factorization approximation and quantum white noise limit are often employed in the derivation of the GKS-Lindblad-type master equation from a microscopic model~\cite{SZ97,WM09}. Other methods are directly relevant to the control of OQSs: the efficacy of dynamical decoupling is important for open-loop control~(see Section~\ref{sec_ddf}) , while the validity of quantum unravelling theory~(see Section~\ref{sec_quantum_unravelling}) is a question of importance for closed-loop control, and also for metrology. In this report we will consider these notions, and many others, in detail. We formalize existing concepts, introduce (and formalize) new ones, and establish the connections between them all. Some are candidate 
definitions for quantum Markovianity, and many of these 
have previously (if less formally) been put forward as such. Others, including those 
related to quantum unravellings, are not suitable as definitions of quantum Markovianity but are intimately related to concepts that could be used as definitions. The great majority of the concepts we consider are provably distinct from all others. That is, there are strict hierarchies in the relations we establish in this report. The elucidation of this complicated space of ideas relating to quantum Markovianity, illustrated in Fig.~\ref{fighie}, is the first main aim of this paper. 

Interest in distinguishing  Markovianity  and  non-Markovianity  for OQSs  has grown enormously and consistently over the last decade or so~\cite{RHP14,BLP16}.  There are at least four reasons for this. First, advances in manufacturing of materials and devices on the quantum scale have made it possible to engineer, and even to control, the coupling of a quantum system to its environment~\cite{MSS01,TB05}. Second, there has been increasing interest in understanding the role of decoherence in quantum phenomena relevant to biology~\cite{LCC13}. In both of these cases, accurate and efficient modelling of such precisely defined systems requires careful consideration of which, if any, notions of Markovianity can be fruitfully applied, and whether there are degrees of non-Markovianity. Third, the compelling long-term aim of building a quantum computer creates the need to efficiently characterize the effect of the environment on a quantum register, and to understand how best to minimize its damage to the computation. The Markovianity versus non-Markovianity question is of vital importance in designing quantum error mitigation schemes~\cite{KLV00,KL05}. Finally, concepts and results from quantum information theory have generated new ideas about how to make the distinction between Markovian and non-Markovian evolution in OQSs (see, e.g., Refs~\cite{BLP09,RHP10}).

Recent formalizations of the Markovian/non-Markovian distinction, from the quantum information science perspective, can be roughly divided into two broad approaches. In the first approach, quantum Markovianity is interpreted in terms of information constantly flowing out from the system. Thus, two initially perfectly distinguishable system states will gradually and continuously become less distinguishable, and eventually, in many cases, indistinguishable. Based on this approach, quantum non-Markovianity corresponds to an increase in the distinguishability of system states, and so may be quantified using a suitable measure of distinguishability such as the trace distance~\cite{BLP09}. The second approach focuses on the divisibility of the system dynamical map: a quantum  system evolution is considered to be Markovian if this map can be divided into a sequence of quantum channels, i.e., completely-positive maps~\cite{RHP10}, analogous to the divisibility of classical Markov processes (see also Section~\ref{sec-cl-CK}). These two approaches are in fact hierarchically related:  quantum Markovianity in the sense of divisibility implies Markovianity in the sense of decreasing system-state distinguishability, but not vice versa~\cite{MJL14,RHP14,BLP16}. They have led to a number of fruitful results~\cite{RHP14,BLP16}, including quantitative signatures of non-Markovianity that have been experimentally tested~\cite{LLH11,TLL12,FKF14}.

Both of these approaches are based solely on the evolution of the quantum state (density operator) of the system, possibly in conjunction with a non-interacting ancilla. Thus, they ignore many features of the system--environment interaction, and the environment's correlations with the system,  which are of  direct relevance to understanding non-Markovianity more broadly. This is in contrast to  other approaches reviewed in this report, such as those based on various {\em physical} interventions on the system or environment that could be performed to test for information backflow. In fact, a general consensus as to the `correct' approach to quantum Markovianity is still very much missing, and it remains a controversial issue: the term `quantum non-Markovianity' has very different physical meanings for different individuals and communities, causing confusion in discussions and in the literature. Thus the second main aim of this report is to defuse the situation by arguing strongly that there is no single concept that deserves the name quantum Markovianity. 

Not only do we advocate avoiding attribution of any precise meaning to the term quantum Markovianity; 
we also propose names and clear definitions for a large number of physical concepts that either have been, or could reasonably be, used to define quantum Markovianity. (Almost all of these concepts appear in 
the hierarchy of Fig.~\ref{fighie}, although some are discussed in the text only.)  If individuals were to recognize the importance of the distinctions we make, and adopt our terminology, much misunderstanding could be avoided. This would be the case even if different communities were still to prefer to identify just one of the concepts we define with quantum Markovianity.  We do not side with any community; we regard quantum Markovianity as very much context-dependent. That is, the evolution of an OQS can be said to be Markovian or non-Markovian only when one has identified which of our conditions is relevant to the context of the work in question, and tested whether the evolution satisfies or fails that condition.  

To avoid any semantic ambiguity, we rigorously define all the Markovianity conditions we introduce 
using a unified framework which is very general. We do not restrict our attention to particular types of systems and baths, nor to the derivation of specific master equations under particular `Markov approximations'.  Thus, the formal definitions of Markovianity in this report, and the hierarchical relations between them, are universally applicable. Further, while we do not aim for the mathematical rigour of, say, axiomatic quantum field theory---some derivations, for example, being strictly rigorous only for finite Hilbert spaces---we expect our results to be valid in all non-relativistic scenarios of physical interest. 


\subsection{Outline of the paper}
\label{sec_outline}

We begin by introducing the standard formalism for describing the dynamics of a general OQS in Section~\ref{sec_oqs}. This serves as the basic ingredient for rigorously defining various Markovian-related concepts in Section~\ref{sec_fqnm}. These concepts fall into two fundamental categories. The first category corresponds to the boxes falling within the dotted boundary in Fig.~\ref{fighie}: those that require explicit consideration of the interaction between the system and the environment, thus providing  clear physical pictures as to the role of the environment in the corresponding definitions of Markovianity. This category, covered in Sections~\ref{sec_FA}--\ref{sec_quantum_unravelling}, includes concepts related to the factorization approximation,    quantum white noise,  temporal  correlation functions,  interventions such as dynamical decoupling,  correlation-breaking operations, and  pure-state unravellings of the evolution. A number of the concepts in this category are defined more generally than hitherto in the literature, and others are introduced here for the first time. The second category, by contrast, comprises concepts formulated solely in terms of the dynamics of the reduced state of the system. The corresponding definitions of Markovianity, all covered in Section~\ref{sub_Mathematician}, have no explicit physical role for the environment of the system. They include the definitions based on divisibility and decreasing system distinguishability mentioned earlier, as well as definitions based on the description of the system evolution via dynamical semigroups, GKS-Lindblad-type master equations, and stochastic Schr\"odinger equations. 

In Section~\ref{sec_hierarchy}, piece by piece, we sort out the relations between all the concepts we have defined, which, as summarized in Fig.~\ref{fighie}, form a nontrivial hierarchy. This figure clearly illustrates the key point we have already emphasized: `Markovianity' in the quantum case is a term which has various different possible physical interpretations, resulting in different mathematical characterizations. Any rigorous discussion of quantum non-Markovianity must therefore begin with a clear understanding of which definition of Markovianity is being used.

As mentioned above, classical Markovianity  is well defined and studied. It is therefore of interest to consider the classical analogue of the complex hierarchy of quantum concepts of Markovianity in Fig.~\ref{fighie}. Unsurprisingly, as discussed in Section~\ref{sec_acsp},  the corresponding classical hierarchy is much simpler:  some concepts in the quantum case are either inapplicable in the classical case, or simply merge with each other. Through this classical analogy, one can also better understand how the more complex hierarchy of quantum Markovian concepts in Fig.~\ref{fighie} emerges from the simpler classical hierarchy in Fig.~\ref{fig_cl}. 

Finally, we summarize and discuss the main results  in Section~\ref{sec_conclusion}.

\section{Open quantum systems}
\label{sec_oqs}

\subsection{System--bath evolution and interaction frames} 
\label{sec_oqs:IF}

We use the standard approach to the dynamics of OQSs,  in which the system and its environment together are considered as an isolated composite system. The total Hilbert space is $\mathbb H_{\rm se} = \mathbb H_{\rm s} \otimes \mathbb H_{\rm e}$, where $\mathbb H_{\rm s}$ and $\mathbb H_{\rm e}$ are the Hilbert spaces for the system and the environment respectively.
The total Hamiltonian can be written in the form
	\begin{align}
	\label{hse}
		\hat H_{\rm tot} = \hat H_0 + \hat H_{\rm int}\ .
	\end{align}
Here 
\begin{align}\label{h0}
	\hat{H}_0 = \hat{H}_{\rm s} \otimes {\hat I}_{\rm e}+ \hat I_{\rm s} \otimes \hat{H}_{\rm e} \ ,
\end{align}
where $\hat I$ represents the  identity operator, $\hat{H}_{\rm s}$ and $\hat{H}_{\rm e}$ are the system and the environment Hamiltonians, respectively, and $\hat{H}_{\rm int}$ is an interaction term, which can be written as
 	\begin{align}
 	\label{hint}
 		\hat H_{\rm int} = \sum_k \hat C_k \otimes \hat \Xi_k\ .	
 	\end{align}
The sum here is typically taken to be finite, and can always be taken thus for a finite-dimensional $\mathbb H_{\rm s}$. 

Note that any of the terms in $\hat H_{\rm tot}$ can, in general, be time-dependent. Note also that we do not require  $\hat H_{\rm tot}$ to be an element of ${\mathfrak B}({\mathbb H}_{\rm se})$, the set of all bounded operators on $\mathbb H_{\rm se}$. Indeed, we even allow Hamiltonians that only exist as a singular limit of regular models~\cite{GJ09}, as required to describe `quantum white noise'~\cite{HP84,GC85}. 
A simpler class of singular Hamiltonians is those describing kicked systems, with a Hamiltonian of the form $\hat K_0 + \sum_{j=1}^{\infty} \delta(t-t_j) \hat K_j$, with all $\hat K_j$ Hermitian. For the latter type of system it is natural to consider the evolution only at the times $t_j$, and we will formulate our definitions, where appropriate, to apply to discrete as well as to continuous time. Thus, for example, the constraint $t\geq t_0$ will refer to either to the discrete set $\{t_j:j=0,1,2,\dots\}$ or to the continuous interval $[t_0,\infty)$, as appropriate.

For a given physical system, the splitting of $\hat{H}_{\rm  tot}$ in \erf{hse} is not unique. Any Hermitian operators of the form $\hat D \otimes \hat I$ or $\hat I \otimes \hat \Delta$ can be added to $\hat{H}_{\rm int}$, while maintaining the form in \erf{hint}, by subtracting these terms from $\hat H_{\rm s}\otimes \hat I_{\rm e}$ and $\hat I_{\rm s}\otimes \hat H_{\rm e}$ respectively. This is usefully exploited in standard derivations of master equations in quantum optics where, for example, $\sum_k ( \hat C_k \otimes \hat I) \tre[\rhoe(t_0)\hat \Xi_k] $ is subtracted from $\hat{H}_{\rm int}$ and added to $\hat{H}_{\rm s}\otimes \hat I_{\rm e}$~\cite{CH09}. Once a particular split into $\hat H_0$ and $\hat{H}_{\rm int}$ has been chosen, the former can be removed by moving to the ``interaction frame'', in which the ``free evolution'' generated by $\hat H_0$ is ``rotated away'', leaving only a (changed) interaction term, $\intfr{\hat{H}}_{\rm tot} = \intfr{\hat H}_{\rm int}$~(hence the name~\cite{WM09}). This is also a standard procedure in describing and deriving approximations to OQS dynamics, and will be explained in more detail below. While various choices of how to express the system--bath Hamiltonian might make it simpler to analyse a given OQS, any definition of Markovianity should be independent of this choice. That is indeed the case 
 for all of the definitions we discuss.

The evolution generated by $\hat H_{\rm tot}$, from an initial time $t_0$, can be described by a one-parameter family $\{\hat{U}_{t_0}^{t}:t\geq t_0\}$ of unitary operators $\hat{U}_{t_0}^{t} \in {\mathfrak{U}(\hse)}$. (Here $\mathfrak{U}({\mathbb H})\subset {\mathfrak B}({\mathbb H})$ denotes the set of all unitary operators on the Hilbert space $\mathbb H$). Specifically,
\begin{align}
	\label{uforcs}
	\hat{U}_{t_0}^{t} =  {\bf  T} \left\{ \exp \left[ - i \int_{t_0}^{t} \hat{H}_{\rm tot} (\tau)\  \dtau \right] \right\} \ ,
\end{align}
where $\hbar$ is set to one, $\bf T$ is the time ordering operator,  and we have explicitly indicated the possible time-dependence of $\hat{H}_{\rm tot}$. Note that from the one-parameter family of unitaries $\{\hat{U}_{t_0}^{t}, \  t \geq t_0 \}$ we can define the unitary evolution between any two times by $\hat U_{t_1}^{t_2} = \hat U_{t_0}^{t_2} \hat U_{t_0}^{t_1\dagger}$. To describe the evolution in an interaction frame we first define, for any $t\geq t_0$, 
\begin{align}
	\label{uforcs0}
	\hat{U}_{0}{}_{t_0}^{t} =  
	{\bf T} \left\{ \exp \left[ - i \int_{t_0}^{t} \hat{H}_0 (\tau)\  \dtau \right] \right\} 
				 = \hat{V}_{t_0}^t{}  \otimes \hat{W}_{t_0}^t{} \ .
\end{align} 
We say {\em an} interaction frame, rather than {\em the} interaction frame, because the unitary operators here, $\hat{V}_{t_0}^t{}$ acting on the system and $\hat{W}_{t_0}^t{}$ acting on the environment, are non-unique due to the above arbitrariness in defining $\hat H_{\rm int}$. The unitary evolution in the chosen interaction frame is then described by $\intfr{\hat U} = \hat U_0\dg\  \hat{U}$, for any initial and final times. This unitary evolution is generated by 
	\begin{align} 
		\intfr{\hat H}_{\rm tot}(t) = 
		\intfr{\hat H}_{\rm int}(t) := \hat{U}_0{}_{t_0}^{t\dagger} \hat{H}_{\rm int} (t)  
		\hat{U}_0{}_{t_0}^{t} \ .
	\label{deftildeH}
	\end{align} 

The time $t_0$ is special because we assume throughout this paper, as is standard, that the combined state at this initial time is factorizable. That is,
\begin{align}
\label{initialfactorization}
	\rhose(t_0) =  \rhos(t_0) \otimes \rhoe(t_0) \ .
\end{align}
Note that, unlike other operators, we do not put a hat on the state or density operator. In the \sch\ picture, this state then evolves according to
\begin{align}
	\label{dynamics_total_state}
	\rhose (t) = {\cal U}_{t_0}^t \big[ \rhos(t_0) \otimes \rhoe(t_0) \big] \ ,  \quad t \geq t_0 \ .
\end{align}
Here ${\cal U}_{t_1}^{t_2} \hat X := \hat{U}^{t_2}_{t_1} \hat X  \hat{U}^{t_2 \dagger}_{t_1} $ for any operator $\hat X \in \fbse$. This ${\cal U}$ is an example of a superoperator, mapping operators to operators. In this report, we always represent superoperators by calligraphic letters, and only use calligraphic letters for superoperators. Moreover if we have previously defined a unitary operator using some letter, such as $\hat W$, then the corresponding calligraphic ${\cal W}$ denotes the corresponding unitary map. For a bounded Hamiltonian $\hat H_{\rm tot} \in \fbse$, Eq.~\eqref{dynamics_total_state} corresponds to the dynamical differential equation ${\rm d}{\rho}_{\rm se} (t)/{\rm d}t = -i[\hat H_{\rm tot}(t),\rhose (t)]$. 

\subsection{Schr\"{o}dinger picture evolution}
\label{sec_oqs:SP}

The initially uncorrelated system and environment will typically become correlated, and often entangled, due to the interaction term $\hat{H}_{\rm int}$ in the Hamiltonian. Although~\eqref{dynamics_total_state} formally contains all the information about the total state, the complexity and  size of a typical  environment  will make an exact analysis intractable. However, it is often the evolution of the system, rather than the total state, that is of interest, described by taking  the partial trace over $\he$ in~\erf{dynamics_total_state}:
\begin{align}
	\label{dm1}
	\rhos (t) =  \tre \left[ {\cal U}_{t_0}^{t}  [ \rhos(t_0) \otimes \rhoe(t_0) ]  \right] \ .
\end{align}
Even if the initial system and bath states were pure, any entanglement induced by the interaction will make $\rhos(t)$ mixed for $t>t_0$, corresponding to decoherence. It is worth emphasizing that this decoherence effect arises from focusing on the system alone; the evolution of the system and bath considered as a whole remains reversible, as it is described by the unitary transformation in Eq.~\eqref{dm1}. By making  suitable approximations, the evolution of the reduced state $\rhos(t)$ may become relatively simple to model. This was the original context for the introduction of `Markov' approximations and models of various kinds for OQSs~\cite{GZ04,CH09}. 

Since \erf{dm1} is valid for any initial system state $\rhos(t_0)$, we can more formally describe 
the system  evolution in terms of a one-parameter family $\{ \ce_{t_0}^t, \  t \geq t_0 \}$ of \emph{dynamical maps}. That is, $\rhos (t) = {\cal E}_{t_0}^t \rhos(t_0)$, where ${\cal E}_{t_0}^t$ is a \emph{dynamical map} for the system, defined by 
\begin{align}
\label{def_dynamical_map}
	{\cal E}_{t_0}^t \hat X := \tre \left[ {\cal U}_{t_0}^{t}  [\hat X  \otimes \rhoe(t_0) ]  \right]  
\end{align}
for all $\hat X \in \fbs$. A dynamical map defined in this way is a linear map acting on $\fbs$, which maps one quantum state to another. It is guaranteed to preserve the trace and positivity of its argument, and hence is called positive and trace-preserving. In fact, by construction, the dynamical map has an even stronger property than positivity, namely complete positivity~(CP). This distinction is a unique feature for quantum systems (the transition matrix for classical systems is required only to have positive elements), 
due to the fact that the system could be entangled with some ancilla system. Technically, CP means  that $({\cal E}_{t_0}^t\otimes {\cal I}_a) [\rho_{\rm sa}]$ is also a quantum state for any joint system-ancilla density operator $\rho_{\rm sa}$, where ${\cal I}_a$ denotes the identity map acting on the ancilla~\cite{BP02}.  Conversely, any  completely positive and trace-preserving~(CPTP)  linear map can be formally represented as a dynamical map with the form of Eq.~\eqref{def_dynamical_map}, with a suitable unitary evolution and environment state constructed via Stinespring's dilation theorem~\cite{FS55}. 

If we wished to consider the \sch\ picture in the interaction frame~\cite{WM09}, we would replace ${\cal U}$ in the above by $\intfr{\cal U} = {\cal U}_0\dg\  {\cal U}$ where ${\cal U}_0$ is defined in Eq.~\eqref{uforcs0}. Thus, 
\begin{align}
	\label{IFdynamics_total_state}
	\intfr \rho_{\rm se} (t) = \intfr{\cal U}{}_{t_0}^t \left[ \rhos(t_0) \otimes \rhoe(t_0) \right] \ ,  \quad t \geq t_0 \ .
\end{align}
We can similarly define  $\intfr{\cal E}{}_{t_0}^t= {\cal V}_{t_0}^{t\dagger} {\cal E}_{t_0}^t $ and $\intfr\rho_{\rm s}(t)= \intfr{\cal E}{}_{t_0}^t \rhos(t_0)$. Under some conditions, as will be discussed, it is possible to derive a differential evolution equation for the system state.  It is most common, in quantum optics at least, to do this in the interaction frame, so that one has 
\begin{align} \label{ME1}
	{\rm d}\intfr{\rho}_{\rm s}(t)/{\rm d}t = {\cal L}_t \intfr\rho_{\rm s}(t).
\end{align}
This is commonly known as a {\em master equation}~(ME).

\subsection{Heisenberg picture evolution}
\label{sec_oqs:HP}

Instead of the above description of evolution in the \sch\ picture (whether in the interaction frame or not), we can use the \hei\ picture (again, in the interaction frame or not~\cite{WM09}). This flexibility has a number of advantages, as we will see. In the \hei\ picture, the state does not evolve, but operators, for both the system and the bath, do evolve, such that all expectation values are the same in both pictures. Using a check in place of a hat to indicate operators in the \hei\ picture, the \hei\ evolution of an operator is thus given by
\begin{align}
	\label{Hdynamics_total_state}
	\check{X} (t) = {\cal U}_{t_0}^{t\dagger} \left[ \hat{X}(t) \right] \ ,  \quad t \geq t_0 \ .
\end{align}
Similarly to the case in the \sch\ picture, for a bounded Hamiltonian $\hat H_{\rm tot}$, this corresponds to a dynamical differential equation ${\rm d}{\check{X}}(t)/{\rm d}t = i[\check H_{\rm tot}(t),\check{X}(t)] + \partial {\check{X}}(t)/\partial t$. Here the second term is due to any explicit time-dependence of the operator $\hat X(t)$ in the \sch\ picture. 
 
Once again, if we work in an interaction frame, we replace ${\cal U}$ with $\intfr{\cal U}$ and  $\check X(t)$ with $\intfr{\check X}(t)$. Note that if $\hat{A}$ is a system operator, that is, $\hat{A} = \hat{A}_{\rm s}\otimes \hat{I}_{\rm e}$, the interaction between system and environment will mean that we cannot in general write $\check{A}(t)$ in such a product form on the joint Hilbert space $\hs \otimes\he$. 
This failure of factorization is the \hei\ picture counterpart to the system--bath correlation leading to decoherence in the \sch\ picture. Note, though, that here it is manifested by environment operators, acting on $\he$, `contaminating' a system operator in $\fbs$. This, in very general terms, is the idea of {\em quantum noise}. The uncertainty principle implies that there must be uncertainty in at least some of the environmental operators at time $t_0$, and if \erf{initialfactorization} holds then this uncertainty is uncorrelated with any properties of the system at time $t_0$. Thus the appearance of this uncertainty in system operators at a later time can be thought of as quantum noise. It is the exact counterpart to classical dynamical noise in a Newtonian model (such as in Langevin's  original approach~\cite{LP08}), when one focuses on the system variables.

\subsection{Relevance to the rest of the paper}
\label{sec_oqs:RR}

This completes our coverage of general open quantum systems. All of the material is relevant to 
the remainder of the paper, and much of it is fundamental for defining different concepts of quantum Markovianity and non-Markovianity in Section~\ref{sec_fqnm} and for deriving the hierarchical relations between them in Section~\ref{sec_hierarchy}. For the definitions that depend upon the environment (Sections~\ref{sec_FA}-\ref{sec_quantum_unravelling}), we define the dynamics of an OQS by the pair $\{ {\cal U}_{t_0}^t\}$ and $\rhoe(t_0)$, where $\{{\cal U}_{t_0}^t\}$ is understood as the one-parameter family of system--environment unitary maps indexed by times $t\geq t_0$. This pair generates a corresponding one-parameter  family of system dynamical maps, $\{ {\cal E}_{t_0}^t\}$, via ${\cal E}_{t_0}^t \hat X  = \tre[{\cal U}_{t_0}^t[\hat X\otimes \rhoe(t_0)]]$ as per Eq.~(\ref{def_dynamical_map}). For the  definitions that depend only on the system state evolution (Section~\ref{sub_Mathematician}), we define the dynamics of an OQS by $\{ {\cal E}_{t_0}^t\}$ alone.  


\section{Formalizations of quantum Markovianity}
\label{sec_fqnm}

\subsection{Factorization approximation}
\label{sec_FA}

A very common approach to approximating the dynamics of open quantum systems is based on the idea that if the system--environment interaction is weak, the effect of the system on the environment may be negligible.  A natural expression of this idea in the \sch\ picture is that at any time the system--environment state $\rhose(t)$ is well approximated by a product state. That is,
\begin{align} 
\label{LaxFactor}
	\rhose	(t)  = \cu_{t_0}^t [ \rhos(t_0) \otimes \rhoe(t_0) ] \approx \rhos(t) \otimes \Big( e^{-i \hat H_{\rm e} (t - t_0)} \rhoe(t_0) e^{ i \hat H_{\rm e} (t - t_0)} \Big) \ , 
\end{align}
where $\rhos(t) = \tre \left[ \rhose(t) \right]$, and the environment Hamiltonian $\hat H_{\rm e}$ is assumed to be time-independent for simplicity. In reality $\rhose(t)$ is generally a correlated state due to the interaction term in the total Hamiltonian in Eq.~\eqref{hse}. The plausibility of employing Eq.~\eqref{LaxFactor} is usually justified in the \emph{weak-coupling} limit~\cite{LM66}, for an environment that is much larger than the system, typically with enormously many degrees of freedom. Under the factorization approximation (FA), the environment state is independent of the system; it cannot store any information about the system. Hence this approximation immediately suggests {\em memoryless} evolution: only the system state itself carries information about its past. This is why the FA is a plausible concept of Markovianity~\cite{GZ04,BP02}. 

To our knowledge, the employment of a factorization approximation dates back to the 1953 paper by Wangsness and Bloch~\cite{WB53} for deriving a master equation describing the dynamics of nuclear induction.  Although they did not specify a formula, or name the approximation,  they use it in their derivation and describe it thus: ``the molecular system is in sufficient contact with a `heat reservoir' which re-establishes equilibrium conditions more rapidly than they would be upset by the sole action of the nuclei.''  As pointed out a decade later by Argyres and Kelley~\cite{AK64}, ``assumptions were made at a certain strategic point of the calculation, that the spin and bath systems are uncorrelated.'' About this time, Lax~\cite{LM63} made explicit the approximation (\ref{LaxFactor}), named it `factorization',  and identified it with ``Markovian character"  in the context of deriving  the quantum regression formula (see Section~\ref{sub_qrf}). Similar approximations were employed by Risken {\it et. al}~\cite{RSW66}, and Mollow~\cite{MR69}. We are aware that this approximation is now often known as the ``Born approximation'' (see e.g. the text book~\cite{BP02}), a term which has its origins in scattering theory~\cite{BM26}. However, it is worth pointing out that the term ``Born approximation'' has been used with diverse meanings, even restricting to the context of OQS dynamics~\cite{AK64,AG69}. 
 Therefore, to respect historical precedent, and to avoid potential confusion, we will exclusively call it the \emph{factorization approximation}.

For the purpose of deriving rigorous relations to other Markovian concepts, we formalize the FA as follows:
\begin{tcolorbox}
\begin{df}[Factorization Approximation]
\label{def_FaA}
The OQS dynamics $\{{\cal U}_{t_0}^t\}$ and $\rhoe(t_0)$ satisfies the factorization approximation if and only if for all $t > t_0$, there exists a unitary operator $\hat W_{t_0}^t \in \fue$ such that 
\begin{align}
\label{def_FA}
	{\cal U}_{t_0}^t[ \hat X\otimes \rhoe(t_0)] = ({\cal E}_{t_0}^t \hat X )\otimes \trhoe(t) \ ,
\end{align}
where $\trhoe(t) = {\cal W}_{t_0}^t[\rhoe(t_0)]$ for all $\hat X \in \fbs$. 
\end{df}
\end{tcolorbox}

Recall that, as will remain implicit in our definitions, ${\cal E}_{t_0}^t \hat X = \tre \left[ {\cal U}_{t_0}^{t}  [ \hat X \otimes \rhoe(t_0) ]  \right]$ (see Section~\ref{sec_oqs:SP}), and ${\cal W}_{t_0}^t$ is the unitary map on the environment corresponding to $W_{t_0}^t$ (see Section~\ref{sec_oqs:IF}). This unitary map could be removed by working in an appropriate interaction frame (see Section~\ref{sec_oqs}); the fact that ${\cal W}_{t_0}^t$ is allowed to be arbitrary is because of the arbitrariness in the decomposition of the total Hamiltonian (\ref{hse}), as discussed in Section~\ref{sec_oqs}. Unlike in \erf{LaxFactor}, we have made it explicit in \erf{def_FA} that ${\cal W}_{t_0}^t$, and hence the environment state, is independent of $\rhos(t_0)$. We have also replaced the approximate equality in \erf{LaxFactor} by an equality in \erf{def_FA}. These alterations make the FA a strong condition, with various implications, as shown in Fig.~\ref{fighie}. 

Even though we have stated \erf{def_FA} with an equality, we continue to call it the factorization {\em approximation} because it can never be exact. The left-hand-side of the equality describes unitary evolution whereas the right-hand-side does not, unless ${\cal E}_{t_0}^t$ is always (that is, for all $t$) a unitary map. Readers familiar with the derivation of master equations may be thinking that \erf{def_FA} is typically only used at a specific point in such derivations, and is not meant to be taken literally. See, for example, Refs.~\cite{SZ97,GZ04,WM09,CH09}. Nevertheless, since it has been stated explicitly~\cite{LM63,BP02}, and used as an assumption capturing the notion of Markovianity~\cite{LM63}, we believe it is important to include it in our hierarchy. Other notions, relating to the possibility of making an {\em effective} factorization approximation, will be discussed in Section~\ref{sec_composability}.

\subsection{Quantum white noise}
\label{sec_quantum_white_noise}

Recall that in Section~\ref{sec_oqs:HP} we introduced the idea of quantum noise in a very general sense in the \hei\ picture. It is possible, and useful, to be more specific about quantum noise in a dynamical sense. Say we work in a particular interaction frame such that, in \erf{deftildeH},  
	\begin{align} \label{intpicH}
		\intfr{\hat H}_{\rm tot}(t) = \sum_k \intfr{\hat c}_k(t) \otimes \intfr{\hat \xi}_k (t) \ ,
	\end{align}
where the $\intfr{\hat c}_k$ are traceless system operators, and the $\intfr{\hat \xi}_k$ are Hermitian bath operators with zero mean (with respect to the time-independent environment state, $\rhoe = \rhoe(t_0)$). For a finite-dimensional $\hs$, such a frame always exists, and is unique, which makes $\intfr{\hat H}_{\rm tot}(t)$ unique also.\footnote{That is not, of course, to say that the particular decomposition of $\intfr{\hat H}_{\rm tot}(t)$ in \erf{intpicH} is unique.} Moving to the \hei\ picture in this frame, and assuming that $\intfr{\hat H}_{\rm tot}(t)$ is bounded, we can describe the evolution of an arbitrary system operator $\hat A$ (assumed constant in the \sch\ picture) by the differential equation   
	\begin{align} \label{HPeeaso}
		{\rm  d}{\intfr{\check A}}(t)/{\rm d}t = i\sum_k [\intfr{\check c}_k(t),\intfr{\check A}(t)] \otimes  \intfr{\check \xi}_k (t) \ .
	\end{align}
The tensor product in Eq.~\eqref{HPeeaso} makes implicit use of the fact that we can define a system Hilbert space $\intfr{\mathbb H}_{\rm s}(t)$, such that any system operator $\intfr{\check A}(t)$ acts only on this Hilbert space,\footnote{The existence of such a Hilbert space is guaranteed by the fact that the algebra for system operators at time $t$ is isometric to that at time $t_0$~\cite{HZ11}.} 
while environment operators at that time act on an independent Hilbert space $\intfr{\mathbb H}_{\rm e}(t)$. But the time-derivative of $\intfr{\check A}(t)$ in \erf{HPeeaso} includes, in general, bath operators $\intfr{\check \xi}_k (t)$ and so must be defined as acting on $\intfr{\mathbb H}_{\rm s}(t)\otimes\intfr{\mathbb H}_{\rm e}(t)$. In the weak-coupling limit~\cite{LM66} (see also Section~\ref{sec_FA}), the effect of the interaction on the bath operators can be expected to be small, so that to zeroth order $\intfr{\check \xi}_k (t) \approx \intfr{\hat \xi}_k (t)$. The statistics of the latter --- the \sch\ picture bath operator $\intfr{\hat \xi}_k (t)$ in the interaction frame --- are determined solely by $\rhoe(t_0)$, and so can be considered to be {\em quantum noise} from the environment, independent of the system. In the special case of Gaussian quantum statistics~\cite{WPG12}, the noise is completely characterized by the discrete set of two-time correlation functions,
	\begin{align} \label{ttcfbo}
		G_{kk'}^\pm(t,t') 
		=  \tr\left[ \rhoe(t_0)\half\left(\intfr{\hat \xi}_k (t)  \intfr{\hat \xi}_{k'}(t') \pm  \intfr{\hat \xi}_{k'}(t') \intfr{\hat \xi}_k (t) \right)\right]  \ .
	\end{align}
In some cases, $G^\pm_{kk'}(t,t')$ depends only on the time-difference $\tau=t-t'$, in which case we write $G^\pm_{kk'}(\tau)$. In such cases we can identify 
\begin{align} \label{psdqn}
S_{k}(\omega) = \int_{-\infty}^{\infty} e^{-i\omega\tau} G^+_{kk}(\tau)  d\tau
\end{align}
as the {\em power spectral density} of the quantum noise $\intfr{\hat \xi}_k$.

Typical  derivations of master equation from a microscopic model assume not only the factorization approximation, but also a short bath-correlation time~\cite{CH09,GZ04}. The bath-correlation time(s) are the characteristic decay times for the two-time correlation functions $G^\pm_{k,k'}(\tau)$ of \erf{ttcfbo}, assumed here, for simplicity, to depend only on the time difference $\tau = t-t'$. In the limit where all of these correlation times are short compared to the characteristic system evolution time (in this frame), the bath correlations can be taken to be $\delta$-correlated; that is, $G^\pm_{k,k'}(\tau) \propto \delta(\tau)$. In this limit the power spectral density of the quantum noise, the $S_k(\omega)$ in \erf{psdqn}, become flat functions of frequency. This is the limit of {\em quantum white noise} (uniform power across all frequencies, like white light).  

One formal treatment of such noise, which makes the Gaussian assumption and takes the white-noise limit at the start, is the so-called quantum optics {\em input--output} theory of Gardiner and Collett~\cite{GC85}. Their presentation makes clear that in this limit no additional factorization approximation is necessary to derive a master equation of the form of Eq.~(\ref{ME1}). However, the Gaussian assumption is not the most general one. There exist quantum noise operators $\hat\xi'(t)$ where $G^\pm_{k,k'}(\tau) \propto \delta(\tau)$ as $\tau \rightarrow 0$, but which differ from Gaussian statistics at all higher orders. This general theory was worked out independently of, and just prior to, Ref.~\cite{GC85}, by the mathematical physicists Hudson and Parthasarathy~\cite{HP84}.  Their equation is formulated for a bosonic environment, but it can be shown to also describe a fermionic environment~\cite{HP86}. Using notation based on the more recent formulation by Gough and James~\cite{GJ09}, we can define the most general type of quantum white noise dynamics as follows. (This theory is rather specialized, and most readers would do well to take this definition, and the remainder of this section, on trust, rather than trying to understand its details.  However, it is worth emphasizing that the following equations do correspond to a Hamiltonian model, as shown, for example, in Refs.~\cite{GC85,Gre01}.) 
\begin{tcolorbox}
\begin{df}[Quantum White Noise]
\label{def_QWN}
The OQS dynamics $\{{\cal U}_{t_0}^t\}$ and $\rhoe(t_0)$ 
is describable as quantum white noise if and only if for all $t > t_0$, there exists a $\hat W_{t_0}^t  \in \fue$ such that
\begin{align}
	{\rm d} \intfr{\hat{U}}{}_{t_0}^{t} & \equiv \intfr{\hat{U}}{}_{t_0}^{t + {\rm d}t} - \intfr{\hat{U}}{}_{t_0}^{t} =
	\hat{G}{\rm d}t \ \intfr{\hat{U}}{}_{t_0}^{t} \  , \label{Utpdt} \\
	\hat{G}{\rm d}t  & = 
	 -( i\hat H  -  \smallfrac{1}{2} \hat{\bm L}^\dagger \hat{\bm L} ) \  {\rm d}t - \hat{\bm L}^\dagger \underline{\hat{S}} {\rm d}\hat{\bm B}(t) + {\rm d}\hat{\bm B}^\dagger (t) \hat{\bm L} +  {\rm tr} \left[  ( \underline{\hat {S} } - \underline{\hat I} ) {\rm d} \underline{ \hat \Lambda}  \right] 
	 \ ,  \label{HPEqn} \\
 \textrm{where} \ &  \hat H = \hat H\dg \ , \   \underline{\hat{S}}\dg \underline{\hat{S}} = \underline{\hat{S}}\ \underline{\hat{S}}\dg = \underline{\hat{I}} \ \textrm{and},\  
 \forall t' > t_0\ , \forall  j, k \in \{1,\ldots,K\}\ , \nn \\ \ & \hat{B}_j (t) =  \int_{0}^t \hat{b}_j (s) {\rm d}s, \ \hat{\Lambda}_{jk} (t) = \int_{0}^t \hat{b}^\dagger_j (s) \hat{b}_k (s) {\rm d}s,  \label{wqnBL} \\
& [ \hat{b}_j (t),  \hat{b}_k (t') ] = 0 \ , \  [  \hat{b}_j (t), \hat{b}^\dagger_k (t') ] = \delta_{jk} \delta(t - t') \ , \label{singcomm}  \\
\ & \hat {b}_j(t) \rhoe(t_0)  = \beta_j(t) \rhoe(t_0)  \label{coherentinput} \ .
\end{align}
\end{df}
\end{tcolorbox}
Recall from Section~\ref{sec_oqs:IF} that $\intfr{\hat{U}}$ is the system--bath unitary map in the interaction frame defined for the free evolution $\hat U_0 = \hat V\otimes \hat W$. In the above, ${\rm d}\hat{\bm B}(t)  = \hat {\bm B}(t + {\rm d}t) - \hat {\bm B}(t)$ is a $K$-dimensional column vector of bath operators, while ${\rm d}\hat{\bm B}\dg(t)$ is a $K$-dimensional row vector whose elements are the Hermitian adjoint of those of ${\rm d}\hat{\bm B}(t)$. Similar remarks hold for $\hat{\bm L}$, except this is a $K$-vector of system operators. Similarly, we have operator-valued second-order tensors ($K\times K$-matrices) $\underline{\hat{S}}$, $\underline{\hat \Lambda}$ and $\underline{\hat{I}}$, the last being simply the identity (in all respects), and the operator-valued scalar $\hat H$. The elements ${\rm d}\hat B_k(t)$ and ${\rm d}\hat\Lambda_{jk}$ are infinitesimal integrals, as per Eq.~\eqref{wqnBL}, over functions of bath field operators $\hat b_k(t)$. The latter are similar in nature to the interaction frame bath operators $\intfr{\hat\xi}_k (t)$ discussed above; here we omit the prime marking interaction-frame operators to ease the notational burden.
To further ease the notational burden we also have omitted the tensor product sign between, and used a flexible ordering for, system and bath operators, enabling us to use the natural vectorial and tensorial structure of the collections of operators. Note that tr means trace in this tensorial space, not in the space $\mathfrak{B} (\mathbb{H}'_{\rm se})$.
The operators in $\hat{\bm L}$ and $\underline{\hat{S}}$, and the scalar $\hat H$, may be time-dependent in the interaction frame for the system, but we have suppressed explicit notation for that possibility.

The writing of \erf{Utpdt} as the difference between $\hat U$ at two infinitesimally different times, rather than as a time-derivative of $\hat U$, signifies that $\hat G{\rm d}t\hat U$ is a quantum \ito increment~\cite{HP84}. This is the most convenient form for this type of evolution with white noise~\cite{WM09}.\footnote{This is so even though it makes it less apparent that \erf{Utpdt} does in fact correspond to unitary evolution: the operator $\hat G dt$, despite its role in \erf{Utpdt}, is not simply an anti-Hermitian operator like its first term, $-i\hat H{\rm d}t$, in Eq.~(\ref{HPEqn}).} This subtlety arises because of  the singularity of the commutation relations for the ${\rm d}B_k(t)$ terms, where the product of two such terms at the same time can be as large as a first order infinitesimal, ${\rm d}t$, rather than a second-order infinitesimal $({\rm d}t)^2$.  Indeed, using the relation ${\rm d}\hat B_k(t)=\hat b_k(t) dt$, when $t=t'$, the $\delta$-function in \erf{singcomm} is `cancelled' by one ${\rm d}t$ to yield $[{\rm d}\hat B_k(t),{\rm d}\hat B_{k}\dg(t)] = {\rm d}t$. It is for this reason that the ${\rm d}B_k(t)$ terms are sometimes called quantum Wiener increments, in analogy with the classical Wiener increment ${\rm d}W(t)$ which will be discussed in Section~\ref{sec-cl-CWN}. Note, however, that \erf{HPEqn} also 
contains terms bilinear in ${\rm d}B_k(t)$, which never occurs classically because 
they would collapse to a deterministic term (since ${\rm d}W_j(t){\rm d}W_k(t) = \delta_{jk}{\rm d}t$ in the \ito calculus~\cite{GC09}). These terms  are actually more closely related to a classical Poisson increment ${\rm d}N(t) \in \{0,1\}$, the other type of classical white noise, as will be discussed in Section~\ref{sec-cl-CWN}. 

The Hudson--Parthasarathy dynamics as an instantiation of the idea of quantum Markovianity can be most 
easily appreciated in the \hei\ picture. The quantum \ito increment for the \hei\ picture evolution in the interaction frame, of an arbitrary system operator $\intfr{\check A}(t)$  is 
\begin{align}
\label{QSDE1}
{\rm d}\intfr{\check A}(t) := & (\intfr{{\cal U}}{}_{t}^{t+{\rm d}t} - {\cal I})\dg \intfr{\check A}(t)  \\
= & \left(  
\check {\bm L}^\dagger(t) {\check A}'(t)\check{\bm L}(t) - \half \left\{\check{\bm L}^\dagger(t) \check{\bm L}(t),{\check A}'(t)\right\} 
- i [ \check{ A}'(t), \check{H}(t)  ]   \right) {\rm d}t + {\rm d} \hat{\bm B}^\dagger(t)\underline{\check S}^\dagger(t) [ \check{A}'(t), \check{\bm L}(t)] \nn \\
& + [ \check{\bm L}^\dagger(t),  \check{A}'(t) ] \underline{\check{S}}(t){\rm d}\hat{\bm B}(t) + {\rm tr} [ (  \underline{\check{S}}^\dagger(t) \check{A}'(t) \underline{\check{S}}(t) - \check{A}'(t) ) {\rm d} \underline{\hat\Lambda}^\dagger(t) ] \ , 
\end{align} 
where  $\{\hat X,\hat Y\}$  denotes the anticommutator  $\hat X\hat Y + \hat Y\hat X$.
Note that the quantum white noise operators ${\rm d}\hat{\bm B}(t)$ and ${\rm d}\underline{\hat\Lambda}(t)$   appearing here are {\em not} in the \hei\ picture, hence their being capped by a hat, not a check. [The distinction between $\hat{\bm b}(t)$ and $\check{\bm b}(t)$ accounts for the less-than-obvious unitarity of \erf{Utpdt}.]  That is, the statistics of these noise operators are determined by the initial bath state $\rhoe(t_0)$, completely independent from the system. This is exact, given the $\delta$-function bath correlation functions implied by \erfand{singcomm}{coherentinput}, not an approximation. Thus the increment, in the \hei\ picture, of an arbitrary system operator at time $t$ depends only upon itself, and other system operators, at the same time, and upon bath operators that are $\delta$-correlated and strictly independent from the system at that, or any earlier, time. It is thus very natural to describe \erf{QSDE1} as a Markovian quantum stochastic differential equation, as indeed is done in Refs.~\cite{HP84,GC85}.

\subsection{Past--future independence}
\label{sec_PFI}

Quantum white noise, as introduced above, is the natural counterpart to classical white noise, which will be covered in Section~\ref{sec-cl-CWN}. However, classical Markovianity (as defined in Section~\ref{introctq}) is more general than classical white noise. The latter gives 
multi-time probabilities that obey \erf{def_classical_Markovianity}, but in addition guarantees that the single-time probability \blk $P(\bm x_t)$ evolves differentiably in time. In contrast, classical Markovianity can be formulated for discrete time evolution, and also in situations where, even though time is treated as continuous, the system configuration changes so `violently' that no differential equation for $P(\bm x_t)$ can be derived (an example of such `violent' evolution would be completely discarding the system at some predetermined time and `loading' a new system).  The same should be true for quantum Markovianity, i.e.,  it should not be restricted to cases where a differential equation for $\rho_{\rm s}(t)$ exists (as it does for QWN, as will be discussed in Section~\ref{sec_ME}).  In this section we introduce just such a broader concept of quantum Markovianity, which we call  \emph{past--future independence}~(PFI). 

\begin{figure}[htbp]
	\includegraphics[width=0.9\linewidth]{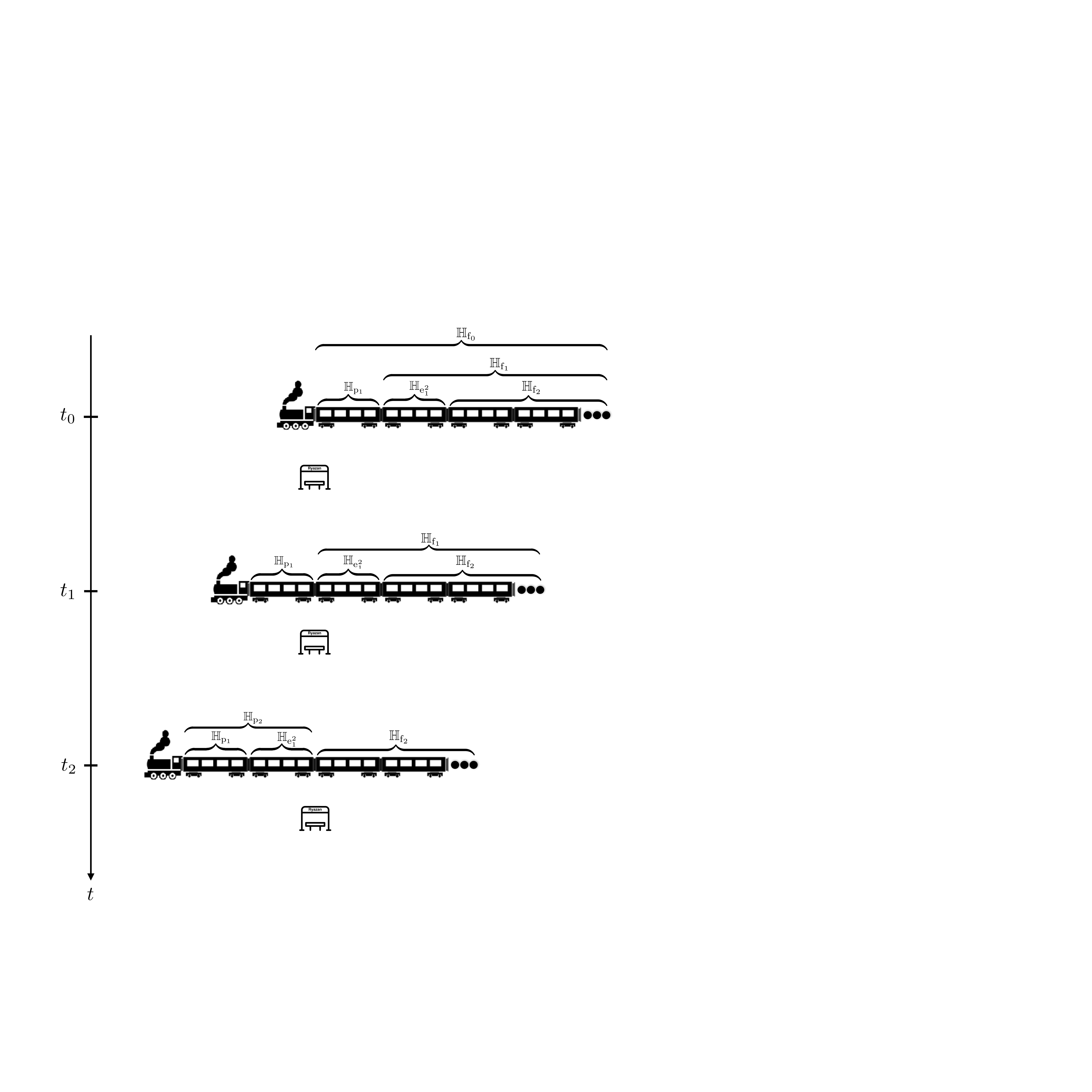}
	\caption{A train model for the concept of past--future independence. The system is represented by the train station and the environment is interacting with it like a passing train, which consists of indefinitely many independent segments, like carriages, and can always be divided into two parts: the past part $\mathbb H_{\rm p_{n}}$ and the future part $\mathbb H_{\rm f_n}$ with respect to the time $t_n$. As shown in the figure, at time $t_1$, the past part $\mathbb H_{\rm p_1}$ has already passed by (interacted with) the station (system), and would never come back again (thus there is no influence on the system evolution from the past part at later times). The future part $\mathbb H_{\rm f_1}$ can only interact with the system after time $t_1$.}
	\label{fig_PFI}
\end{figure}

The intuition behind PFI is that a Markovian environment can be thought of as a sequence of initially uncorrelated components such that each of them independently interacts with the system in sequence. This can be visualized as a `one-way' train, passing through a station (representing the system), as illustrated in Fig.~\ref{fig_PFI}. Relative to any time $t$, the environment can be divided into a past part (having already passed the station), and a future part (incoming toward to the station). 
This imposes a tensor structure of $\he$ at any time, including the existence of a `current' part relating to an interval $[t_1,t_2)$, which is what remains when the past part at time $t_1$ and the future part at time $t_2$ are excluded. 
Moreover,  we can impose dynamical restrictions relating this to the system--bath evolution, in addition to restrictions on the initial bath correlations. The dynamical rules on each part of $\he$  and $\hs$ should be specified. By doing so, we formalize the concept of the PFI  as follows:
\blk
\begin{tcolorbox}
\begin{df}[Past--Future Independence]
\label{def_PFI}

The OQS dynamics $\{{\cal U}_{t_0}^t\}$ and $\rhoe(t_0)$ exhibits past--future independence if and only if for all $  t_2 \geq  t_1 \geq t_0 $, the environment factorizes into three subsystems, $ {\mathbb H}_{\rm e} = {\mathbb H}_{\rm p_1} \otimes {\mathbb H}_{\rm e_1^2} \otimes {\mathbb H}_{\rm f_2}$, such that

\begin{enumerate}[label=(\roman*)]
	\item \label{pfi_con0}  the `current' part of the environment will become part of the `past' at the end of the interval:
	 \begin{align} \label{growingpast}
	 	\hp{2}=\hp{1}\otimes\hn{1}{2} \;
	\end{align} 
	\item \label{pfi_con1} the initial environment state factorizes as
	\begin{align}
	\label{qwndf1}
		\rhoe(t_0)  = \rho_{\rm p_1} (t_0) \otimes \rho_{\rm e_1^2}(t_0) \otimes  \rho_{\rm f_2}(t_0) \ ,
	\end{align}
	where $\rho_{\rm p_1} (t_0) \in {\mathfrak B} ( \hp{1} )$, $\rho_{\rm e_1^2} (t_0) \in {\mathfrak B} ( \hn{1}{2} )$  and $\rho_{\rm f_2}(t_0)  \in {\mathfrak B} ( \hf{2} )$\; ;
	\item \label{pfi_con2}  the unitary evolution of the system and the environment  in the interval $[t_1,t_2)$  factorizes as
	\begin{align}
	\label{qwndf2}
		\hat{U}_{t_1}^{t_2} =\hat{U}_{t_1}^{t_2} ( {\rm p_1} ) \otimes \hat{U}_{t_1}^{t_2} ( {\rm s}, {\rm e}_1^2 ) \otimes \hat{U}_{t_1}^{t_2} ( {\rm f}_2 )  \ 
	\end{align}
with $\hat{U} ( {\rm p_1} ) \in {\mathfrak U} ( {\mathbb H}_{\rm p_1} ) $, $ \hat{U}( \rm{s}, {\rm e}_1^2 ) \in {\mathfrak U} ( \hs \otimes {\mathbb H}_{ \rm e_1^2 } ) $, and $ \hat{U} ( {\rm f}_2 ) \in {\mathfrak U}  ( {\mathbb H}_{\rm f_2} )$. %
\end{enumerate}
\end{df}
\end{tcolorbox}

We now explain this definition in detail. $\hep{1}$ is the Hilbert space for the part that has been involved in the interaction with the system from time $t_0$ to $t_1$. It is like the passed `carriages' of the train~(see Fig.~\ref{fig_PFI}), and is thus called the past ($\rm  p$) part. From time $t_1$ to $t_2$, the system will interact with a new part of the environment, which, as shown in Fig.~\ref{fig_PFI}, can be understood as a carriage of the train that passes the station within that time interval. The corresponding Hilbert space is denoted by $\mathbb H_{\rm e_1^2}$. Similarly, $\hef{2}$ is the Hilbert space of the part of $\he$ that has not yet (at time $t_2$) interacted with the system. It is like the `carriages' in Fig.~\ref{fig_PFI} that are still to arrive at time $t_2$, and is called the future ($\rm f$) part. 

The division of $\he$  depends on the times $t_1$ and $t_2$. If $t_1 = t_0$, there is in fact no p part as the interaction has not even started and thus obviously we have $\hef{0} = \he$, as indicated in Fig.~\ref{fig_PFI}. Furthermore, as required by condition~\ref{pfi_con0}, the p part with respect to time $t_2$ consists of  $\mathbb H_{\rm p_1}$ and $\mathbb H_{\rm e_1^2}$ since both of them have already interacted with the system after $t_2$. It can be easily shown from condition~\ref{pfi_con0} that $\mathbb H_{\rm f_1} = \mathbb H_{\rm e_1^2} \otimes \mathbb H_{\rm f_2}$. That is, for increasing $t_k$ the size of  $\hep{k}$ increases, as per \erf{growingpast}, while that of $\hef{k}$ decreases correspondingly. Thus, for any time $t_k \geq t_0$ the environment  Hilbert space can always be decomposed as $\mathbb H_{\rm e} = \mathbb H_{ {\rm p}_k} \otimes \mathbb H_{ {\rm f}_k}$ (hence the name for this model). 

Condition~\ref{pfi_con1} requires that subsystems in the past component of the environment, relative to time $t_1$ or $t_2$\blk,  are initially uncorrelated with any subsystem in the future component of the environment relative to that time. Thus, correlations between the `carriages' in Fig.~\ref{fig_PFI} can only arise via interaction with the system. Further, such interactions are causally constrained by conditions~\ref{pfi_con2}. 
In particular, condition~\ref{pfi_con2} requires that, on any given time interval $[t_1,t_2)$, the system can only interact with the `current'  part of the environment, $\hn{1}{2}$, while the past $\hp{1}$ and future $\hf{2}$ parts of the environment must evolve 
independently. 
 
To demonstrate that these conditions preserve  independence between past and future parts, let $t_k> t_j> t_0$. Then 
from Eqs.~\eqref{qwndf1} and~\eqref{qwndf2} one has
 	\begin{align}
 	\label{qwneq11}
 		\rhose(t_j)  = \hat{U}_{t_0}^{t_j} \  \rhos(t_0) \otimes \rhoe(t_0) \ \hat{U}_{t_0}^{t_j \dagger}
 		 = \rho_{ {\rm s p}_j} (t_j) \otimes \rho_{ { \rm f}_j } (t_j) \ ,
 	\end{align}
where $\rho_{ {\rm sp}_j} (t_j):={\rm Tr}_{f_j}[\rhose(t_j)] \in {\mathfrak B}( {\mathbb H}_{ {\rm sp}_j} )$, and $\rho_{ {\rm f}_j}(t_j) := {\rm Tr}_{sp_j}[\rhose(t_j)] \in \mathfrak B ( {\mathbb H}_{ {\rm f}_j} )$. Thus, by time $t_j$, the system  has only become correlated with the part of the bath described by~$\hep{j}$.
From time $t_j$ to $t_k$ it then follows that
\begin{align}
	\label{qwnc1}
	\rhose(t_k)  = \hat{U}_{t_j}^{t_k} \rhose(t_j) \hat{U}_{t_j}^{t_k \dagger}  =   \rho_{ {\rm sp}_k} (t_k) \otimes \rho_{ {\rm f}_k}(t_k) \ ,
\end{align}
where the terms are defined analogously to those in \erf{qwneq11}. Indeed, the right-hand side of Eq.~\eqref{qwnc1} is of the same form as that of Eq.~\eqref{qwneq11}, with $t_j$ advanced to $t_k$. The necessity of imposing condition~\ref{pfi_con2} is even clearer when moving to the \hei \ picture.  In particular,  for any system operator $\hat A$, it implies that 
\begin{align}
	\label{qwnc2}
	\hat{U}_{t_j}^{t_k\dagger} (\hat{A} \otimes \hat I_{\rm e} )\  \hat{U}_{t_j}^{t_k}
	= \hat{U}_{t_j}^{t_k\dagger} (\hat{A} \otimes \hat I_{ {\rm p}_j} \otimes \hat I_{{\rm e}_j^k}\otimes \hat I_{ {\rm f}_k} )\  \hat{U}_{t_j}^{t_k}
	= \hat I_{ {\rm p}_j} \otimes \hat{A}_{ {\rm se}_j^k} \otimes \hat I_{ {\rm f}_k} \ ,
\end{align}
where $\hat{A}_{{\rm se}_j^k} \in {\mathfrak B}( \hn{j}{k} )$. 
Eq.~\eqref{qwnc2} shows that the `contamination' in the system operator after time $t_j$, i.e., the `quantum noise' discussed in Section~\ref{sec_oqs:HP}, comes only from the future part of the environment at time $t_j$, up to time $t_k$. There are no contributions from the past (already interacted) part at $t_j$, nor from the future (yet to interact) part \blk at time $t_k$.

It is worth emphasizing that all these three conditions together are necessary to define past--future independence. If any one of them fails, the system evolution could be strongly non-Markovian, by any of the usual measures. In the case of continuous time, where $t_k$ can take any real value $\geq t_0$, these conditions are not strictly physical. To say that the part of the environment interacting with the system up to time $t_k$ is completely uncorrelated with the part interacting after time $t_k$ implies a bath correlation time that is strictly zero. Indeed, this assumption (zero bath correlation time) is required to define QWN, which is a special case of PFI. Any physical system will have a finite bath correlation time, but when this is short compared to any interesting system evolution, PFI can be reasonably said to hold. For more details see the earlier discussion in Section~\ref{sec_quantum_white_noise}.

In contrast to the continuous time case, if we consider only discrete times $\tau_k$, with $t > \cdots >\tau_{k+1} >\tau_k >\cdots >t_0 $, then the PFI can be an arbitrarily good approximation to the true evolution of the system. If the total unitary evolution is such that, in each time interval $(\tau_k,\tau_{k+1})$, the system interacts with a different `carriage' in the environment, and each interaction is strictly finished within each interval, and the carriages are initially uncorrelated, then the PFI will hold exactly at the times $\tau_k$. A specific case of such an interaction is the `collision model'~\cite{ZSBH02,SZSGB02,ZB05,RHP14}, so called because the interaction of the system in each interval can be thought of as a collision with an environment `particle'. We note that Rau~\cite{Rau63} introduced this type of model in 1963, albeit without specifying any name for it. In the collision model, it is assumed that each environment `particle' (or `carriage' in our metaphor) is in the same initial state $\rhoe$, independent of the environment in preceding intervals.\footnote{It can also be generalized to allow for non-Markovian dynamics, by introducing correlations in the bath~\cite{DL08,RFZB12,CPG13}, thus breaking the PFI condition~\ref{pfi_con1}.}

\subsection{Correlation functions}
\label{sec_correlation_function}

The concepts of quantum Markovianity in the preceding sections were all formalized in terms of the dynamics of the system and environment, a compelling approach in consideration of the difficulty, as discussed in Section~\ref{intro}, in directly generalizing classical Markovianity to the quantum regime. However, classically it is common to characterize the joint probability distribution $P(\bm x_t, \bm x_{t_n},\ldots,\bm x_{t_0})$ of the dynamical process by calculating the correlation functions of system variables at different times, which are often  also experimentally measurable. Indeed, there is certainly a strong link between (classical) Markovianity and correlation functions (see Section~\ref{sec-cl-CRF}). This immediately motivates another approach to defining quantum Markovianity, through multi-time quantum correlation functions. This type of definition or interpretation of quantum Markovianity was long employed implicitly in some earlier literature, as we will discuss in following sections. The story begins with the simplest case of the two-time correlation function.

\subsubsection{Quantum regression formula}
\label{sub_qrf}

For the two-time correlation function of classical systems in quasi-equilibrium, Onsager put forth his famous postulate~\cite{OL131,OL231} that ``the average regression of fluctuations will obey the same laws as the corresponding macroscopic irreversible processes''~\cite{OL231}. That is, the average time correlation of two macroscopic quantities of the system, $\la A(t)B(0) \ra$, will follow the same decay law as that of $\la A(t) \ra$. Lax then showed in the quantum regime that, even for the nonequilibrium case, a similar relation holds under certain assumptions~\cite{LM63}. This relation, originally expressed in terms of differential equations, has been reformulated in the context of quantum optics and is now well known as the `quantum regression formula'~(QRF)~\cite{GZ04,HC991}.  As we will see, it  simplifies the calculation of correlation functions by reducing the relevant Hilbert space from $\hs \otimes \he$ to $\hs$. In this report, we generalize the definition of QRF in Ref.~\cite{HC991}, and state it  in the following  frame-independent form:  

\begin{tcolorbox}
\begin{df}[Quantum Regression Formula]
\label{def_QRF}
The OQS dynamics $\{{\cal U}_{t_0}^t\}$ and $\rhoe(t_0)$ satisfies the quantum regression formula if and only if for all times $t_2 \geq t_1 \geq t_0$, there exists a unitary operator $\hat W_{t_0}^{t_1} \in \fue$ such that the correlation function of  any two system operators $\hat{A}$ and $\hat{B}$ can be calculated via
\begin{align}
	\langle \check{B}(t_2)  \check{A} (t_1) \rangle = \trs \bigg[ \hat{B}\ \tce_{t_1}^{t_2} \big[ \hat{A} \rhos(t_1) \big]   \bigg]  \ , \label{qrf1} \\
	\langle \check{A}(t_1)  \check{B} (t_2) \rangle = \trs \bigg[ \hat{B}\ \tce_{t_1}^{t_2} \big[  \rhos(t_1) \hat{A} \big]   \bigg]  \ , \label{qrf2}
\end{align}
where $\rhos(t_1) = \ce_{t_1}^{t_0}\rhos(t_0)$, and 
\begin{align}
\label{general_dynamical_map}
	\tce_{t_1}^{t_2} \hat X  := \tre \left[   \cu_{t_1}^{t_2} [  \hat X \otimes \trhoe(t_1) ]  \right] 
\end{align}
for all $\hat X \in {\fbs}$, with  $\trhoe(t_1) = \cw_{t_0}^{t_1} \rhoe(t_0) $.

\end{df}
\end{tcolorbox}

Comparing Eqs.~\eqref{def_dynamical_map} and~\eqref{general_dynamical_map}, it is seen that $\tce_{t_1}^{t_2}$ is a generalized dynamical map for  the evolution of the system from $t_1$ to $t_2$, which corresponds to replacing the joint state $\rhose(t_1)$ at time $t_1$ by a factorized state $\rho_s(t_1)\otimes \tilde{\rho}_{\rm e}(t_1)$. Hence QRF is conceptually related to (but weaker than) the factorization approximation defined in  Eq.~\eqref{def_FA}.  In particular, it captures a similar memoryless aspect of the system evolution, but is restricted to the physical context of two-time correlation functions. 

The original Onsager postulate has been be proved to be equivalent to the classical Markovian condition in Eq.~\eqref{def_classical_Markovianity} for Gaussian processes~\cite{OM53,OM253}. Its quantum version, QRF, was proved by Lax, first under  the weak-coupling assumption~\cite{LM63}, and later under the assumption of a white noise model, i.e., an environment with a flat power spectrum, employing the quantum Langevin equation~\cite{LM66}. QRF has been a convenient tool for calculating correlation functions, especially in quantum optics~\cite{CH09,GZ04}. Specifically, by computing correlation functions of particular operators which are related to properties of the output optical fields in the QWN limit (the `past' state of the environment in terms of Section~\ref{sec_PFI}), it is possible to compute experimentally measurable correlation functions of homodyne photocurrents derived from measuring those output fields. More generally, the two-time correlation functions in \erfand{qrf1}{qrf2} can be given a simple operational meaning for any Hermitian operators $\hat A$ and $\hat B$: adding them together and dividing by two gives the measurable correlation of the result of a {\em weak} measurement~\cite{AAV88} of $\hat{A}$ at time $t_1$ with that of a (weak or strong) measurement of $\hat{B}$ at time $t_2$. The preceding sentence is true regardless of whether the QRF holds, but it is obviously easier to calculate the desired correlation function if it does. 

We should note that some authors deny the very existence of a QRF. In Ref.~\cite{TP86}, Talkner made the criticism that ``[the QRF] can only give correct results in zeroth order in the damping constant according to the weak coupling limit'' and is thus ``quite different from the classical case where large damping does not exclude a Markovian modelling". Ford and O'Connell have even claimed that ``there is no quantum regression theorem''~\cite{FO96}, based on an explicit calculation for quantum Brownian motion in which a quantum oscillator is coupled to an environment consisting of harmonic oscillators with a non-flat power spectrum, showing that the QRF failed to coincide with the exact solution of the correlation function.

The arguments made in~\cite{TP86,FO96} are correct. Like all Markovian assumptions considered thus far, QRF is never strictly true for continuous time~\cite{GCC11}. 
As pointed out by Lax~\cite{LM00}, QRF is an \emph{idealization}, accurate in physical scenarios that can be modelled by, for example, a white noise power spectrum. Thus, it is the range of validity of such models that is the real issue of the debate between Lax and Ford~\cite{FO100}. We will not pursue the debate here, but simply accept that, just as for the classical case, no precise definition of Markovianity will be satisfied exactly for any physical system. That does not prevent these definitions from holding in models that, although idealized, may nevertheless be very good approximations in practice.

\subsubsection{General quantum regression formula}
\label{sec_gqrf}

It is possible to generalize the regression formula for calculating multi-time correlation functions. These are also of practical interest in quantum optics. For example, the calculation of the two-time correlation function for photon-counting requires a  correlation function involving three operators, albeit only at two times~\cite{WM09}. A general quantum regression formula~(GQRF) is, as will be shown later, a strictly stronger concept of quantum Markovianity. The GQRF does not allow any possible time order of operators to be considered. Rather, it only allows time orderings such that a calculation in terms of the system alone may be performed, using the family of maps $\tce$ as in QRF~\footnote{\label{footgqrf} Some authors (e.g.~\cite{GAV14}) defining (G)QRF using the system propagator ${\cal G}_{t_1}^{t_2}$ in place of $\tce_{t_1}^{t_2}$. Here ${{\cal G}_{t_1}^{t_2}}$ is a two-parameter family of linear TP (but not necessarily CP) maps such that 
${\cal E}_{t_0}^{t_2}\rho =  {\cal G}_{t_1}^{t_2}{\cal E}_{t_0}^{t_1}\rho$, for all $\rho \in \fbs$ 
 (one may take ${\cal G}^{t_2}_{t_1} = \ce ^{t_2}_{t_0} {\ce'} ^{t_1}_{t_0}$, where here $\phi'$ denotes the map corresponding to the Moore-Penrose pseudo-inverse of map $\phi$~\cite{EJM07}). While such an alternate definition of (G)QRF might be considered a Markovian concept,  it would not be elementary to derive other Markovian concepts from it. This is in contrast with our definition, which we also consider to be better motivated physically.  One might also consider using an arbitrary CPTP map ${\cal Q}^{t_2}_{t_1}$ in place of $\tce_{t_1}^{t_2}$, but again we regard this as lacking a good physical motivation. See Carmichael~\cite{HC991} (Sec.~1.5) for a discussion on the quantum regression formula / theorem, its history, and its relation to Markovianity; also see Sec.~7.3.3 of that book for more on the role of system--bath correlations in the GQRF.}. 
Specifically, we define GQRF as follows: 
\begin{tcolorbox}
\begin{df}[General Quantum Regression Formula]  
\label{def_GQRF}
The OQS dynamics $\{{\cal U}_{t_0}^t\}$ and $\rhoe(t_0)$ satisfies the general quantum regression formula if and only if the time-ordered correlation function for any set of system operators~$\{ \hat{A}_j(t_j), \hat{B}_j(t_j) \}_{j=0}^{n} $ can be calculated via
	\begin{align}
		\label{gqrf}
		 \langle \check{A}_0(t_0) \check{A}_1(t_1) \ldots \check{A}_{n}(t_n) \check{B}_n(t_n) \ldots \check{B}_{1}(t_{1})\check{B}_0(t_0) \rangle 
		=  \trs[ {\cal C}_{n} \tce_{t_{n-1}}^{t_n} \ldots {\cal C}_1\ce_{t_0}^{t_1} {\cal C}_{0} \rhos(t_0) ]  \ ,
	\end{align}
where ${\cal C}_j$ is the superoperator defined on $\fbs$ by
	\begin{align}
	\label{gqrf-NA}
		{\cal C}_j \hat X = \hat{B}_j \hat X \hat{A}_j 
	\end{align}
for all $\hat X \in \fbs$, and  $ \tce_{t_j}^{t_{j+1}} $ is the generalized dynamical map defined in Eq.~\eqref{general_dynamical_map}.
\end{df}
\end{tcolorbox}
Note that this definition is consistent with QRF for the two-time case, and also with Gardiner's multi-time formula in Ref.~\cite{GZ04} (which corresponds to taking one of $\hat{A}_j(t_j)$ and $\hat{B}_j(t_j)$ to be an identity operator, for each value of $j$). A further discussion of GQRF from other perspectives is given in Section~\ref{sec_revisit_GQRF}.  The generalized regression formula has been connected with quantum Markovianity in a number of works~\cite{Lin79,LL01,GSB14}. Further, as we will show in Section~\ref{sec_acsp},  an analogous  classical regression formula turns out to be  equivalent to the classical definition of  Markovianity.

\subsection{System interventions}
\label{sec-SI}

From an operational perspective, the multitime statistics of system operators, as discussed in the previous section, can be regarded as the description of the effect of (measurement) operations performed on the system at different times. The intuition behind classical Markovianity, that the future evolution depends only on the present state, suggests that the future evolution, given the present state, should be independent of any earlier interventions on the system. This motivates a concept of quantum Markovianity that corresponds to the dynamical map for the system, from any time $t$ onwards, being independent of any system interventions at times prior to $t$. This operational approach to quantum Markovianity is not only conceptually natural, but is also tightly related to applications. Specifically, quantum control technologies typically involve performing interventions (including unitary operations and measurements) on a quantum system with some goal in mind. The question of how to model such interventions, and whether they will be effective, is linked to whether the system evolution is Markovian in the operational sense described here. In the following sections, we will discuss this idea in two concrete scenarios.

\subsubsection{Failure of  dynamical decoupling} 
\label{sec_ddf}

As discussed in Section~\ref{sec_oqs}, due to the interaction with the environment an OQS will typically gradually lose its purity, a process known as decoherence. This is one of the main obstacles for the applications of quantum technology~\cite{CLP95}. Fortunately, there are many methods that have been proposed and developed to protect the quantum system from decoherence~\cite{CMD01,MKT00,LCW98}. Among them, one particularly relevant to this report is the dynamical decoupling (DD) method~\cite{VL98,VKL99}.

The idea of DD is to apply a sequence of pre-defined unitary operations on the system at specific times, referred to as the control sequence, such that the decoherence  is almost cancelled at the end of the sequence. Consider the simplest example:  the spin echo effect introduced by Hahn~\cite{ELH50}.   This concerns the dephasing of a single qubit in the $\s{z}$ basis due to a random Hamiltonian $\kappa \s{z}$, where $\kappa$ is a constant-in-time variable, arising from the environment, with some probability distribution. If the system Hamiltonian is also proportional to $\s{z}$ then  one could eliminate both the self-evolution and the decoherence between time $t_0$ and $t_2$ by applying the operator $\hat{U} = \s{x}$ at times $t_1 = t_2/2$ and $t_2$. These operations flip the sign of $\s{z}$ at the intermediate time and so the evolution in the second half of the period reverses that in the first half. This works because the environment is extremely non-Markovian --- the fact that the Hamiltonian is constant in the interaction frame means that the correlation time of the bath is infinite. (For the definition of the bath correlation time in general, see Section~\ref{sec_quantum_white_noise}.)  More generally, for DD to work it is necessary that 
the time scales of the control sequence (the duration of each unitary operation, and the time-separation between 
operations) are much smaller than the correlation time of the bath~\cite{VKL99}. 
Owing to its effectiveness and applicability, DD has been widely applied in quantum science and technology~\cite{BUV09,DWR10,BGY11}.

The dynamics of a DD control sequence from time $t_0$ to $t$ can be described by a map~${\cal H}_{t_0}^t$ (where the letter ${\cal H}$ is  chosen in honour of Hahn),  defined on $\fbs$ by:
\begin{equation} \label{DDmap}
{\cal H}_{t_0}^{t} \hat X =  \tre \left[ {\cal U}_{t_n}^{t} ({\cal V}_n \otimes {\cal I}_{\rm e}) \ldots\  \clu{1}{2} ({\cal V}_1 \otimes {\cal I}_{\rm e})\  \clu{0}{1} [ \hat X \otimes \rhoe(t_0) ] \right]	\ .
\end{equation}
Here $\hat X \in \fbs$, ${\cal I}_{\rm e}$ is the identity map on $\fbe$, and $\{ {\cal V}_j \}_{j=1}^n $ is a set of unitary maps on $\fbs$ representing the controls. These controls are interventions on the system which, in this case, happen to be unitary. For `memoryless' dynamics, the past interventions are irrelevant to the future dynamics, as argued above. 
This would result in the failure of DD as per the  box below, which we call only a (sufficient) condition for the failure of DD because there might be other ways that DD could fail. 
\begin{tcolorbox}
	\begin{crit}[Condition for Failure of Dynamical Decoupling]
		\label{def_CFDD}
		The OQS dynamics $\{{\cal U}_{t_0}^t\}$ and $\rhoe(t_0)$ suffers failure of dynamical decoupling (FDD) if,  for any $n$  and all $\{ t_k \}_{k=1}^n$, there exist a set of CPTP maps on $\fbs$,  $\{ {\cal Q}_{t_k}^{t_{k+1}} \}$, such that for 
		all unitary control sequences $\{ \hat V_k \}_{k=1}^n$, with $\hat V_k \in \fus$, 
		\begin{align}
			\label{def_ccdf}
			{\cal H }_{t_0}^{t}   =  {\cal Q}_{t_n}^{t} {\cal V}_{t_n} \dots\ {\cal Q}_{t_1}^{t_2} {\cal V}_{t_1} {\cal E}_{t_0}^{t_1}\ ,
		\end{align}
		where the Hahn map ${\cal H}_{t_0}^{t}$  is defined in \erf{DDmap}. 
	\end{crit}
\end{tcolorbox}

The condition (\ref{def_ccdf}) implies a  {\it complete} failure of DD because any decoherence is caused by the dynamical maps ${\cal Q}$, and, if Eq.~(\ref{def_ccdf}) holds, then DD does nothing to eliminate such decoherence.\footnote{The reader might think we could define a stronger condition, such as requiring that the purity of the system be unaffected by any unitary controls.  However, this is not possible. The controls do still affect the state of the system, and with a Markovian bath (in the sense that the correlation time of the bath approaches to zero, for example the QWN dynamics in Section~\ref{sec_quantum_white_noise}), it could still be possible, for example, to transfer population from a high-decoherence subspace to a low-decoherence subspace, within the system's Hilbert space. This could be a useful decoherence-minimization control strategy, but it is not the mechanism of DD, which specifically works by taking advantage of a nonzero bath correlation time.}
As noted above, DD cannot eliminate quantum noise at frequencies 
higher than the frequency of the control operations (`pulses'). In the limit of quantum white noise, 
where there is noise at all frequencies, DD will fail, in the manner of \erf{def_ccdf}, no matter how many pulses are employed in a given time. 

We will show in subsection~\ref{theorems5and6} that FDD is indeed a logical consequence of the assumption of QWN, and of the more general concept of PFI. Note that this is consistent with the results of Ref.~\cite{AHFB14}, that, under some assumptions that seem physically reasonable, DD is always possible. This is because the QWN Hamiltonian, describing a quantum field, violates the assumptions of Ref.~\cite{AHFB14}, as does any unitary satisfying PFI. The relation of DD to quantum Markovianity has been recently discussed~\cite{ACC15,ABFH17,GH17}, and was linked to issues in quantum foundations in Ref.~\cite{AHFB14}. Finally, it is worth emphasizing that because we give only a sufficient condition for FDD, we cannot make any logical implications from FDD to other concepts of Markovianity in Fig~\ref{fighie}. However, given any concept of Markovianity that implies FDD, we can make a converse deduction. That is, success, however limited, of DD is a meaningful concept of non-Markovianity.

\subsubsection{Reinterpreting the general quantum regression formula}
\label{sec_revisit_GQRF}
	
Inspired by the preceding section on DD,  we now revisit the general quantum regression formula, defined in Section~\ref{sec_gqrf}, from a more operational perspective. First, Eq.~\eqref{gqrf} can be rewritten in the equivalent form
	\begin{align}
	\label{gqrfs1}
		\tr [{\cal C}_n{\cal U}^{t_n}_{t_{n-1}}\dots {\cal C}_1{\cal U}^{t_1}_{t_0}{\cal C}_0\rhose(t_0) ]=  \trs[ {\cal C}_{n} \tce_{t_{n-1}}^{t_n} \cdots {\cal C}_1\tce_{t_0}^{t_1} {\cal C}_{0} \rhos(t_0) ] \  .		
	\end{align}
Second, note that \emph{any} linear map on the system can be written in the form ${\cal M} \hat X=\sum_r \hat A^{(r)} \hat X \hat B^{(r)}$ for some set of corresponding operators $\{\hat A^{(r)},\hat B^{(r)}\}$. Recalling that ${\cal C}_j  \hat X := \hat A_j  \hat X \hat B_j$ for arbitrary system operators $\hat A_j$ and $\hat B_j$, Eq.~\eqref{gqrfs1} is then equivalent to the condition
	\begin{align}
	\label{gqrfs2}
		\rhos(t_n | {\cal M}_{0}, \ldots, {\cal M}_{n-1}) := \tre \left[  \cu_{t_{n-1}}^{t_n} {\cal M}_{n-1} \ldots  \cu_{t_{1}}^{t_0} {\cal M}_{0} \rhose(t_0) \right] = \tce_{t_{n-1}}^{t_n} {\cal M}_{n-1} \ldots \tce_{t_0}^{t_1} {\cal M}_0 \rhos(t_0) \ ,
	\end{align}
for arbitrary linear maps ${\cal M}_{0},\dots,{\cal M}_{n-1}$ on $\fbs$ applied to the system at respective time $t_0, \ldots t_{n-1}$. It is straightforward to show, moreover, that equivalence with GQRF still holds under the restriction that these maps ${\cal M}_j$ are completely positive,\footnote{Any general linear map $\cal M$ can be written as a linear combination ${\cal M} = {\cal M}_{+} + i {\cal M}_{-}$, where ${\cal M}_{+}( \hat X):=\half({\cal M} \hat X + [{\cal M} \hat X^\dagger]^\dagger)$ and ${\cal M}_{-}( \hat X):=\frac{1}{2i}({\cal M} \hat X - [{\cal M} \hat X^\dagger]^\dagger)$ map Hermitian operators to Hermitian operators.  Further, any map $\cal N$ that preserves Hermiticity can be written as the difference of two completely positive maps ${\cal N}={\cal N}^1-{\cal N}^2$~\cite{CM75}. Thus, a general linear map $\cal M$ can always be written as the linear combination, ${\cal M}= ({\cal M}_+^1-{\cal M}_+^2)+i({\cal M}_-^1-{\cal M}_-^2)$, of four CP maps. Noting that Eq.~\eqref{gqrfs2} is linear in the maps ${\cal M}_j$, it immediately follows that it is valid for all linear maps if and only if it is valid for all CP maps.} thus corresponding to physical operations that act locally on the system. 

It follows that the interpretation of GQRF is also related to FDD. Our condition~\erf{def_ccdf} for the latter says, roughly, that unitary interventions on the system in the past have no effect on its future evolution, given its present state. The former relaxes the condition of unitary interventions by allowing arbitrary interventions describable by CP maps, which could include conditional operations (measurements with a particular result), as well as unconditional operations (CPTP maps)~\footnote{Given the similarity of the two concepts, one might think that if we had made \erf{def_ccdf} a {\em definition} of FDD, then it would have the same implications as GQRF when one considers the case of no interventions. This is not the case, however, as we have made \erf{def_ccdf} as weak as possible by requiring only that the system dynamics be described by a family of CPTP maps~$\{ {\cal Q}_{t_k}^{t_{k+1}} \}$. By contrast, in GQRF, we imposed the strong restriction that these CPTP maps be the maps $\{\tce_{t_k}^{t_{k+1}}\}$ generated on this interval from the unitary evolution starting with an independently evolved environment state $\trhoe(t_k)$.}. This concept of Markovianity from GQRF is quite similar to the one used by Lindblad in Ref.~\cite{Lin79}, and Accardi, Frigerio, and Lewis in Ref.~\cite{AFL82}.   
It can be shown to be equivalent to the definition of quantum Markovianity recently suggested by Pollack \ea~\cite{PRC15,PRC18}, based on ``causal interventions'' (see also Ref.~\cite{CS16}).  In particular, such interventions correspond to the special case that ${\cal M}_{n-1}$ in Eq.~\eqref{gqrfs2} is a replacement map, with ${\cal M}_{n-1}  \hat X \propto \rho'$ for some fixed system state $\rho'$, and the remaining ${\cal M}_j$ being CPTP maps. 

\subsection{Environment interventions}
\label{sec_ei}

\begin{figure}
	\centering
	\subfigure[Information flow in OQS dynamics]{
		\includegraphics[width=0.8\linewidth]{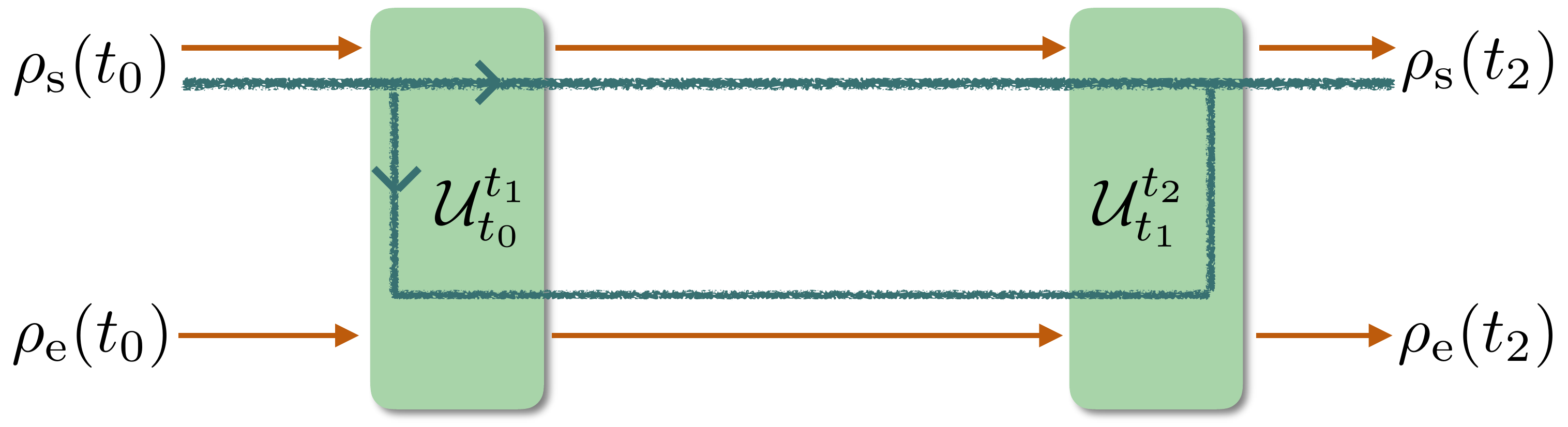}
		\label{fig_channel1}
	}
	\subfigure[Information flow blocked due to a decorrelating intervention upon the environment]{
		\includegraphics[width=0.8\linewidth]{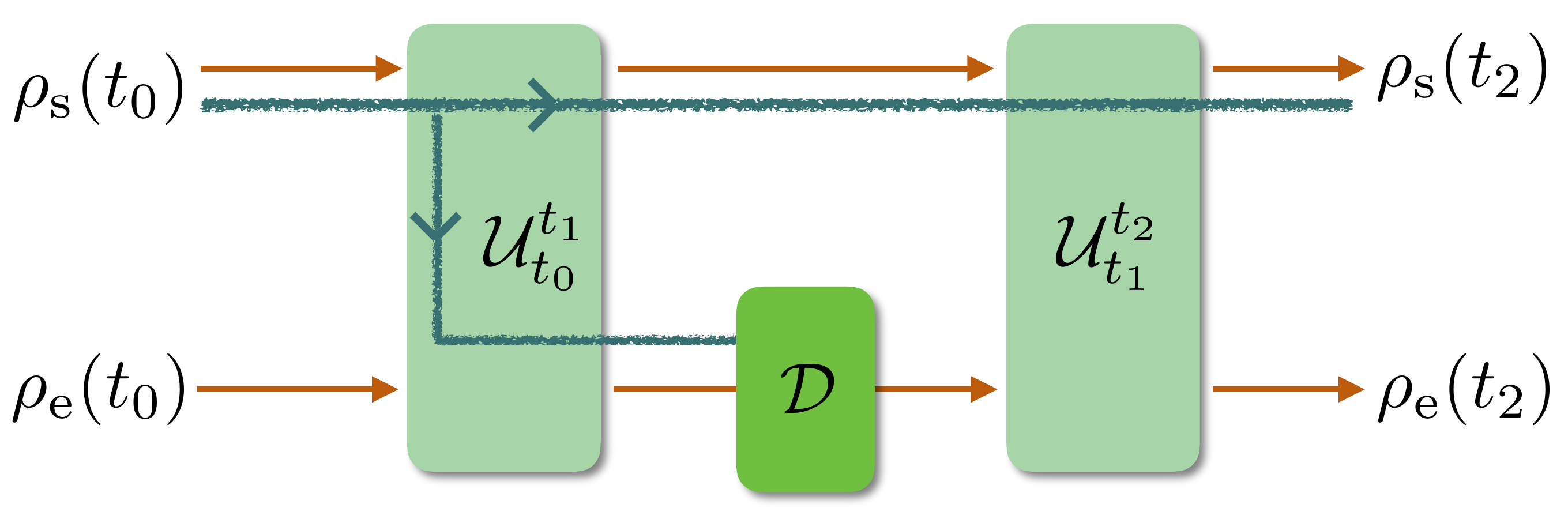}
		\label{fig_channel2}
	}
	\caption{OQS dynamics under an intervention on the environment. The dark orange lines indicate the direction of the dynamics, and the thick blue lines indicate information flow from the system to the environment and back again. The intervention is chosen to remove a given type of system--environment correlation, via a decorrelating quantum channel $\cal D$. If the system evolution is invariant under this intervention, then it is not influenced by a flow of information from the correlation back into the system, and may be associated with a corresponding type of quantum Markovianity. We consider three types of decorrelating channel in the main text: environment-reset channels, factorization channels, and entanglement-breaking channels.
	}
\end{figure}

An information-based approach to the study of quantum systems has been extensively employed in parallel with the progress in the thriving field of quantum information~\cite{BEZ00,KM02,NC10}. 
In general there will be information flow in both directions between the system and the environment. This allows the system evolution to be influenced by information about its earlier state encoded in the environment, as shown in Fig.~\ref{fig_channel1}. Intuitively, this is a non-Markovian flow of information. This possibility may, however, be excluded by passing the environment alone through a certain artificially introduced channel, which breaks its correlations with the system, as illustrated in Fig.~\ref{fig_channel2}. If this hypothetical intervention on the environment were to make no difference to the system evolution, we could conclude that the system evolution was Markovian to begin with. This formulation of Markovianity in terms of interventions on the environment is thus complementary to that involving interventions on the system as discussed in Section~\ref{sec-SI}. 

A mathematical description of the above is as follows. Consider the evolution of the system from time $t_0$ up to $t_2$, in the case where a decorrelating quantum channel ${\cal D}$ has been applied to the environment at an intermediate time $t_1$ (see Fig.~\ref{fig_channel2}). Say this environment intervention removes a particular type of system--environment correlation without affecting the final system state, i.e.,
 	\begin{align}
 		\label{decorr}
		\tre\left[{\cal U}^{t_2}_{t_1}({\cal I}_{\rm s} \otimes{\cal D})\ {\cal U}^{t_1}_{t_0}\ \rhose(t_0)\right] =\ce^{t_2}_{t_0}\ \rho_s(t_0)\ ,		
 	\end{align}
where ${\cal I}_{\rm s}$ is the identity map on $\fbs$, and $\ce^t_{t_0}$ is the dynamical map for system evolution in Eq.~\eqref{def_dynamical_map}. Then there is no information backflow into the system via such a correlation. We will first consider two natural classes of completely decorrelating channels, giving rise to two corresponding definitions of quantum Markovianity. In addition, we will consider a decorrelating channel that does not remove all correlations between system and bath, but does remove all quantum correlations. This last is achieved by an entanglement-breaking channel, under which there may still be information backflow, but no {\it quantum} information backflow. This cannot be considered a definition of quantum Markovianity, but is an interesting concept in its own right, with relations to other concepts in our hierarchy.

Before proceeding, it is worth pointing out that the terminology `information flow' has been widely used in the study of quantum non-Markovianity, but based on a variety of different criteria. For example, in Ref.~\cite{BLP09}  a change in the distinguishability of system states has been interpreted as information flow, and quantified by trace distance. Entanglement measures~\cite{LFS12}, entropic measures~\cite{FKF14,HKSK14} and quantum Fisher information~\cite{LWS10} are also considered to quantify information flow. Moreover, an ensemble-based interpretation of information flow has been proposed in Ref.~\cite{BJA16}. However, these approaches are either a necessary consequence of, or equivalent to, divisibility of the dynamical map~(see Section~\ref{sec_divisibility}), which is very different from the explicit considerations of information in terms of environment interventions that we employ here. Indeed, these approaches do not rely on actual system--environment interactions and correlations at all. The hierarchy relations between them will be discussed in Sections~\ref{sub-hierarchy-EI} and~\ref{sec_eid}.

\subsubsection{Composability}
\label{sec_composability}

The first type of decorrelating channel we introduce is a~\emph{factorization channel}, denoted by $\cal F$, through which the environment is reset to the initial state~$\rhoe(t_0)$ modulo a local  unitary transformation (corresponding, for example, to an interaction frame). That is, there exists a unitary operator $\hat{W}_{t_0}^t \in \fue$ such that 
\begin{align}
\label{channel_f}
{\cal F}_t \, \hat X := \tre[\hat X]\,  \cw_{t_0}^t \rhoe(t_0) =: \tre[ \hat X ]\, \tilde{\rho}_{\rm e}(t)   
\end{align}
for all $\hat X \in \fbe$. The action of ${\cal F}_t$ on a joint system-environment state is thus 
\begin{equation}
\label{action_f}
  ( {\cal I}_{\rm s} \otimes {\cal F}_t) \, \rhose(t) = \rhos(t) \otimes \cw_{t_0}^t \rhoe(t_0) \ .
\end{equation} \blk
Such a resetting of the environment is closely related to the factorization approximation discussed in Section~\ref{sec_FA}, where the latter is equivalent to 
\[  \rhose(t) = ({\cal I}_{\rm s}\otimes {\cal F}_t) \, \rhose(t)   =  \rhos(t)\otimes \tilde\rho_{\rm e}(t) \ . \] 

The environment-reset channel clearly removes all correlations between the system and environment. Hence, as discussed at the beginning of this section, there can be no information backflow if the system evolution is invariant under such a reset. This corresponds to ${\cal D}= {\cal F}_{t_1}$ in Eq.~(\ref{decorr}),
leading to the following concept of quantum Markovianity, for which we have coined the term composability:   

\begin{tcolorbox}
\begin{df}[Composability]
\label{df_composability}
The OQS dynamics $\{\cu_{t_0}^t \}$ and $\rhoe(t_0)$ is composable if and only if for all times $t_2 > t_1 > t_0$, there exists a $\hat W_{t_0}^{t_1} \in \fue$ such that 
\begin{align}
	\label{com}
	{\cal E}_{t_0}^{t_2} \hat X = \tre \left[  \cu_{t_1}^{t_2}  ({\cal I}_{\rm s} \otimes { \cal F }_{t_1} )   \,\cu_{t_0}^{t_1} [ \hat X \otimes \rhoe(t_0) ]   \right]  
\end{align}
for all $\hat X \in \fbs$, where ${\cal F}_{t}$ is defined in Eq.~\eqref{channel_f}. 
\end{df}
\end{tcolorbox} 
The term `composability' is motivated by the more condensed but equivalent definition that
\begin{align}
\label{com_2t}
	{\cal E}_{t_0}^{t_2}  = \tce_{t_1}^{t_2} \  \ce_{t_0}^{t_1}  \ ,
\end{align}
where $\tce_{t_1}^{t_2}$ is the generalized dynamical map defined in Eq.~\eqref{general_dynamical_map} in terms of the unitarily evolved environment state $\tilde{\rho}_{\rm e}(t)$. Composability has a natural physical interpretation: that the system evolves {\em as if} the factorization approximation held. That is, it is possible to make the FA (\ref{def_FA}) at any time $t$, and the future evolution of the system will be the same as if that approximation had not been made. This does not mean that the FA must hold for composability to hold, which is fortunate because, as discussed in Section~\ref{sec_FA}, the FA is unphysical. In fact, composability follows naturally from PFI, which is physically reasonable. Thinking in terms of the train model discussed in Section~\ref{sec_PFI}, the unitary map ${\cal W}$ on the environment moves the train along so that the system continually interacts with a new part of the environment. The system becomes correlated with the past carriages but is uncorrelated with the future carriages. Imagine now a second train that begins in the same state, and moves in parallel and in synchrony, but does not interact with the system (this describes an environment evolving only under the unitary map ${\cal W}$). Now say the factorization channel ${\cal F}$ is applied to the first train at time $t$. This changes its state into that of the second train, enforcing the FA at that time. But this has no effect on the future (yet-to-interact) carriages, and the future evolution of the system will remain the same as before, as expressed by~\erf{com}.

By recursively applying Eq.~\eqref{com_2t} to a time sequence $t > t_n > \cdots > t_1 > t_0$, it follows that composability is equivalent to
\begin{equation}
\label{com_mt}
	\ce_{t_0}^t= \tce_{t_n}^t\ \tce_{t_{n-1}}^{t_n}\cdots\ \tce_{t_0}^{t_1} \ .
\end{equation}
This condition may be recognized as a special case of the general quantum regression formula as given in Eq.~(\ref{gqrfs1}), corresponding to ${\cal C}_j\equiv {\cal I}_{\rm s}$. Thus, composability is a weaker formulation of quantum Markovianity than GQRF.  Composability is of particular interest in that it is in fact rather centrally placed within the hierarchy of definitions of quantum Markovianity, as will be seen in Section~\ref{sec_hierarchy} (see also Fig.~\ref{fighie}).

\subsubsection{No  information backflow}
\label{sec_NIB}

While composability is a sufficient condition for no backflow of information from the environment to the system, it assumes a specific form, ${\cal F}$, for the decorrelating channel ${\cal D}$. Thus a natural generalization of the definition (which in this case means to weaken this concept) is to 
allow ${\cal D}$ to be \emph{any} channel that removes all correlations between the system and environment at time $t_1$. We call these {\it replacement channels}, meaning that there exists an environment state $\sigma_{\rm e}(t)$ such that 
\begin{align}
\label{channel_r}
	{\cal R}_t \, \hat X := \tre[\hat X]\, \sigma_{\rm e}(t) 
\end{align}
for all $\hat X \in \fbe$. The action of ${\cal R}_t$ on a joint system-bath state follows as 
\begin{equation}
\label{action_r}
	({\cal I}_{\rm s} \otimes {\cal R}_t ) \, \rhose(t)  =  \rhos(t) \otimes \sigma_{\rm e}(t)  \ .
\end{equation} \blk
Note that $\sigma_{\rm e}(t)$ here is an \emph{arbitrary} environment state, which is not necessarily unitarily equivalent to the initial state $\rhoe(t_0)$. Similarly to the factorization channel, the replacement channel also implies that the system evolution from time $t$ to any later time is described by a CPTP map (see Section~2). Replacing ${\cal D}$ by ${\cal R}_{t_1}$ in Eq.~\eqref{decorr}, we formulate the idea of no information backflow as follows:
\begin{tcolorbox}
\begin{df}[No Information Backflow]
\label{dfnib}
The OQS dynamics $\{\cu_{t_0}^t \}$ and $\rhoe(t_0)$  exhibits no information backflow if and only if for all times $t_2 > t_1 > t_0$, there exists an environment state $\sigma_{\rm e}(t_1)$ such that
\begin{align}
	\label{nib}
	{\cal E}_{t_0}^{t_2}  \hat X   = \tre \left[ \cu_{t_1}^{t_2} ( {\cal I}_{\rm s} \otimes {\cal R}_{t_1})  [\  \cu_{t_0}^{t_1} [ \hat X \otimes \rhoe(t_0) ] ] \right] \ 
\end{align}
for all $\hat X \in \fbs$, where ${\cal R}_{t}$ is defined in Eq.~\eqref{channel_r}. \blk
\end{df}
\end{tcolorbox} 

Note from this definition that one can define another CPTP map  ${\cal Q}_{t_1}^{t_2}$, with respect to $\sigma_{\rm e}(t_1)$, as
\begin{equation}
\label{nibdm}
	{\cal Q}_{t_1}^{t_2}  \hat X = \tre \left[  \cu_{t_1}^{t_2} \left[  \hat X \otimes \sigma_{\rm e}(t_1)  \right] \right] 
\end{equation}
for all $\hat X \in \fbs$ . A formula similar to Eq.~\eqref{com_2t} then naturally follows from Eq.~\eqref{nib}:
\begin{align}
\label{nib_2t}
	{\cal E}_{t_0}^{t_2} = {\cal Q}_{t_1}^{t_2} \  {\cal E}_{t_0}^{t_1} \ .
\end{align}
Furthermore,by recursively applying Eq.~\eqref{nib} for a time sequence $t > t_n > \cdots > t_1 > t_0$, it follows that, similarly to Eq.~\eqref{com_mt},
\begin{align}
\label{nib_mt}
	{\cal E}_{t_0}^{t} = {\cal Q}_{t_n}^{t} {\cal Q}_{t_{n-1}}^{t_n} \ldots  {\cal E}_{t_0}^{t_1} \ .
\end{align}
Thus no information backflow (NIB) captures, in a more general way than composability, the idea that information lost from the system, whether stored in the bath or in the system--bath correlation, should never affect the system dynamics at later times. In fact, it is the most general  characterization of this idea of quantum Markovianity. We note finally that the multi-time formula for NIB, Eq.~(\ref{nib_mt}), can be obtained as a special case of FDD, as per Eq~\eqref{def_ccdf}, by setting all system unitary operators $\hat V_{k}$ to the identity operator.

\subsubsection{No quantum information backflow}
\label{sec_NQIB}

The factorization and replacement channels remove all correlations, classical and quantum, between the system and the environment. However, it is of interest to consider the class of decorrelating channels that remove only the {\it quantum} correlations. By this we mean channels that remove all entanglement between the system and environment at a given time $t$.  These are just the well known entanglement-breaking channels~\cite{HSR03}, which we denote as ${\cal B}$, through which the combined state  is transformed into a separable state.  Such a channel has the general form of a measure-and-prepare operation on the environment~\cite{HSR03}: 
\begin{equation}
\label{channel_e}
	{\cal B}_t \hat X  :=  \sum_k \tre[\hat F_k(t) \hat X]\ \sigma_{\rm e}^k(t) 
\end{equation}
for all $\hat X \in \fbe$, with $\{ \hat{F}_k(t) \}$ being a positive-operator-valued measure (POVM) on the environment and $\sigma_{\rm e}^k(t)$ being the re-prepared environment state for measurement result~$k$. The action of this channel on a joint system--environment state is: 
\begin{align}
\label{action_e}
	({\cal I}_{\rm s} \otimes {\cal B}_t )  \rhose(t) = \sum_{k} p_k(t) \ \rhos^k(t) \otimes \sigma_{\rm e}^k(t) \ ,
\end{align}
where $p_k(t)$ and $\rhos^k(t)$ are implicitly defined via $p_k(t) \rhos^k(t) =\tre[ \rhose(t) \hat{F}_k(t) ]$. 

If the system evolution is not affected after applying channel ${\cal B}_t$, then,  following the same style of argument as for NIB, any {\it quantum} information that flowed from the system to the environment before time $t$ does not flow back after time $t$. Formally, we define no quantum information  backflow~(NQIB) as follows:
\begin{tcolorbox}
\begin{df}[No Quantum Information Backflow]
\label{def_NQIB}
The OQS dynamics $\{\cu_{t_0}^t \}$ and $\rhoe(t_0)$  exhibits  no quantum information backflow if and only if for all times $t_2 > t_1 > t_0$, there exists an entanglement-breaking channel ${\cal B}_{t}$  such that
\begin{align}
\label{eq:dfNQIB}
	\ce_{t_0}^{t_2} \hat X = \tre \big[  \cu_{t_1}^{t_2} \  ({\cal I}_{\rm s}\otimes {\cal B}_{t_1}) \  [ \cu_{t_0}^{t_1}  [  \hat X \otimes \rhoe(t_0) ] ] \big] 
\end{align}
for all $\hat X \in \fbs$, where ${\cal B}_{t}$ is defined in Eq.~\eqref{channel_e}.
\end{df}
\end{tcolorbox}

Note that NQIB does not prohibit classical information backflow, and hence we do not propose it as a good definition of Markovianity for quantum systems.\footnote{We note it could be argued that no quantum information backflow (NQIB) captures a  `quantum' concept of memorylessness, and  thus is related to quantum Markovianity. In particular, an OQS satisfying NIQB cannot access any quantum memory from its environment, but can access classical memory.} However, it is an obvious concept to consider, given NIB, and is related to other concepts of interest in the context of quantum Markovianity. In particular, the definition of NQIB, implies that the system dynamics is not affected, on average, by employing at least one sort of measurement on the environment.  That is, information about the system flowing to the environment could be collected by a measurement on the environment, providing better knowledge about the system state than the dynamical map provides, while remaining consistent with that map on average. Doing this repeatedly is the essential idea of quantum unravellings~\cite{CH09}, as will be discussed next.

\subsection{Quantum unravellings}
\label{sec_quantum_unravelling}

As discussed in Section~\ref{sec_oqs} and elsewhere, the loss of purity that is typical for the state $\rhos$ of an OQS is due to tracing over the environment. This amounts to throwing away any information that has leaked from the system into the environment (or into correlations with it). However, such information need not be thrown away. Rather, it is possible to perform measurements on the environment and thereby regain some, or even all, of the lost information, thereby producing a more pure system state conditioned upon that information. Moreover, as discussed in the preceding section, under a Markovian-type assumption it is possible to do this without altering the future evolution of the system, when averaging over the result obtained. Finally, it may be possible to repeat such measurements at arbitrarily close times (a process we call monitoring), thereby producing a stochastic {\em quantum trajectory}~\cite{Car93} for the conditional system state, while still not affecting the average evolution of the OQS. In the best situation, if the system begins in a pure state then its conditional states will remain pure. In this case we refer to the set of such quantum trajectories, for some particular choice of measurement, as a {\em pure-state unravelling} (PU) of the OQS dynamics. The idea is that the average evolution of $\rhos(t)$ can be regarded as a knitting together of pure state trajectories, which can be unravelled by a suitable monitoring of the environment~\cite{Car93,CW11}. 

One notable application for quantum unravellings is quantum feedback control technology, which plays an important role in engineering quantum systems~\cite{WM09}. Roughly, through the monitoring on the environment, the outward-flowing information is collected by the observer, who can estimate the system state in real time. This conditioned state represents the observer's best knowledge of the system, based on which the observer can design control signals for the system to manipulate it in an optimal way. Technically, this requires high-speed feedback such that the delay time in applying the control signal is much smaller than the response time of the system. This is feasible with some current technologies, and thus the idea of feedback control has been successfully applied to various different quantum systems~\cite{CIX11,RCD12,HDT12}.

The possibility of PU arises most naturally if the system and environment, in the absence of measurement, are in an entangled pure state. This will always be the case if the initial system and bath states are pure. Requiring $\rhos(t_0)$ to be pure is not a restriction on the OQS dynamics (these are defined by $\rhoe(t_0)$ and $\{\cu_{t_0}^t \}$), and is unavoidable if we want the conditioned system state to be pure at all times.  However, requiring $\rhoe(t_0)$ to be pure would be a significant restriction on the OQS dynamics. Fortunately, such a requirement is not strictly necessary. Because we are considering measurements on the environment, we can allow a measurement at time $t_0$ to put the environment into a pure state $\ket{\phi_{i}(t_0)}_{\rm e}$. If we
restrict  this measurement to being performed in the basis that diagonalizes the initial bath state, $\rhoe(t_0) = \sum_i \wp_i \ket{\phi_{i}(t_0)}_{\rm e}\bra{\phi_{i}(t_0)}$, then the measurement will not change this unconditioned environment state, and hence will not change the average evolution of the OQS. Any subsequent rank-one measurement (see e.g.~\cite{WM09}) of the environment will then collapse the system into a pure state $\ket{\psi_{r|i}(t)}_{\rm s}$: 
\beq \label{single-time-collapse}
{}_{\rm e}\bra{r}\hat{U}_{t_0}^t\ket{\psi(t_0)}_{\rm s}\ket{\phi_i(t_0)}_{\rm e} = \sqrt{p_{r|i}(t)}\,\ket{\psi_{r|i}(t)}_{\rm s}.
\eeq
Here $\{\ket{r}\}_r$ forms a basis for the environment, 
$r$ labels the result of the measurement in that basis at time $t$, 
and $p_r(t)$ is the probability of obtaining that result.  It is trivial to verify that 
the system state, averaged over all possible results $r$ and $i$, is the same as it would have 
been with no measurement: 
\beq
\sum_i \wp_i \sum_r p_{r|i}(t)  \ket{\psi_{r|i}(t)}_{\rm s}\bra{\psi_{r|i}(t)} 
= {\cal E}_{t_0}^t \ket{\psi(t_0)}_{\rm s}\bra{\psi(t_0)}.
\eeq

As the above argument shows, the existence of a random pure conditioned system state at an arbitrary 
time $t$, that reproduces the dynamical map from $t_0$ to $t$, is universal for open quantum systems. 
It therefore has no relation to Markovianity. 
But what we require for a pure {\em unravelling} is a 
much stronger condition: that it is possible to induce such a conditioned state 
for each of an arbitrary set of times $t_0 < t_1 \dots < t_n$, by measuring the environment at 
all these times, all the while leaving the average system state unchanged. 
Both  the \blk initial measurement, and any further measurements on the bath at later times, will, in general, have unpredictable outcomes $r_0, r_1, \dots , r_n$, 
\blk hence making the unravelling a stochastic process.  
We will denote the collection of results up to and including at a given time $t_j$ 
by a single index ${\bf r}_j = r_0, r_1, \dots , r_j$. 
Under the conditions being considered, this indexes an ensemble of pure states 
$\{\wp_{{\bf r}_j}(t_j)\, \hat\pi_{{\bf r}_j}(t_j)\}_{{\bf r}_j}$ at any of the 
times $t_j$. Here we use  
the notation $\hat\pi$ to represent a rank-one projector on $\hs$, \blk 
and the weighting   $\wp$  which multiplies it is the probability with which it occurs in the 
ensemble. 
 
To state our formal definition of pure unravellings, we introduce one more piece of notation,  $\cm_j = \{{\cal M}_{r_j}^j\}$, for the bath measurement at time $t_j$. Here ${\cal M}_{r_j}^j$ is an operation~\cite{WM09} (a completely positive trace-decreasing map on $\fbe$) that acts on the bath state when result $r_j$ is obtained at time $t_j$, reducing the trace of the state by a factor equal to the probability of that result being obtained. The corresponding POVM is given by $\{ {\cal M}_{r_j}^{j\dagger}\,\hat I\}$.  Note that the  POVM elements need not be rank-one. 
We refer to the collection $\{\cm_0, \cm_1,\cdots, \cm_n\}$ (which may be chosen adaptively, dependent upon the result of measurements at earlier times) as a \textit{measurement scheme}.

Since we are allowing non-projective measurements, we will  replace the earlier restriction, that the measurement at time $t_0$ be in the diagonal basis of $\rhoe(t_0)$, by the less restrictive requirement that it not change the initial state on average: ${\cal M}^0\rhoe(t_0) = \rhoe(t_0)$. Here ${\cal M}^j := \sum_{r_j}{\cal M}_{r_j}^j$ denotes the CPTP map corresponding to averaging over all  measurement results. 
 Thus we can formalize PU as follows: 
\begin{tcolorbox}
\begin{df}[Pure Unravelling] 
\label{dfpu}
The OQS dynamics $\{\cu_{t_0}^t \}$ and $\rhoe(t_0)$ admits a pure unravelling description if and only if, for any initial pure system state $\hat\pi(t_0) = \ket{\psi(t_0)}_{\rm s}\bra{\psi(t_0)}$, and \blk for any finite sequence of times $t_n >\cdots >t_1 >t_0 $, there exists at least one measurement scheme $\{\cm_0, \cm_1,\cdots, \cm_n\}$ on the environment  which,  at each time \blk $t_j$ in the sequence, induces a conditional pure system state  ensemble~$\{ \wp_{{\bf r}_j}(t_j) \,  \hat{\pi}_{{\bf r}_j}(t_j) \}_{{\bf r}_j}$   such that
	\begin{align}
	\label{eq:dfPU}
		 \sum_{{\bf r}_j} \wp_{{\bf r}_j} (t_j) \  \hat{\pi}_{{\bf r}_j} (t_j) = \ce_{t_0}^{t_j} \  \hat\pi(t_0) \ , 
	\end{align}
and for which ${\cal M}^0\rhoe(t_0) = \rhoe(t_0)$. Formally, a pure unravelling induces, for a given initial state $\hat\pi(t_0)$, an ensemble $\{\Upsilon(\hat{\pi}(t_0),{\bf r}_n)\}_{{\bf r}_n}$ of {\em quantum trajectories}, each of the form  
$$\Upsilon(\hat{\pi}(t_0),{\bf r}_n) \equiv [\hat{\pi}(t_0) \to \wp_{{r}_1}(t_1)\ \hat{\pi}_{{r}_1}(t_1) \to \ \dots \ \to \wp_{{\bf r}_j}(t_j)\ \hat{\pi}_{{\bf r}_j}(t_j) \to \ \dots\ \to 
	 \wp_{{\bf r}_n}(t_n)\ \hat{\pi}_{{\bf r}_n}(t_n) ]\ .$$  
\end{df}
\end{tcolorbox}

From the definition, it is easy to see that an OQS satisfying past--future independence immediately admits a pure unravelling description. Since a pure  unravelling, as just defined, allows one to conditionally purify the initial environment state, the PFI assumption means that at any time $t_j$ the system and the past part ${\rm p}_j$ of the environment  will be in a pure entangled state.  
Hence a suitable measurement on the ${\rm p}_j$  part of the environment will purify the system at time $t_j$. But this indirect extraction of information about the system can be performed without affecting the future ${\rm f}_j$ part of the environment, and hence will not affect the  the system's future dynamics on average.  

The idea of PU is also clearly related to that of NQIB, as discussed in the preceding section, since the measurements on the environment that project the system into a pure state at each time $t_j$ clearly breaks any entanglement between the system and the environment. Note, however, that for our concept of a pure-state unravelling it is {\em not} necessary to assume that the measurements performed on the environment realize an entanglement-breaking channel. The measurements must break entanglement for the actual system--bath state (which may be conditioned on past measurement results), but there is no requirement that it break the entanglement for an arbitrary system--bath state, as there would be if it were required to be an entanglement-breaking channel. The logical relation between PU and NQIB will be discussed in more detail in subsection~\ref{sub-unravellings-related}.

For the case of continuous time, we can consider the limit where $t_n$ is held fixed, equal to $T$, say, while $n$ is increased so that the time between measurements decreases towards zero. In this limit, we obtain the typical idea of a quantum trajectory: a conditioned system state that is always pure, and evolves stochastically in time. There are different types of such evolutions, including quantum jumps~\cite{CSVR89,DCM92,GPZ92,Bar93}, and quantum diffusion~\cite{Bel89,BS92,Car93,HM93}. Of course, in reality, any measurement on the bath yielding a meaningful result will take a finite time, referred to as the measurement time  $\tau_{\rm m}$.  Recall that for the system to be described by quantum white noise (see Section~\ref{sec_quantum_white_noise}), it is necessary that the bath-correlation time $\tau_{\rm e}$ be much shorter than relevant system time scales $\tau_{\rm s}$. In this case, a continuous-time limit of pure state unravellings makes sense as long as $\tau_{\rm e} \ll \tau_{\rm m} \ll \tau_{\rm s}$. The first of these inequalities  is necessary for the measurement not to disrupt the future evolution of the system, while the latter is necessary to be able to say that the system state is pure at all times. 

The reader has probably already noticed that the possible single-time conditioned system pure state in \erf{single-time-collapse} depends on the choice of basis $\{\ket{r}\}$ in which the environment is measured. This is the phenomenon ubiquitous to pure bipartite entangled states first noted by Einstein, Podolsky, and Rosen~\cite{EPR35} and subsequently called `steering' by \sch~\cite{Sch35}. For some OQSs, this dependence on the measurement choice can be generalized to multiple-time measurements, giving rise to different pure-state unravellings. For example, in the context of quantum optics, we can characterize different measurement schemes by the type of quantum trajectories they induce in the emitting system: counting photons in emitted field induces quantum jumps in the system, while homodyne detection (interfering the emitted field with a strong `local oscillator' field prior to detection) induces quantum diffusion~\cite{WM09} The latter example actually includes infinitely many different types of quantum diffusion, depending on the phase of the local oscillator field. 

For the purpose of relating to other concepts in our Markovian hierarchy, it is useful to introduce a separate concept from PU above, namely many pure unravellings~(MPU). By this we mean the existence of infinitely many distinct PUs, which we formalize as follows: 

\begin{tcolorbox}
\begin{df}[Many Pure Unravellings]\label{dfmpu}
	The OQS dynamics $\{\cu_{t_0}^t \}$ and $\rhoe(t_0)$ admits many pure unravelling descriptions if and only if PU is satisfied and, moreover, for some  
	initial system state $\hat{\pi}(t_0)$,  there exists an infinite-cardinality set \blk $\mathfrak{S}$ of pure unravellings, each element  of which induces an ensemble $\{\Upsilon(\hat{\pi}(t_0),{\bf r}_n)\}_{{\bf r}_n}$  of quantum trajectories $\Upsilon(\hat{\pi}(t_0),{\bf r}_n)$ (see Definition~\ref{dfpu}) which is distinct from the ensemble induced by every other unravelling in $\mathfrak{S}$. 
	%
	%
\end{df}
\end{tcolorbox}

As we will see in Section~\ref{sub-unravellings-related}, PU and MPU have different relations to concepts in our quantum Markovianity hierarchy. The latter, of course, is stronger, and relates to stronger concepts of quantum Markovianity. 
Another related concept, introduced in Section~\ref{sec_MCWF}, is that of the existence of a stochastic simulation scheme using pure states to efficiently simulate the dynamics of an OQS~\cite{DCM92,MC96}. 
This is a distinct concept because an ensemble of pure-state trajectories may be useful as a numerical tool for simulating an OQS without corresponding to anything physically realizable by a measurement scheme on the bath. By contrast, the physical realizability of quantum unravellings has technological applications (as discussed earlier), as well as being important conceptually. 

\subsection{Concepts deriving from the dynamical map}
\label{sub_Mathematician}

In the preceding subsections, we have explored  possible definitions of  quantum Markovianity based on the character of the interaction between the system and the environment, as defined by $\rhoe(t_0)$ and $\cu_{t_0}^t$. In this subsection we take a different approach, focusing on the mathematical structure of the dynamical map $\ce^t_{t_0}$, without any reference to physical details of the system--environment interaction. This approach has been more favoured by some groups of physicists, especially in mathematical physics and quantum information. It leads to definitions of quantum Markovianity that are mathematically simpler. However, they correspond to classical concepts which are strictly weaker than the classical definition of Markovianity in Eq.~\eqref{def_classical_Markovianity}.

\subsubsection{Divisibility}
\label{sec_divisibility}

Divisibility of a dynamical map corresponds to the property that the evolution of the system between two times $t_2\geq t_1$ can be described by a quantum information channel, i.e., by some CPTP map that takes any state $\rhos(t_1)$ at time $t_1$ to the corresponding state at time $t_2$. Thus, no memory about the system state prior to $t_1$ is required to describe its later evolution, and in this sense divisibility may be viewed as a Markovian property.

Wolf and Cirac originally considered only a particular subclass of divisible dynamical maps to be `Markovian' ~\cite{WC08}~(see also Section~\ref{sec-semigroup}). Rivas {\it et al.} subsequently suggested taking divisibility itself as a definition of quantum Markovianity~\cite{RHP10,RHP14}. Formally, divisibility is defined as follows:

\begin{tcolorbox}
\begin{df}[Divisibility]
\label{dfdivisibility}
The dynamical map  $\{{\cal E}_{t_0}^t\}$  is divisible if and only if for all times $t_2 \geq t_1 \geq t_0$, there exists a CPTP map $\cq_{t_1}^{t_2}$ such that
\begin{align}
\label{divisible}
	{\cal E}_{t_0}^{t_2}  = {\cal Q}_{t_1}^{t_2} \  {\cal E}_{t_0}^{t_1}  \ .
\end{align}
\end{df}
\end{tcolorbox}

It is seen that divisibility formally generalizes the notion of composability in Eq.~\eqref{com}. Recursively applying this definition to consecutive times \ $t_n>t_{n-1}>\dots>t_1>t_0$, it further follows that divisibility is equivalent to the existence of a corresponding set of CPTP maps $\{ \cq_{t_i}^{t_{i+1}} \}_{i=1}^{n-1}$ satisfying
\begin{equation}
\label{mdiv}
	\ce_{t_{0}}^{t_{n}} = \cq_{t_{n-1}}^{t_{n}} \ \cq_{t_{n-2}}^{t_{n-1}} \cdots\ \ce_{t_0}^{t_{1}}  \ ,
\end{equation}
similarly to Eq.~\eqref{com_mt} for composability and Eq.~(\ref{nib_mt}) for no information backflow (where the latter equations further restrict the form of $\cq_{t_i}^{t_{i+1}}$).

Finally, to avoid possible confusion, it  is worth mentioning that  divisibility of the dynamical map should  be distinguished from the related  property of \emph{infinite divisibility}~\cite{WC08,HA01}:  a CPTP map ${\cal M}$ is infinitely divisible if and only if for every positive integer $n$  it has the form ${\cal M} = ({\cal M}_n)^n$, for example, when the dynamical map corresponds to a continuous-in-time dynamical semigroup (defined below).

\subsubsection{Dynamical semigroups}
\label{sec-semigroup} 

Up until the 1970s, the study of the dynamics of open quantum systems was dominated by the use of time-homogeneous master equations. In this context, a concept of quantum Markovianity which was natural to consider was that of the \emph{dynamical semigroup}. This terminology was introduced by Kossakowski {\it et al.}~\cite{KA72,GKS76}, but the same idea was employed under different names at that time. For example, Davies~\cite{DE69} called processes with this property ``quantum stochastic processes'' and Accardi~\cite{AL76} called them ``stationary noncommutative Markov processes''. In this report, we follow Ref.~\cite{GKS76} in referring to a \emph{one parameter} semigroup as a dynamical semigroup, but generalize such semigroups to include discrete times, as follows:

 \begin{tcolorbox}
\begin{df}[Dynamical semigroups]
\label{df:Dynamical semigroups}
The dynamical map $\{ \ce_{t_0}^t\}$ form a dynamical semigroup if and only if  for any $r,s \geq 0$, it satisfies
 \begin{align}
 \label{semigroup_con}
 	\ce_{t_0}^{t_0 + r + s} = \ce_{t_0}^{t_0 + r} \ce_{t_0}^{t_0 + s} \ .
 \end{align}
\end{df}
 \end{tcolorbox}

In particular, noting for $t=t_0$ that the dynamical map reduces to ${\cal E}_{t_0}^{t_0} = {\cal I}_{\rm s}$, i.e., to the identity map on $\fbs$, the composition condition~\eqref{semigroup_con} implies that ${\cal S}_t : = \ce_{t_0}^{t_0 + t}$ generates a one-parameter commutative semigroup $\{ {\cal S}_t \}$, with
\begin{align}
\label{semigroup_con2}
	{\cal S}_{t_1 + t_2} = {\cal S}_{t_1} {\cal S}_{t_2} = {\cal S}_{t_2} {\cal S}_{t_1}\ , \qquad {\cal S}_0 = {\cal I}_{\rm s} \ .
\end{align}
Note that Definition~\ref{df:Dynamical semigroups} also includes discrete time dynamics, with a fixed time interval $\Delta$ between evolutions such that $t_k-t_0 = k \Delta \in \{0,\Delta, 2\Delta, \dots\}$, with $k \in {\mathbb Z}^{+}$. Letting ${\cal S}:={\cal S}_{\Delta}$, it then follows that Eq.~\eqref{semigroup_con2} is satisfied with
\begin{align}
	{\cal S}_{t_n} = ({\cal S})^n \, .
\end{align}

Another way to state the concept of a dynamical semigroup is that the evolution of the system state is time-homogeneous: $\rhos(t_2) = \ce_{t_0}^{t_0 + (t_2 - t_1)} \rhos(t_1)$. 
Some mathematical physicists have regarded the semigroup property as a definition of quantum Markovianity. For example, both Davies~\cite{DE69} and Kossakowski~\cite{KA72} treated Eq.~\eqref{semigroup_con} as the quantum generalization of the classical Chapman-Kolmogorov equation (see also Section~\ref{sec_cke}), with Kossakowski explicitly identifying it as the condition for a ``quantum Markov process''. 

\subsubsection{GKS-Lindblad-type master equations}
\label{sec_ME} 

For the case that the semigroup in Definition~\ref{df:Dynamical semigroups} is continuous in time, one can take $t_1=t$ and $t_2=dt$ in Eq.~(\ref{semigroup_con2}) to give $\frac{{\rm d} {\cal S}_t}{{\rm d} t} = {\cal L}\, {\cal S}_t$,
whenever the limit ${\cal L}:=\lim_{\tau\rightarrow 0^+} ({\cal S}_\tau-{\cal I}_{\rm s})/\tau$ exists (see, e.g., Ref.~\cite{Lin76} for sufficient conditions; a simple example where these conditions fail is given in Section~\ref{sec-semi-ME}). The time-independent linear map ${\cal L}$ on $\fbs$ is called the \emph{generator} of the semigroup (since ${\cal S}_t=e^{{\cal L} t}$).  The system state ${\rhos} (t) ={\cal S}_{t-t_0} {\rhos} (t_0)$ then obeys the  time-homogeneous  master equation 
\begin{align} \label{generator2}
		\frac{{\rm d} \rhos(t)}{{\rm d} t} = {\cal L} \,\rhos(t)\ .
\end{align}
The explicit form of ${\cal L}$ for such master equations was characterised by Gorini {\it el al.} for finite dimensional systems~\cite{GKS76}, and by Lindblad for both finite and infinite dimensional systems (restricted to a bounded generator)~\cite{Lin76}. For this reason, ${\cal L}$ is often called the Lindblad superoperator, or the Lindbladian. The equation for $\rhos(t)$ is, correspondingly, a \emph{strict GKS-Lindblad master equation}: 
\begin{tcolorbox}
\begin{df}[Strict GKS-Lindblad master equation]
\label{df:Strict GKS-Lindblad master equation}
The dynamical map $\{ \ce_{t_0}^t\}$ admits a strict GKS-Lindblad master equation if and only if the dynamics of the system can be written as
\begin{align}
\label{strictLindblad}
	\frac{{\rm d} \rhos(t)}{{\rm d} t} = {\cal L} \rhos(t) := - i [ \hat H, \rhos(t)] + \sum_k {\cal D}[ \hat A_k ] \rhos(t) \,
\end{align}
where $\hat H$ is an Hermitian operator (the effective system Hamiltonian) and $\hat A_k$ are coupling operators (also called Lindblad operators). The superoperator $\cal D$ is defined as~\cite{WM09}
\begin{align}
\label{superD}
	{\cal D}[ \hat X] \rho : = \hat X \rho \hat X^\dagger - \smallfrac{1}{2} ( \hat X^\dagger \hat X \rho + \rho \hat X^\dagger \hat X )
\end{align}
for all operators $\hat X \in \fbs$.
\end{df}
\end{tcolorbox}
\blk
Equations of the form of \erf{strictLindblad} have a long history~\cite{Zwan64,Red65,GG68} in the study of 
open quantum systems prior to the work of Gorini {\it el al.} and Lindblad in 1976. Under the type of `Markovian' approximations discussed in Section~\ref{sec_FA}, together with the so-called rotating-wave approximation~\cite{SZ97}, the standard techniques for deriving a time-independent master equation for $\rhos(t)$ typically yield an equation of the GKS-Lindblad form. 
However, there are exceptions. For instance, the so-called quantum Brownian motion master equation~\cite{Sch61} is a time-independent equation for $\rhos(t)$ which is not of the GKS-Lindblad form. It can be proven that all such exceptions 
correspond to evolution that is not completely positive~\cite{GKS76,RH12}.

In its most general conception, however, a quantum master equation is any time-differential form of the dynamical map of an OQS. Strict GKS-Lindblad equations are thus only one class of master equations. In the generic case, following the method developed by Nakajima~\cite{NS58} and Zwanzig~\cite{RZ60}, a master equation for an OQS can be formally written as 
\begin{align}
\label{general_ME}
	\frac{{\rm d} \rhos(t)}{{\rm d} t} =   \int_{t_0}^t {\cal K}(t, \tau)   \rhos(\tau) {\rm d} \tau  \ ,
\end{align}
where ${\cal K}(t,\tau)$, the kernel superoperator on $\fbs$, can obviously include non-Markovian memory effects, if any, on the system dynamics. Under certain conditions, Eq.~\eqref{general_ME} can be reduced to a time-local differential equation: 
\begin{align}
\label{timelocal}
	\frac{d\rhos(t) }{dt} = {\cal L}_t  \rhos(t)\ ,
\end{align}
where ${\cal L}_t$ is a (in general time-dependent) superoperator on $\fbs$ \cite{STH77,CS79,EJM07,CK10}. A simple sufficient condition for the existence of such a form is that for all $t$, $\ce^t_{t_0}$ is differentiable and has a left-inverse $(\ce^t_{t_0})^{-1}$.  Then, since $(\ce^t_{t_0})^{-1}\ \ce^t_{t_0}={\cal I}_{\rm s}$, this implies that ${\cal L}_t=(d\ce^t_{t_0}/dt)\ (\ce^t_{t_0})^{-1}$.  Following the methods of Gorini {\it et al.} for dynamical semigroups \cite{GKS76}, it may be shown that any time-local master equation can be put in the `canonical' form~\cite{MJL14}
\begin{equation}\label{TDME}
	\frac{d\rhos}{dt} = -i [\hat H(t), \rhos]  + \sum_k \gamma_k(t) {\cal D} [ \hat C_k ] \rhos 
\end{equation}
where $\hat H(t) \in \fbs$ is Hermitian, the $ \hat C_k(t)$ are traceless orthonormal operators on $\fbs$, i.e., 
\begin{align}
	\tr [ \hat{C}_k (t) ] = 0\ ,\qquad	\tr [  \hat{C}_j(t)^\dagger \hat{C}_k(t) ] = \delta_{jk} \ ,
\end{align} 
and the $\gamma_k(t)$ are  real numbers, called the canonical decoherence rates. The set of canonical decoherence rates is uniquely determined by the dynamical map (and is invariant under any unitary transformation $\rhos(t)\rightarrow {\cal V}(t)\rhos(t)$), and has been shown to be valuable in characterising various measures of quantum non-Markovianity~\cite{MJL14} (see also Section~\ref{sec_conclusion}). Note that these `rates' need not be nonnegative, even though $\ce^t_{t_0}$ is a completely positive map~\cite{MJL14,EJM07,MH08}. However, for the time-independent case, with $\gamma_k(t)\equiv \gamma_k(t_0)$, the rates are always nonnegative and the time-local master equation~(\ref{TDME}) reduces to a strict GKS-Lindblad master equation as per Definition~\ref{df:Strict GKS-Lindblad master equation}.

This suggests a natural generalization of strict GKS-Lindblad equations, to a larger class of master equations that some  consider to be Markovian (see, e.g., Ref.~\cite{RHP14,BLP16}): 
\begin{tcolorbox}
\begin{df}[Time-dependent GKS-Lindblad equation] 
\label{df:time-dependent Lindblad equation}
	The dynamical map $\{ \ce_{t_0}^t\}$ admits a time-dependent GKS-Lindblad master equation if and only if the evolution is described by an equation of the form of \erf{TDME} for which the canonical decoherence rates  are nonnegative, i.e., 
	\begin{equation}
		\label{decoh}
		\gamma_k(t) \geq 0
	\end{equation}
	for all $k$ and times $t\geq t_0$.
\end{df}
\end{tcolorbox}
If this concept of Markovianity is adopted, an OQS with at least one negative canonical decoherence rate, $\gamma_k(t)<0$, then corresponds to a form of quantum non-Markovianity. Note that, since a time-dependent GKS-Lindblad equation generates a completely positive map over any infinitesimal time interval \cite{GKS76,MJL14}, it is natural to regard it as the differential form of divisibility as per Eq.~(\ref{divisible}). \blk

\subsubsection{System state distinguishability}
\label{sec-system-state-dist}

Breuer {\it et al.} have suggested defining the evolution of an OQS to be Markovian if and only if  system states become less distinguishable over any time period \cite{BLP09,BLP16}.  Noting that the maximum probability of correctly distinguishing between any two quantum states $\rho$ and $\sigma$,  having respective prior probabilities $w$ and $1-w$, is given by the Helstrom formula $P_{\rm disting} = \half\left(1+ \tr \left|w\rho-(1-w)\sigma\right| \right)$\cite{HC69}, this leads to \cite{BLP16,CDA11}: 

\begin{tcolorbox}
\begin{df}[Decreasing system distinguishability]
\label{dfdisting}
The dynamical map $\{\ce_{t_0}^t\}$ satisfies decreasing system distinguishability if and only if the maximum probability of correctly distinguishing between any two system states is non-increasing in time, i.e., if and only if 
	\begin{align}
		\label{disting}
		\tr\left|w\rhos(t_1)-(1-w)\rhos'(t_1)\right|\geq \tr\left|w\rhos(t_2)-(1-w)\rhos'(t_2)\right| 
	\end{align}
for all $w\in(0,1)$, \blk initial states $\rhos(t_0)$, $\rhos'(t_0)$, and times $t_2>t_1\geq t_0$.
\end{df}
\end{tcolorbox}

The case of fixed $w=\half$ was initially considered by Breuer {\it et al.}~\cite{BLP09},  and gives a criterion for decreasing system distinguishability in terms of decreasing trace distance which is easy to evaluate.  Decreasing system distinguishability \blk has been broadly interpreted in terms of `information' flowing from the system into its environment, but not vice versa, implying that the environment cannot store and later feed back information to the system about its earlier history \cite{BLP09,BLP16}.  
Note, however, that Definition~\ref{dfdisting} is, as we will see, strictly weaker than our more general Definition~\ref{dfnib} of no information backflow (NIB) in  section~\ref{sec_NIB}.

A stronger argument for considering decreasing system distinguishability as a signature of quantum Markovianity is that, similarly to divisibility, it mirrors a related property of classical Markovianity~\cite{VSL11}. In particular, for any classical Markovian process as per Eq.~(\ref{def_classical_Markovianity}), the probability of correctly distinguishing between any two probability densities, $P(\bm x_{t})$ and $P'(\bm x_{t})$, at a given time $t$,  is nonincreasing with time~\cite{VSL11}.
Hence, just as an increase in distinguishability for the classical case implies classical non-Markovianity (although not vice versa), an increase of system state distinguishability in the quantum case may be taken to imply quantum non-Markovianity (but, again, not vice versa).


\subsubsection{Monte Carlo wave function simulations}
\label{sec_MCWF}

The concept of pure-state unravellings (see Section~\ref{sec_quantum_unravelling}) was introduced almost simultaneously with the closely related concept of the Monte Carlo wave function (MCWF) simulation method~\cite{DCM92,DumZolRit92} (see Ref.~\cite{Wis96} for a review). The former concept (unravellings) emphasized that the stochastic quantum trajectory for a pure state represented the evolution of its state conditioned on an ideal continuous measurement. The latter (MCWF) instead emphasizes the use of an ensemble of such trajectories to reproduce the properties of the system state $\rhos(t)$ at any time. The MCWF method was originally introduced~\cite{DCM92,DumZolRit92,MolCasDal93} to simulate systems obeying a (possibly time-dependent) GKS-Lindblad-type master equation such as \erf{TDME}. The solution $\rhos(t)$ of the master equation is approximated by the ensemble average ${\rm E}[\ket{\psi(t)}\bra{\psi(t)}]$ over a finite number $M\gg 1 $ of numerical realizations of the stochastic pure-state evolution. 
The original MCWF method~\cite{DCM92,DumZolRit92} involved $\ket{\psi(t)}$ undergoing smooth evolution interspersed with quantum jumps, which corresponds (in quantum optics) to a photon counting unravelling as discussed in Section~\ref{sec_quantum_unravelling}. However, it is equally possible, for any GKS-Lindblad-type master equation, to use a MCWF method with everywhere continuous but non-differentiable evolution for $\ket{\psi(t)}$, corresponding to the diffusive unravellings discussed in Section~\ref{sec_quantum_unravelling}.

It might seem counter-productive to replace a single deterministic equation by an ensemble of stochastic equations. However, there 
can be advantages for systems which require a Hilbert space $\hs$ of large dimension $N$ in order to be represented accurately. 
In such cases, storing the state matrix $\rhos$ requires of order $N^{2}$ real numbers, whereas storing the state vector $\ket{\psi}$ requires only of order $N$. Because of the stochastic evolution of $\ket{\psi(t)}$, further advantages also exist in some cases, via the {\em moving basis} method~\cite{RTI96}. 
This uses the fact that $\ket{\psi(t)}$ at any given time $t$ may be describable using a smaller Hilbert space, of dimension $n$, with the (often small) additional expense of changing the Hilbert space over time as $\ket{\psi(t)}$ changes. For large $N$, the time to directly compute the evolution of the state matrix via the master equation can scale as badly as $N^{4}$. This compares to the time to compute the ensemble of state vectors via the quantum trajectory which scales, at worse, as $n^2 M$, or just $n^2$ if parallel processors are available. Even though one requires a number of realizations $M\gg 1$, reasonable results may be obtainable with $M \ll (N/n)^2$. For extremely large $N$ it may be impossible even to store the state matrix on an available computer. In this case the MCWF method may still be useful,  if one only wishes to calculate certain system averages, rather than the entire state matrix, via ${\rm E}[\bra{\psi(t)}\hat A\ket{\psi(t)}] = {\rm Tr}[\rhos(t)\hat A]$. Areas where this technique has great application include the quantized motion of atoms undergoing spontaneous emission~\cite{DumZolRit92}; analysis of laser cooling schemes~\cite{CM95}; and simulations of many-body quantum systems~\cite{FMN01}.

Although the MCWF method was originally developed for GKS-Lindblad-type master equations, it has been adapted in various ways for more general (``non-Markovian'') evolution~\cite{IA94,DGS98,YDGS99,GamWis02b,Breuer04b,Piilo08,Suess14,QYBY14}. Using the same notation as that for PU (see Section~\ref{sec_quantum_unravelling}), we can describe the general MCWF method by the following:
\begin{tcolorbox}
\begin{ds}[Monte Carlo wave function simulation] 
\label{def-MCWFS}
The dynamical map  $\{ \ce_{t_0}^t\}$ can be simulated by a Monte Carlo wave function (MCWF) method if,  for any initial pure system state $\hat\pi(t_0) = \ket{\psi(t_0)}_{\rm s}\bra{\psi(t_0)}$, and for any finite sequence of times $t_n >\cdots >t_1 >t_0 $, it is possible to numerically generate, in parallel, an ensemble of $M$ pure state trajectories 
$\Upsilon(\hat{\pi}(t_0),{\bm r}_n)$,  each of the form  
\begin{align*}
\Upsilon(\hat{\pi}(t_0),{\bm r}_n) \equiv [& \hat{\pi}(t_0) \to \wp_{{r}_1}(t_1)\ \hat{\pi}_{{r}_1}(t_1) \to \ \dots \ \to \wp_{{\bm r}_j}(t_j)\ \hat{\pi}_{{\bm r}_j}(t_j) 
\to \ \dots\ \to 
	 \wp_{{\bm r}_n}(t_n)\ \hat{\pi}_{{\bm r}_n}(t_n) ]\ ,	
\end{align*}
such that,  at each time $t_j$ in the sequence, 
\begin{align}
	\lim_{M \to \infty} {\rm E}_M[  \hat{\pi}_{{\bm r}_j} (t_j) ]  =  \ce_{t_0}^{t_j} \ \hat\pi(t_0) \ ,
\end{align}
where $E_M[\cdot]$ denotes an equally-weighted average over the numerically-generated trajectories.
Moreover, the method of generation cannot rely on explicit knowledge of the ensemble average, $\rhos(t) \equiv \ce_{t_0}^{t} \ \hat\pi(t_0)$. 
\end{ds}
\end{tcolorbox}
The last condition is necessary because otherwise it would be completely trivial to perform a MCWF simulation: simply diagonalize $\rhos(t_j)$ as an ensemble $\{ {\wp_{r_j}(t_j) , \hat \pi_{r_j}(t_j)}\}$, 
for each $t_j$, and choose $\hat\pi_{{\bf r}_j}(t_j)=\hat\pi_{r_j}(t_j)$  and 
$\wp_{{\bf r}_j}(t_j) = \wp_{r_1}(t_1)\times \wp_{r_2}(t_2) \times \dots \times \wp_{r_j}(t_j)$.  It may not always be obvious, given a simulation method, 
to say whether it relies on explicit knowledge of $\rhos(t)$. 
This is one of the reasons we 
have called the above a {\em description} rather than a {\em definition} of the MCWF 
method, as reflected by the shape of the box for this concept in Fig.~\ref{fighie}. 
Another reason is that some generalizations of the MCWF method involve an 
auxiliary quantum system~\cite{IA94,Breuer04b}, and it is a matter of 
taste as to whether it should be necessary to generate, actually or potentially, 
a pure state of the system as initially defined in order to call the technique 
a MCWF method. 

Despite the resultant vagueness in our characterization of the MCWF concept, 
we believe it is worth introducing in the context of a discussion of Markovianity and non-Markovianity, for two reasons. First, it is implied by concepts we have already introduced, namely divisibility (Section~\ref{sec_divisibility}) and the existence of a pure-state unravelling (Section~\ref{sec_quantum_unravelling}), as will be shown in Sections~\ref{sec_hie_divisibility_ME} and~\ref{sec-QWN-ME-PU}, respectively. 
Thus it links two branches of our hierarchy.  However, it implies 
neither of those two concepts.  In particular, it does not imply the existence 
of a pure-state unravelling, and so is a distinct concept from PU. This last point 
(our second reason) is often overlooked or misunderstood, and is 
of great importance for interpreting pure-state 
trajectories in the non-Markovian case~\cite{GamWis03,GamWis04,HJ08}. 

Finally, we note an additional subtlety: very soon after the introduction of the MCWF method for Lindblad-type master equations it was noted that they could be used not only for simulating 
the properties of $\rhos(t)$ at a single time, but also for calculating two-time correlation functions and spectra (Fourier transforms two-time correlation functions)~\cite{MDT93,BKP97,PK98}. 
This required the generalization of evolving for each trajectory, 
in parallel with $\ket{\psi(t)}$, an additional state-vector. However, it was implicit in those papers that the quantum regression formula (QRF) holds. Thus, if we were to include this generalization as part of the MCWF method we could no longer say that MCWF is implied by divisibility; we would need the much stronger (and qualitatively different) assumption of QRF, from Definition~\ref{def_QRF}. Whether this form of MCWF would still be implied by PU we leave as an open question.

\section{Building the hierarchy}
\label{sec_hierarchy}

With all Markov-related concepts well defined in Section~\ref{sec_fqnm}, we are now ready to provide a hierarchy view on quantum non-Markovianity. To obtain an overall impression, we summarize all relations between these concepts in Fig.~\ref{fighie}. This hierarchy figure consists of many blocks representing different concepts. Concepts bearing certain physical similarities are coloured in comparable hues. All these concepts fall into two categories: those based on modelling the interaction between the system and the bath, and those deriving from the dynamical map, as we have discussed at length in previous sections.

The following discussions on the hierarchy of quantum non-Markovianity are thus mainly divided into two parts: the system--environment approach (Section~\ref{sub:the_system_en}) and the dynamical map approach~(Section~\ref{sub:the_dynamical_approach}). The bridging of the two approaches then follows (Section~\ref{sec:bridge}).
We provide analytical proofs for logical relations between these concepts, and counterexamples proving the nonexistence of converse relations where possible. We discuss conjectures for several unresolved relations in~\ref{app-conj}.

\subsection{The system--environment approach} 
\label{sub:the_system_en}

In accordance with the hierarchy shown in Figure~\ref{fighie}, we first go through the system--environment approach, starting with the factorization approximation. 

\subsubsection{Factorization approximation and the quantum regression formula} 
\indent

\begin{theo}[label = Fa-QRF]{}{}

	The factorization approximation (FA) in Definition~\ref{def_FaA} is a sufficient but not necessary condition for the quantum regression formula (QRF) in Definition~\ref{def_QRF}. 
\end{theo}

\begin{proof}
	We first show the sufficient direction always holds. Given any any two system operators $\hat A$ and $\hat B$, the correlation function $\la \check B(t_2) \check A(t_1)  \ra$ may be rewritten, using the FA $\rhose(t_1) = \rhos(t_1) \otimes \trhoe(t_1) = \rhos(t_1) \otimes (\cw_{t_0}^{t_1} [\rhoe(t_0)] )$ in Eq~\eqref{def_FA}, as 
	\begin{align}
		\la \check B(t_2) \check A(t_1)  \ra & = \trse \left[  (\cu_{t_0}^{t_2 \dagger} \hat B) (\cu_{t_0}^{t_1 \dagger} \hat A ) \rhose(t_0)   \right] \nonumber \\
		& = \trse \left[  \hat B \hat U_{t_1}^{t_2} \hat A \rhose(t_1) \hat U_{t_1}^{t_2 \dagger}  \right] \overset{\rm FA}=
		\trse \left[  \hat B \hat U_{t_1}^{t_2} \hat A \rhos(t_1) \otimes \trhoe(t_1) \hat U_{t_1}^{t_2 \dagger}  \right]   \nn \\
		& = \trs \left[  \hat B \tce_{t_1}^{t_2} \hat A \rhose(t_1) \right]  \label{fa-to-qrf} \ .
	\end{align}
	Here $\cu_{t_0}^{t_1 \dagger}$ is defined to be the dual map of $\cu_{t_0}^{t_1}$, i.e., $\cu_{t_0}^{t_1 \dagger}\hat X= \hat U_{t_0}^{t_1 \dagger} \hat X \hat U_{t_0}^{t_1 }$ for all $\hat X \in \fbs$. A similar proof holds for the correlation function of $\la \check A(t_1) \check B(t_2) \ra$ and is given in~\ref{app-fa}. 
	
	The necessary direction, however, does not hold in general. An explicit counter-example can be found in a model introduced by Accardi, Frigerio, and Lewis~\cite{AFL82}, which we thus call  the AFL model.  We present this model, in the form in which it was recently rediscovered~\cite{AHFB14},  in~\ref{app-AFL}, where we show it satisfies QRF, but fails FA. Furthermore, as we later show in the proof of Theorem~\ref{PFI-GQRF}, PFI also serves as a counter-example here. That is, an OQS satisfying PFI will also satisfy QRF, but does not necessarily satisfy FA.
\end{proof}

\begin{theo}[label = FA-GQRF]{}{}
	For an OQS with a finite dimensional Hilbert space, the factorization approximation (FA) in Definition~\ref{def_FaA} is a sufficient but not necessary condition for the  generalized quantum regression formula (GQRF) in Definition~\ref{def_GQRF}.
\end{theo}

\indent

Before going to the proof, we first need to point out a limitation of FA (that it only applies to evolution from $t= t_0$), and how the finite dimensional Hilbert space condition of the theorem overcomes this limitation. First, for any system operator $\hat X \in \fbs$, it follows from FA that, for any $t_2 \geq t_1\geq t_0$,
\begin{align}
	\cu_{t_0}^{t_2} [ \hat X \otimes \rhoe(t_0) ] = \cu_{t_1}^{t_2} [ ( \ce_{t_0}^{t_1} \hat X ) \otimes \trhoe(t_1) ] = \ce_{t_0}^{t_2} \hat X \otimes \trhoe(t_2) \label{fa-dim} \ .
\end{align}
The second equality is not in general, however, equivalent to the property
\begin{align}
	\label{gfa}
	\cu^{t_2}_{t_1} [\hat X \otimes \trhoe(t_1)] = \tce^{t_2}_{t_1} \hat X \otimes \trhoe(t_2) \ ,	
\end{align}
which we will call the generalized factorization approximation (GFA). For example, the image of the dynamical map, $\{\ce_{t_0}^{t_1} \hat X\}$, can be a strictly lower-dimensional subset of $\fbs$.  However, we can show that  GFA does hold whenever the dynamical map is invertible, and that this is always the case for a finite dimensional system Hilbert space when FA holds. 

First, define  the `time-reversed' linear map
\begin{align}
	\label{fa-dim-inv1}
	\bar{\ce}_{t_1}^{t_2} \hat X := \tre \left[   \cu_{t_1}^{t_2\dagger}[ \hat X \otimes \trhoe(t_2)  ]  \right]
\end{align}
for all $\hat X \in \fbs$ and $t_2\geq t_1\geq t_0$. It then follows from Eq.~\eqref{fa-dim} that
\begin{align}
	\label{fa-dim-inv2}
	\ce_{t_0}^{t_1} = \bar{\ce}_{t_1}^{t_2} \ce_{t_0}^{t_2} \ .
\end{align}
Letting $t_1 = t_0$, Eq.~\eqref{fa-dim-inv2} further implies 
\begin{align}
	\bar{\ce}_{t_0}^{t_2} \ce_{t_0}^{t_2} = {\cal I}_{\rm s} \ .
\end{align}
That is, FA implies the existence of a left inverse for the dynamical map $\ce_{t_0}^{t}$. If the condition of a finite Hilbert space condition is further imposed, $\bar{\ce}_{t_0}^{t}$ must also be a right inverse~\cite{AR05}, i.e., $\ce_{t_0}^{t}$ is invertible.  

Second, another important observation we need from Eq.~\eqref{fa-dim} is that FA implies composability in Definition~\ref{df_composability}. In fact, by tracing over the bath on both sides of  the second equality of Eq.~\eqref{fa-dim}, we have
\begin{align}
	\tce_{t_1}^{t_2} \ce_{t_0}^{t_1} = \ce_{t_0}^{t_2} \ .
\end{align}
Finally, it then follows when the dynamical map is invertible that
\begin{align}
	\cu_{t_1}^{t_2}  [ \hat X \otimes \trhoe(t_1) ] 
	& = \cu_{t_1}^{t_2} [ \ce_{t_0}^{t_1} \bar{\ce}_{t_0}^{t_1} \hat X  \otimes \trhoe(t_1)] 
	= \ce_{t_0}^{t_2} \bar{\ce}_{t_0}^{t_1} \hat X \otimes \trhoe(t_2) \\
	& = \tce_{t_1}^{t_2} \ce_{t_0}^{t_1} \bar{\ce}_{t_0}^{t_1} \hat X \otimes \trhoe(t_2) \\
	& = \tce_{t_1}^{t_2} \hat X \otimes \trhoe(t_2) \ 
\end{align}
for all $\hat X \in \fbs$, which is the desired GFA property in Eq.~\eqref{gfa}.
\blk
We can now proceed to the proof of Theorem~\ref{FA-GQRF}.

\begin{proof}
	We show the sufficient direction for a particular case of four-time correlation functions. The derivation is straightforward to generalize to the general case and to higher orders. Given any system operators $\hat A$, $\hat B$ and $\hat C$, $\hat D$, the calculation of  the correlation function $\la \check C(t_2) \check D(t_3) \check B (t_1) \check A(t_0) \ra$ follows under the FA as
	\begin{align}
		\la \check C (t_2)  \check D (t_3)  \check B (t_1)  \check A (t_0) \ra  
		= \ & \tr_{\rm se} \big[ (\cu_{t_0}^{t_2\dagger} \hat C ) (\cu_{t_0}^{t_3\dagger}  \hat D ) (\cu_{t_0}^{t_1\dagger} \hat B ) \hat A \rhose(t_0) \big] \nonumber \\ 
		= \  & \tr_{\rm se} \bigg[ \hat U_{t_1}^{t_2\dagger} \hat C \hat U_{t_2}^{t_3\dagger}  \hat D \hat U_{t_1}^{t_3}  \hat B \, \cu_{t_0}^{t_1} [ \hat A \rhos(t_0) \otimes \rhoe(t_0) ]  \bigg] \nonumber \\
		\overset{\rm FA}= \  & \tr_{\rm se}  \bigg[ \hat C \hat U_{t_2}^{t_3\dagger}  \hat D \hat U_{t_2}^{t_3} \,
		\cu_{t_1}^{t_2} \big[ \hat B  \ce_{t_0}^{t_1}[ \hat A  \rhos(t_1) ]  \otimes \trhoe(t_1)  \big] \bigg] \label{batogqrf1} \\
		\overset{\rm GFA}=  \ & \tr_{\rm se} \bigg[ \hat D \, \cu_{t_2}^{t_3} \left[ \left(   \tce_{t_1}^{t_2}  \hat B \ce_{t_0}^{t_1} [ \hat A \rhos(t_0) ] \hat C \right)   \otimes \trhoe(t_2) \right] \bigg]  \label{batogqrf2} \\
		\overset{\rm GFA}= \  & \trs \left[ \hat D \tce_{t_2}^{t_3} \big[  (   \tce_{t_1}^{t_2}  \hat B \ce_{t_0}^{t_1} [ \hat A \rhos(t_0) ] ) \hat C \big] \right ] \label{batogqrf3}\ ,
	\end{align}
	where \eq{batogqrf3} is exactly the same as that predicted by GQRF via Eq.~\eqref{gqrf}.
	
	The necessary direction does not hold. This is because any physical process that can be modelled by PFI also satisfies GQRF, as later shown in Theorem~\ref{PFI-GQRF}, where such a process certainly fails FA as the system is correlated with the past part of the bath. 
\end{proof}

\subsubsection{Quantum regression, past-future independence and quantum white noise}
\label{sec_GQRF-PFI-QWN}

Continuing on the discussion of correlation functions, we show that GQRF can  be derived from PFI,  where the latter can in turn be derived from QWN. 

\indent

\begin{theo}[label = PFI-GQRF]{}{}
	Past--future independence (PFI) in Definition~\ref{def_PFI} is a sufficient condition for the general quantum regression formula (GQRF) in Definition~\ref{def_GQRF}.
\end{theo}

\indent

It is an open question as to whether the reverse direction of the theorem also holds, i.e., whether PFI and GQRF are  equivalent concepts of quantum Markovianity. This question is briefly discussed in~\ref{app-conj}, where we conjecture that PFI is in fact strictly stronger than GQRF. 

Before proving  the theorem, we need some important corollaries that follow from the definition of PFI. First, note that from condition~\ref{pfi_con0} in Definition~\ref{def_PFI} that ${\mathbb H}_{\rm e} ={\mathbb H}_{\rm p_2} \otimes {\mathbb H}_{\rm f_2}$ for any time $t_2\geq t_0$, and hence, replacing $t_2$ by $t_1$, that ${\mathbb H}_{\rm e} = {\mathbb H}_{\rm p_1} \otimes {\mathbb H}_{\rm f_1} = {\mathbb H}_{\rm p_1} \otimes {\mathbb H}_{\rm e_1^2} \otimes {\mathbb H}_{\rm f_2}$ for all times $t_2 \geq t_1 \geq t_0$. It follows that
\begin{align} \label{futurefact}
	{\mathbb H}_{\rm f_1} = {\mathbb H}_{\rm e_1^2}\otimes {\mathbb H}_{\rm f_2} \ .
\end{align}
That is, the future part of the environment at any time also admits a tensor product structure, as mentioned in Section~\ref{sec_PFI}. Second, from condition~\ref{pfi_con2} we have
\begin{align}
	\label{cors_PFI_1}
	\hat U_{t_1}^{t_3} = \hat U_{t_2}^{t_3} \hat U_{t_1}^{t_2} 
	= \left(\hat U_{t_2}^{t_3} ({\rm p}_2) \otimes \hat U_{t_2}^{t_3} ({\rm s, e_2^3}) \otimes \hat U_{t_2}^{t_3} ({\rm f_3}) \right) \left(\hat U_{t_1}^{t_2} ({\rm p}_1) \otimes \hat U_{t_1}^{t_2} ({\rm s, e_1^2}) \otimes \hat U_{t_1}^{t_2} ({\rm f_2}) \right) \ . 
\end{align}
%
Ignoring explicit tensor products of identity operators for convenience, we can rewrite Eq.~\eqref{cors_PFI_1} as
\begin{align}
	\label{cors_PFI_2}
	\hat U_{t_1}^{t_3}  &= \left(\hat U_{t_2}^{t_3}({\rm p_2}) U_{t_1}^{t_2}({\rm p_1})\right)\left( \hat U_{t_2}^{t_3} ({\rm s, e_2^3}) \hat U_{t_1}^{t_2} ({\rm s, e_1^2}) \right)\left(  \hat U_{t_2}^{t_3}({\rm f_3}) U_{t_1}^{t_2}({\rm f_2})  \right) \nn \\
	&=  \left(\hat U_{t_2}^{t_3} ({\rm p_1}, {\rm e_1^2}) U_{t_1}^{t_2}({\rm p_1})\right)\left( \hat U_{t_2}^{t_3} ({\rm s, e_2^3}) \hat U_{t_1}^{t_2} ({\rm s, e_1^2}) \right)\left(  \hat U_{t_2}^{t_3}({\rm f_3}) \hat U_{t_1}^{t_2} (\rm e_2^3,\rm f_3) \right)\ ,
\end{align} 
where the second line follows using ${\mathbb H}_{\rm p_2} = {\mathbb H}_{\rm p_1} \otimes {\mathbb H}_{\rm e_1^2}$ from condition~\ref{pfi_con0} and ${\mathbb H}_{\rm f_2} = {\mathbb H}_{\rm e_2^3}\otimes {\mathbb H}_{\rm f_3}$ from Eq.~(\ref{futurefact}).
From condition~\ref{pfi_con2} we also have
\begin{align}
	\label{cors_PFI_3}
	\hat U_{t_1}^{t_3} = \hat U_{t_1}^{t_3} ({\rm p}_1) \otimes \hat U_{t_1}^{t_3} ({\rm s, e_1^3}) \otimes \hat U_{t_1}^{t_3} ({\rm f_3}) \ .
\end{align}
Equation~\eqref{cors_PFI_3} implies that no correlation is established between ${\mathbb H}_{\rm p_1}$ and ${\mathbb H}_{\rm e_1^3}$. Noting ${\mathbb H}_{\rm e_1^2}\subset{\mathbb H}_{\rm e_1^3}$, it then follows from Eq.~\eqref{cors_PFI_2} that
\begin{align}
	\label{fac_PFI_1}
	\hat U_{t_2}^{t_3} ({\rm p_2}) = \hat U_{t_2}^{t_3} ({\rm p_1}, {\rm e_1^2}) = \hat U_{t_2}^{t_3}({\rm p_1}) \otimes \hat U_{t_2}^{t_3}(\rm e_1^2) \ .
\end{align}
That is, the unitary on the past part of the environment Hilbert space is factorizable. By a similar argument it can be shown that the unitary on the future part follows a similar rule:
\begin{align}
	\label{fac_PFI_2}
	\hat U_{t_1}^{t_2} ({\rm f_2}) =  \hat U_{t_1}^{t_2}( {\rm e}_2^3, {\rm f}_3) = \hat U_{t_1}^{t_2} ({\rm e}_2^3) \otimes \hat U_{t_1}^{t_2} ({\rm f}_3) \ .
\end{align}
These factorization properties, Eqs.~\eqref{fac_PFI_1} and~\eqref{fac_PFI_2}, are essential for defining generalized dynamical maps in the following proof. The factorizable unitary on the future part further allows us to generalize condition~\ref{pfi_con1} to all times. In particular, noting that there is no past part with respect to the initial time $t_0$ (see discussion in Section~\ref{sec_PFI}), we have $\hat U_{t_0}^{t_1}({\rm p}_0) = \hat 1$ and ${\mathbb H}_{\rm p_1} = {\mathbb H}_{\rm e_0^1}$. Setting the time interval to be  from  $t_0$ to $t_1$ in Eqs.~\eqref{cors_PFI_3} and~\eqref{fac_PFI_2}, and using  condition~\ref{pfi_con1},  we then have
\begin{align}
	\rhose(t_1)=	\blk{\cal U}_{t_0}^{t_1} (\rhos(t_0) \otimes \rhoe(t_0)) 
	& = \left({\cal U}_{t_0}^{t_1}({\rm s,p_1}) \otimes {\cal U}_{t_0}^{t_1} ({ \rm e_1^2}) \otimes {\cal U}_{t_0}^{t_1} ({\rm f_2}) \right) \left( \rhos(t_0) \otimes \rho_{ \rm p_1}(t_0) \otimes \rho_{\rm e_1^2}(t_0) \otimes \rho_{\rm f_2}(t_0) \right) \label{cors-PFI-gfac0} \\ 
	& = \rho_{\rm sp_1}(t_1) \otimes \rho_{\rm e_1^2}(t_1) \otimes \rho_{\rm f_2}(t_1) \label{cors-PFI-gfac} \ ,
\end{align}
where $\rho_{\rm sp_1}(t) := {\rm Tr}_{\rm f_1}[\rhose(t)]$,  $\rho_{\rm e_1^2}(t) := {\rm Tr}_{\rm sp_1f_2}[\rhose(t)]$ 
and $\rho_{\rm f_2}(t) :=  {\rm Tr}_{\rm sp_2}[\rhose(t)]$. Taking the partial trace over ${\mathbb H}_{\rm s}\otimes{\mathbb H}_{\rm p_1}$, it further follows that
\begin{align}\label{cor-pfi1} 
	\rho_{\rm f_1}(t_1) = \rho_{\rm e_1^2}(t_1) \otimes \rho_{\rm f_2}(t_1) \ .
\end{align}
Armed with these results, we now proceed to the proof of the theorem. 

\begin{proof}
	We first prove  Theorem~\ref{PFI-GQRF} for the simplest case of two-time correlation functions, thus showing that PFI implies QRF.  Given any system operators $\hat{A}$ and $\hat{B}$, the calculation of the two-time correlation function $\me{\check B(t_2)\check A(t_1)}$ from PFI follows for $t_2\geq t_1$ as:
	\begin{align}
		\me{\check B(t_2)\check A(t_1)} 
		& = \trse \left[  (\cu_{t_0}^{t_2\dagger} \hb) (\cu_{t_0}^{t_1\dagger} \ha ) \rhose(t_0) \right] \nonumber\\
		& = \trse \left[ \hb \, \hat {\cal U}_{t_1}^{t_2} \ha \rhose(t_1) \right]  \nonumber \\
		& = \trse \left[  \hat B \bigg( {\cal U}_{t_1}^{t_2}({\rm p_1}) \otimes {\cal U}_{t_1}^{t_2}({\rm s,e_1^2})  \otimes {\cal U}_{t_1}^{t_2}({\rm f_2}) \bigg) \bigg( \ha \rho_{\rm sp_1}(t_1) \otimes  \rho_{\rm e_1^2}(t_1) \otimes \rho_{\rm f_2}(t_1) \bigg)  \right] \nn \\ 
		& = \trs\bigg[\hat B\, \tr_{\rm e_1^2f_2} \left[  \bigg( {\cal U}_{t_1}^{t_2}({\rm s,e_1^2})\otimes {\cal U}_{t_1}^{t_2}({\rm f_2}) \bigg) \left( \ha \rhos(t_1) \otimes \rho_{\rm e_1^2}(t_1) \otimes \rho_{\rm f_2}(t_1)\right)  \right] \bigg] \label{der-pfi-qrf2} \ .
	\end{align}
	Here the third line follows via condition~\ref{pfi_con2} and Eq.~\eqref{cors-PFI-gfac}, and the partial trace is taken over  $\mathbb{H}_{\rm p_1}$ in the last line.
	One further has for any operator $\hat X\in \fbs$, using condition~\ref{pfi_con2} and Eqs.~(\ref{cors-PFI-gfac0}) and (\ref{cors-PFI-gfac}), that
	\begin{align}
		&\tr_{\rm e_1^2f_2} \left[  \bigg( {\cal U}_{t_1}^{t_2}({\rm s,e_1^2})\otimes {\cal U}_{t_1}^{t_2}({\rm f_2}) \bigg) \left( \hat X  \otimes \rho_{\rm e_1^2}(t_1) \otimes \rho_{\rm f_2}(t_1)\right)  \right] \nn \\
		&\qquad\qquad\qquad\qquad\qquad\qquad= \tre \left[  \left( {\cu}_{t_1}^{t_2}( {\rm p_1} ) \otimes {\cu}_{t_1}^{t_2}( {\rm s,e_1^2} ) \otimes {\cu}_{t_1}^{t_2} ( {\rm f_2} )  \right) [ \hat X \otimes \rho_{\rm p_1}(t_0) \otimes \rho_{\rm e_1^2}(t_1) \otimes \rho_{\rm f_2}(t_1) ]\right] \nn \\
		&\qquad\qquad\qquad\qquad\qquad\qquad=\tre \left[ {\cu}_{t_1}^{t_2} [ \hat X \otimes {\cw}_{t_0}^{t_1} \rhoe(t_0) ] \right] \nn \\
		&\qquad\qquad\qquad\qquad\qquad\qquad= \tce_{t_1}^{t_2} \hat X \ , \label{gDM-PFI}
	\end{align}
	where ${\cw}_{t_0}^{t_1}:={\cal I}_{\rm p_1} \otimes {\cu}_{t_0}^{t_1}({\rm e_1^2}) \otimes {\cu}_{t_1}^{t_2}( {\rm f_2})$, and $\tce_{t_1}^{t_2}$ is the corresponding generalized dynamical map defined in Eq.~\eqref{general_dynamical_map}. Note that one can modify this choice of ${\cw}_{t_0}^{t_1}$ by appending an arbitrary unitary evolution on $\rho_{\rm p_1}$, $\hat W_{t_0}^{t_1}({\rm p}_{1}) \in {\mathfrak U}({\mathbb H}_{\rm p_{1}})$, without affecting the generalized dynamical map.
	Comparing Eqs.~(\ref{der-pfi-qrf2}) and (\ref{gDM-PFI}) immediately yields the QRF in Eq.~\eqref{qrf1}, as desired. The corresponding formula in Eq.~(\ref{qrf2}) for  $\me{\ha(t_1)\hb(t_2)}$ may be similarly obtained (see \ref{app-fa}). 
	
	To show more generally that GQRF follows from PFI, we first consider the general four-time case as an example. For  this case one has
	\begin{align}
		& \la \check A_0(t_0) \check A_1(t_1) \check A_2(t_2) \check A_3(t_3) \check B_3(t_3) \check B_2(t_2) \check B_1(t_1) \check B_0(t_0) \ra \nonumber \\ 
		&\qquad\qquad =  \tr [ \, {\cal C}_3 \, {\cal U}_{t_2}^{t_3} \, {\cal C}_2  \, {\cal U}_{t_1}^{t_2} \, {\cal C}_1 \, {\cal U}_{t_0}^{t_1} \, {\cal C}_0 \rhose(t_0) ]  \\
		&\qquad\qquad =   \tr [ \, {\cal C}_3 \, {\cal U}_{t_2}^{t_3} \, {\cal C}_2 \, {\cal U}_{t_1}^{t_2} \,  {\cal C}_1 \left( {\cal U}_{t_0}^{t_1}(\rm s, p_1) \otimes {\cal U}_{t_0}^{t_1}({\rm e_1^2}) \otimes {\cal U}_{t_0}^{t_1} ({\rm e_2^3}) \otimes {\cal U}_{t_0}^{t_1}(\rm f_3) \right) {\cal C}_0 \rhose(t_0) ]  \label{der-pfi-gqrf1}  \ ,
	\end{align}
	where ${\cal C}_j \hat X = \hat B_j \hat X \hat A_j$ as per Eq.~\eqref{gqrf-NA}, and the last line follows from  Eqs.~\eqref{cors_PFI_3} and~\eqref{fac_PFI_2}  similarly to Eq.~\eqref{cors-PFI-gfac0}. 
	Note that we have made a further decomposition of ${\cal U}_{t_0}^{t_1}( {\rm f_1})$.  In fact, from \eq{fac_PFI_2}, one has
	\begin{align}
		\hat U_{t_0}^{t_1}  ( {\rm f_1} ) 
		& = \hat U_{t_0}^{t_1} (  {\rm e}_1^2, {\rm f}_2 ) =  \hat U_{t_0}^{t_1} ( {\rm e}_1^2 ) \otimes \hat U_{t_0}^{t_1} ( {\rm f}_2 ) = \hat U_{t_0}^{t_1} ( {\rm e}_1^2 ) \otimes \hat U_{t_0}^{t_1} ( {\rm e}_2^3 ,{\rm f}_3 )   \\
		& = \hat U_{t_0}^{t_1} ( {\rm e}_1^3 , {\rm f}_3 ) = \hat U_{t_0}^{t_1} (  {\rm e}_1^3)  \otimes \hat U_{t_0}^{t_1} (  {\rm f}_3 )  =  \hat U_{t_0}^{t_1} (  {\rm e}_1^2, {\rm e}_2^3 )  \otimes \hat U_{t_0}^{t_1} ( {\rm f}_3 )  \ .
	\end{align}
	That is, the unitary $\hat U_{t_0}^{t_1} ( {\rm f_1} ) $  does not establish any correlation between different parts of ${\mathbb H}_{\rm f_1}$. Hence we have ${\cal U}_{t_0}^{t_1}( {\rm f_1}) = \cu_{t_0}^{t_1} ({\rm e_1^2}) \otimes \cu_{t_0}^{t_1} ( {\rm e_2^3} ) \otimes \cu_{t_0}^{t_1} (  {\rm f_3} )$. 
	
	Also, note from condition~\ref{pfi_con2} [or equivalently Eq.~\eqref{cors_PFI_3}] that ${\cal U}_{t_2}^{t_3} = {\cal U}_{t_2}^{t_3}({\rm p}_2) \otimes {\cal U}_{t_2}^{t_3}( {\rm s,e_2^3} ) \otimes {\cal U}_{t_2}^{t_3}({\rm f}_3)$. Then by factorizing~${\cal U}_{t_2}^{t_3}({\rm p}_2)$ as per Eq.~\eqref{fac_PFI_1}, we have
	\begin{align}
		{\cal U}_{t_2}^{t_3} & = {\cal U}_{t_2}^{t_3}( {\rm  p_1} ) \otimes {\cal U}_{t_2}^{t_3}( {\rm  e_1^2} ) \otimes {\cal U}_{t_2}^{t_3}( {\rm  s,e_2^3} ) \otimes {\cal U}_{t_2}^{t_3}( {\rm  f_3} ) \,  \label{der-pfi-gqrf-con1} \ .
	\end{align}
	Similarly, using condition~\ref{pfi_con2} and Eq.~(\ref{fac_PFI_2}), one finds
	\begin{align}
		{\cal U}_{t_1}^{t_2} & = {\cal U}_{t_1}^{t_2} ({\rm p_1}) \otimes {\cal U}_{t_1}^{t_2} (\rm s, e_1^2) \otimes {\cal U}_{t_1}^{t_2} ({\rm e_2^3}) \otimes {\cal U}_{t_1}^{t_2} ({\rm f_3})  \label{der-pfi-gqrf-con2} \ . 
	\end{align}
	Further, taking the partial trace of  condition~\ref{pfi_con1} over $\mathbbm{H}_{p_1}$ gives $\rho_{\rm f_1}(t_0)= \rho_{\rm e^2_1}(t_0)\otimes \rho_{\rm f_2}(t_0)$, which applied inductively yields $\rho_{\rm f_1}(t_0)= \rho_{\rm e^2_1}(t_0)\otimes \rho_{\rm e^3_2}(t_0)\otimes \rho_{\rm f_3}(t_0)$. Substituting this back into condition~\ref{pfi_con1}, the initial state can be finally rewritten as
	\begin{align} \label{rhozero}
		\rhose(t_0) = \rhos(t_0) \otimes \rho_{\rm p_1} (t_0) \otimes \rho_{\rm e_1^2}(t_0) \otimes \rho_{\rm e_2^3}(t_0) \otimes \rho_{\rm f_3}(t_0) \,
	\end{align}
	Thus by tracing over ${\mathbb H}_{\rm p_1}$ and ${\mathbb H}_{\rm f_3}$, similarly to Eq.~\eqref{der-pfi-qrf2}, it follows from Eq.~\eqref{der-pfi-gqrf1} that
	\begin{align}
		& \la \check A_0(t_0) \check A_1(t_1) \check A_2(t_2) \check A_3(t_3) \check B_3(t_3) \check B_2(t_2) \check B_1(t_1) \check B_0(t_0) \ra \nonumber \\
		= \; & \tr_{\rm se_1^3} \big[  {\cal C}_3 (  {\cal U}_{t_2}^{t_3}( {\rm  e_1^2} ) \otimes {\cal U}_{t_2}^{t_3}( {\rm  s,e_2^3} ) ) {\cal C}_2  (  {\cal U}_{t_1}^{t_2} (\rm s, e_1^2) \otimes {\cal U}_{t_1}^{t_2} ({\rm e_2^3})  )  {\cal C}_1 {\tr_{\rm p_1}} [ \cu_{t_0}^{t_1} [ {\cal C}_0 \rhos(t_0) \otimes \rho_{\rm p_1}(t_0) ] \otimes \rho_{\rm e_1^2}(t_1) \otimes \rho_{\rm e_2^3}(t_1) ] \big]  \nonumber \\
		= \; & \tr_{\rm se_1^3} \left[  {\cal C}_3  \left(  {\cal U}_{t_2}^{t_3}( {\rm  e_1^2} ) \otimes {\cal U}_{t_2}^{t_3}( {\rm  s,e_2^3} )  \right) {\cal C}_2 \left(  {\cal U}_{t_1}^{t_2} (\rm s, e_1^2) \otimes {\cal U}_{t_1}^{t_2} ({\rm e_2^3})  \right)  {\cal C}_1 \left( \ce_{t_0}^{t_1} [ {\cal C}_0 \rhos(t_0)  ] \otimes \rho_{\rm e_1^2}(t_1) \otimes \rho_{\rm e_2^3}(t_1) \right) \right]  \label{der-pfi-gqrf-c1} \\
		= \; & \tr_{\rm se_2^3}  \left[  {\cal C}_3 \, {\cal U}_{t_2}^{t_3}( {\rm  s,e_2^3} ) \, {\cal C}_2 \tr_{ \rm e_1^2 }  \left[ {\cal U}_{t_1}^{t_2} (\rm s, e_1^2) [  {\cal C}_1 \ce_{t_0}^{t_1} [ {\cal C}_0 \rhos(t_0)  ] \otimes \rho_{\rm e_1^2}(t_1) ] \otimes \left( {\cal U}_{t_1}^{t_2} ({\rm e_2^3}) \rho_{\rm e_2^3}(t_1)  \right) \right] \right]  \nonumber \\
		= \; & \tr_{\rm se_2^3} \left[ {\cal C}_3  \, {\cu}_{t_2}^{t_3}({\rm s,e_2^3}) \left( {\cal C}_2 \, \tce_{t_1}^{t_2} [  {\cal C}_1 \ce_{t_0}^{t_1} [ {\cal C}_0 \rhos(t_0) ] ] \otimes \rho_{\rm e_2^3}(t_2) \right) \right]  \label{der-pfi-gqrf-c2} \\
		= \; & \trs \left[ {\cal C}_3 \, \tce_{t_2}^{t_3} \, {\cal C}_2 \, \tce_{t_1}^{t_2} \, {\cal C}_1 \, \ce_{t_0}^{t_1} {\cal C}_0 \rhos(t_0) \right] \label{der-pfi-gqrf-c3}
	\end{align}

	Noting that the unitary $\cu_{t_2}^{t_3} ({\rm p_1}) \otimes \cu_{t_1}^{t_2} ({\rm p_1})$ from Eqs.~\eqref{der-pfi-gqrf-con1} and~\eqref{der-pfi-gqrf-con2} does not affect the partial trace over $\mathbb H_{\rm p_1}$ in Eq.~\eqref{der-pfi-gqrf1}, the dynamical map $\ce_{t_0}^{t_1}$ is thus defined in a similar way as we did for the general dynamical map in Eq.~\eqref{gDM-PFI}.  
	The generalized dynamical maps in Eqs.~\eqref{der-pfi-gqrf-c2} and~\eqref{der-pfi-gqrf-c3} are also obtained similarly to that in Eq.~\eqref{gDM-PFI}, corresponding to choosing $\hat W_{t_1}^{t_2} := \hat I_{\rm p_1} \otimes \hat U_{t_1}^{t_2}({\rm e_1^2}) \otimes \hat U_{t_1}^{t_2} ({\rm f_2})$ and $\hat W_{t_2}^{t_3} := \hat I_{\rm p_2} \otimes \hat U_{t_2}^{t_3}({\rm e_2^3}) \otimes \hat U_{t_2}^{t_3} ({\rm f_3})$ respectively in Eq.~\eqref{general_dynamical_map}.
	The resulting Eq.~\eqref{der-pfi-gqrf-c3} is just the GQRF in Eq.~\eqref{gqrf}, for the case $n=3$.
	
	For higher-order correlation functions the above method again yields the corresponding GQRF, via inductive generalizations of the tensor product decompositions in Eqs.~(\ref{der-pfi-gqrf1})--(\ref{rhozero}). \blk
\end{proof}

As mentioned in the beginning of Section~\ref{sec_PFI}, PFI is a broader concept of Markovianity, compared to QWN.
Indeed, these two concepts are hierarchically linked by the following theorem.  

\indent

\begin{theo}[label = QWN-PFI]{}{}
	Quantum white noise (QWN) in Definition~\ref{def_QWN} is a sufficient but not a necessary condition for past--future independence (PFI) in Definition~\ref{def_PFI}.
\end{theo}

\begin{proof}
	In Definition~\ref{def_QWN} for QWN it is seen that the algebra of observables for the environment is generated by  $\{b_j(t),b^\dagger_k(t')\}_{j,k,t,t'}$. These satisfy the commutation relations in Eq.~(\ref{singcomm}), and hence correspond to a continuum of bosonic field modes.  Letting $\mathbb{H}_t$ denote the Hilbert space corresponding to those modes having a fixed value of $t$, we may then formally write the environment Hilbert space as a continuous tensor product, $\mathbb{H}_{\rm e}=\otimes_{t\geq t_0}\, \mathbb{H}_t$ (see Ref.~\cite{HP84}). It follows that at any times $t_2> t_1>t_0$ we have the corresponding tensor-product decomposition \beq
	{\mathbb H}_{\rm e} = {\mathbb H}_{\rm p_1} \otimes {\mathbb H}_{\rm e_1^2} \otimes {\mathbb H}_{\rm f_2}
	\eeq
	with ${\mathbb H}_{\rm p_1}:=\otimes_{t\in[t_0,t_1)}\,{\mathbb H}_{t}$,  ${\mathbb H}_{\rm e_1^2}:=\otimes_{t\in[t_1,t_2)}\,{\mathbb H}_{t}$,  ${\mathbb H}_{\rm f_2}:=\otimes_{t\in[t_2,\infty)}\,{\mathbb H}_{t}$. Condition~\ref{pfi_con0} in Eq.~(\ref{growingpast}) for PFI immediately follows.
	
	Moreover, Eq.~(\ref{coherentinput}) for QWN implies that the initial state of the environment may be written in continuous tensor-product form as
	\beq
	\rhoe(t_0) = \otimes_{t} \,|\bm \beta(t)\rangle_t\langle\bm\beta(t) | \ ,
	\eeq
	where $\bm \beta(t)$ is the complex vector with $j$th component equal to $\beta_j(t)$ and $|\bm \beta(t)\rangle_t\in\mathbb H_t$ denotes the corresponding (multicomponent) coherent state defined by $b_j(t) |\bm \beta(t)\rangle_t=\beta_j(t) |\bm \beta(t)\rangle_t$~\cite{GZ04}.
	It immediately follows that condition~\ref{pfi_con1} in Eq.~(\ref{qwndf1}) for PFI holds, with $\rho_{\rm p_1}(t_0):={\rm Tr}_{\rm f_1}[\rhoe(t_0)]$, $\rho_{\rm e_1^2}(t_0):={\rm Tr}_{\rm p_1 f_2}[\rhoe(t_0)]$, and $\rho_{\rm f_2}(t_0):={\rm Tr}_{\rm p_2}[\rhoe(t_0)]$. 
	
	Finally, noting the commutation relations in Eq.~\eqref{singcomm}, the bath modes at different times act on different Hilbert spaces. One therefore can decompose the total unitary $\intfr{{\cal U}}{}_{t}^{t + \dt}$ as
	\begin{align}
		\intfr{{\cal U}}{}_{t}^{t + \dt} = \intfr{{\cal U}}{}_{t}^{t + \dt} ({\rm p}_t) \otimes \intfr{{\cal U}}{}_{t}^{t + \dt}({\rm s,e}_t^{t + \dt})\otimes \intfr{{\cal U}}{}_{t}^{t + \dt} ({\rm f}_{t + \dt}) \ .
	\end{align}
	\blk
	This completes the sufficient direction.
	
	The necessary direction, however, does not hold. For example, PFI includes discrete-time dynamical models  such as the collision models  discussed in section~\ref{sec_PFI}, which are intrinsically incompatible with a differential-in-time equation as per Eq.~\eqref{HPEqn}. 
\end{proof}

\subsubsection{System interventions}
\label{theorems5and6}

In this section, continuing on the discussion of GQRF, we sort out the hierarchy for the concepts categorized as system interventions in Section~\ref{sec-SI}. We begin with an obvious theorem. 

\indent

\begin{theo}[label = GQRF-QRF]{}{}
	The general quantum regression formula (GQRF) in Definition~\ref{def_GQRF} is a sufficient but not necessary condition for quantum regression formula (QRF) in Definition~\ref{def_QRF}. 
\end{theo}

\begin{proof}
	It is trivial to show sufficiency, as one can just set all but two system operators in Eq.~\eqref{gqrf} for GQRF to be the identity operator, which immediately shows QRF is valid. The reverse relation, however, is not true. As a counterexample to the converse direction, we show in~\ref{app-AFL} that the AFL model~\cite{AHFB14} satisfies QRF but fails GQRF. 
\end{proof}

In Section~\ref{sec-SI}, we have shown how GQRF can be understood in an operational way, and introduced the idea of FDD. These two concepts are related by the following theorem.

\indent

\begin{theo}[label = gqtoddf]{}{}
	The general quantum regression formula (GQRF) in Definition~\ref{def_GQRF} is a sufficient condition for the failure of dynamical decoupling (FDD) in Definition~\ref{def_CFDD}.
\end{theo}

\begin{proof}
	First let us consider the correlation function 
	\begin{align}
		\me{ \check A_1(t_1) \dots \check A_n(t_n) \check B_n(t_n) \dots \check B_1(t_1)   }  = \trse \left[    \hb_n \hat U_{t_{n-1}}^{t_n} \dots \hat U_{t_1}^{t_2}\hb_1 \hat U_{t_0}^{t_1} \rhose(t_0) \hat U_{t_0}^{t_1\dagger} \ha_1 \hat U_{t_1}^{t_2\dagger} \dots \hat U_{t_{n-1}}^{t_n\dagger} \ha_n \right] \ ,
	\end{align}
	where $\{\ha_k, \hb_k\}_k$ are system operators. If GQRF holds, then
	\begin{align}
		\label{ddcf-gqrf-pf1}
		\me{ \check A_1(t_1) \dots \check A_n(t_n) \check B_n(t_n) \dots \check B_1(t_1)   }  =
		\trs \left[  \hb_n  \tce_{t_{n-1}}^{t_n} [ \dots \hb_1 [ \ce_{t_0}^{t_1} \rhos(t_0)  ] \ha_1 \dots ] \ha_n  \right] \ .
	\end{align}
	Noting Eq.~\eqref{ddcf-gqrf-pf1}  is true for any choice of system operators $\ha_n$ and $\hb_n$, these equations imply that
	\begin{align}
		\label{ddfpv1}
		\tre \left[  \hat U_{t_{n-1}}^{t_n}\hb_{n-1} \dots \hat U_{t_1}^{t_2}\hb_1 \hat U_{t_0}^{t_1} \rhose(t_0) \hat  U_{t_0}^{t_1\dagger} \ha_1  \hat U_{t_1}^{t_2\dagger} \dots \ha_{n-1}\hat U_{t_{n-1}}^{t_n\dagger}  \right] =
		\tce_{t_{n-1}}^{t_n} [\hb_{n-1} \dots \hb_1[ \ce_{t_0}^{t_1} \rhos(t_0) ] \ha_1  \dots ] \ha_{n-1}] \ .
	\end{align}
	Now consider a dynamical decoupling pulse control sequence as per Eq.~\eqref{DDmap} (replacing $n$ by $n-1$): \blk
	\begin{align}
		{\cal H}_{t_0}^{t_n} [\rhos(t_0)]  = \tre \left[  \hat U_{t_{n-1}}^{t_n}  \hat V_{n-1} \dots \hat V_1 \hat U_{t_0}^{t_1}\rhose(t_0) \hat U_{t_0}^{t_1\dagger} \hat V_1^\dagger \dots \hat V_{n-1}^\dagger   \hat U_{t_{n-1}}^{t_n\dagger}  \right] \ ,
	\end{align}
	where the unitary operation $\hat V_k \equiv \hat V_k \otimes \hat1_{\rm e}$ represents the control pulse on the system at time $t_k$. It then follows from Eq.~\eqref{ddfpv1} that
	\begin{align}
		{\cal H}_{t_0}^{t_n} \blk [\rhos(t_0)]  
		= \tce_{t_{n-1}}^{t_n} \left[ \hat V_{n-1} \blk \dots \bigl[ \hat V_1 [ \ce_{t_0}^{t_1} \rhos(t_0) ]\ V_1^\dagger \bigr] \dots \hat V_{n-1}^\dagger \right]   \ ,
	\end{align}
	which fulfils the definition of FDD as per Eq.~\eqref{def_ccdf} with \blk $\cq_{t_k}^{t_{k+1}}=\tce_{t_k}^{t_{k+1}}$. 
\end{proof}

Note that since FDD is only a criterion, providing a {\em sufficient condition} for the failure of DD, rather than a definition of quantum non-Markovianity {\it per se}, we do not consider the reverse direction here (see also the discussion in section~\ref{sec_ddf}).

An immediate corollary of Theorem~\ref{gqtoddf} is that if dynamical decoupling of the system and bath \emph{is} possible, via some suitable sequence of pulses, then GQRF cannot hold. An example of this is given by the AFL model (see~\ref{app-AFL}). In particular, as shown in Ref.~\cite{AHFB14}, in this model the system, a qubit, can be perfectly decoupled from the bath by employing $\hat \sigma_0 = \hat I$ and $\hat \sigma_x$ pulses. We show in~\ref{app-AFL} that the model does not satisfy GQRF, in accordance with the theorem. Lastly, it is worth pointing out that Theorem~\ref{gqtoddf} is consistent with, but more general, than that in Ref.~\cite{GH17}, where the starting point is QWN in Eq.~\eqref{Utpdt}.

\subsubsection{Environment interventions} 
\label{sub-hierarchy-EI}

In this section, by exploring the relation between QRF and composability, we first show how the two classes of concepts, defined through system interventions and environment interventions (recall Section~\ref{sec_ei}), are connected, and then discuss the hierarchical relation of the latter to NIB and NQIB. 

\indent

\begin{theo}[label = QRFtoCom]{}{}
	The quantum regression formula (QRF) in Definition~\ref{def_QRF} is a sufficient condition for composability in Definition~\ref{df_composability}.
\end{theo}

\begin{proof}
	Consider the correlation function of two system operators $\hat A$ and $\hat B$. A direct calculation shows
	\begin{align}
		\me{ \check A(t_1) \check B(t_2) }  = & \trse \left[ \hat A(t_1) \hat U_{t_0}^{t_2\dagger} \hb \hat U_{t_0}^{t_2}  \rhose(t_0)   \right]  \nonumber \\
		= & \trs \left[ \hb \, \tre [ \, \cu_{t_0}^{t_2} [ \rhos(t_0) \otimes \rhoe(t_0) \hat A(t_1) ]  ] \right]  \label{qrftc1} \ .
	\end{align}
	From QRF, the correlation function can also be rewritten as
	\begin{align}
		\me{ \check A(t_1) \check B(t_2) }  & = \trs \left[ \hb \, \tce_{t_1}^{t_2}[\rhos(t_1) \ha ]   \right] \nonumber \\
		&= \trs \left[ \hb \, \tre \left[ \hat U_{t_1}^{t_2}  \bigg( \ce_{t_0}^{t_1} \rhos(t_0)  \ha \bigg)   \otimes \trhoe(t_1)  \hat U_{t_1}^{t_2\dagger}   \right] \right]  \label{qrftc2} \ . 
	\end{align}
	Note Eq.~\eqref{qrftc2} is equivalent to~\eqref{qrftc1} for any system operator $\hb$, thus we have:
	\begin{align}
		\label{qrftc3}
		\tre \left[ \, \cu_{t_0}^{t_2} \left[ \rhos(t_0) \otimes \rhoe(t_0) \hat A(t_1) \right]  \right]  =  \tre \left[ \hat U_{t_1}^{t_2}  \bigg( \ce_{t_0}^{t_1} \rhos(t_0)  \ha \bigg)   \otimes \trhoe(t_1)  \hat U_{t_1}^{t_2\dagger}   \right] \ .
	\end{align}
	Recalling the definition of the (generalized) dynamical map in Eqs.~\eqref{def_dynamical_map} and~\eqref{general_dynamical_map}, and setting $\hat A = \hat I$, it then follows from Eq.~\eqref{qrftc3} that
	\begin{align}
		\ce_{t_0}^{t_2} = \tce_{t_1}^{t_2} \ce_{t_0}^{t_1}  
	\end{align}
	which is the composability condition in Eq.~\eqref{com_2t}, as required. 
\end{proof}

The following two theorems establish the hierarchy of the concepts introduced in Section~\ref{sec_ei}.

\indent

\begin{theo}[label = ComtoNIB]{}{}
	Composability in Definition~\ref{df_composability} is a sufficient condition for no information backflow (NIB) in Definition~\ref{dfnib}.
\end{theo}

\begin{proof}
	The proof is trivial, since composability as per Eq.~\eqref{com} corresponds to NIB as per Eq.~\eqref{nib} for the particular choice ${\cal R}_t={\cal F}_t$, i.e., $\sigma_{\rm e}(t) = \trhoe(t) = \cw_{t_0}^{t} \rhoe(t_0) $. 
\end{proof}

The discussion of the converse direction to Theorem~\ref{ComtoNIB} is left to~\ref{app-conj}.  We now show that NIB is strictly a special case of NQIB.

\indent

\begin{theo}[label = NIBtoNQIB]{}{}
	No information backflow (NIB) in Definition~\ref{dfnib} is a sufficient but not a necessary condition for no quantum information backflow (NQIB) in Definition~\ref{def_NQIB}.
\end{theo}

\begin{figure}
	\centering
	\subfigure[Measurement-based feedback control]{
		\includegraphics[width=0.6\linewidth]{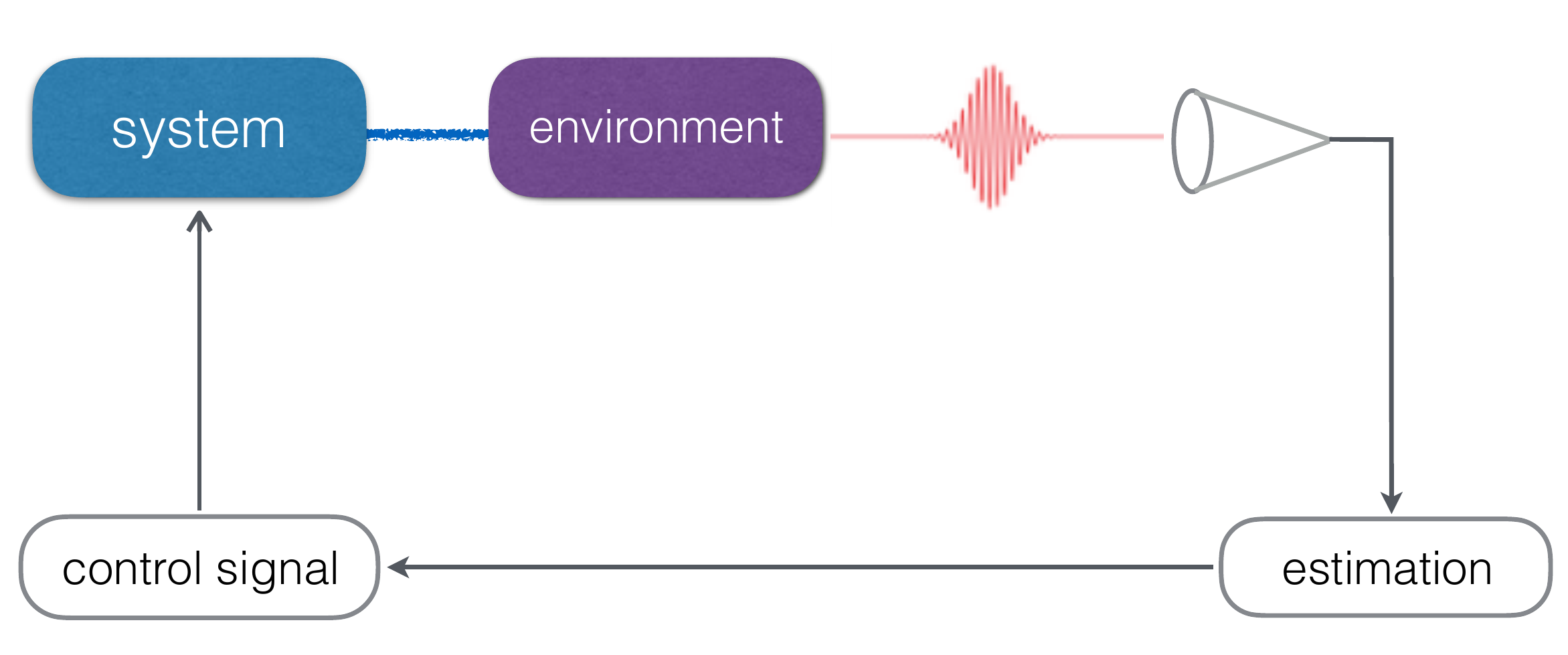}
		\label{fig_fb1}
	}
	\subfigure[Measurement-based feedback control and unravelling]{
		\includegraphics[width=0.6\linewidth]{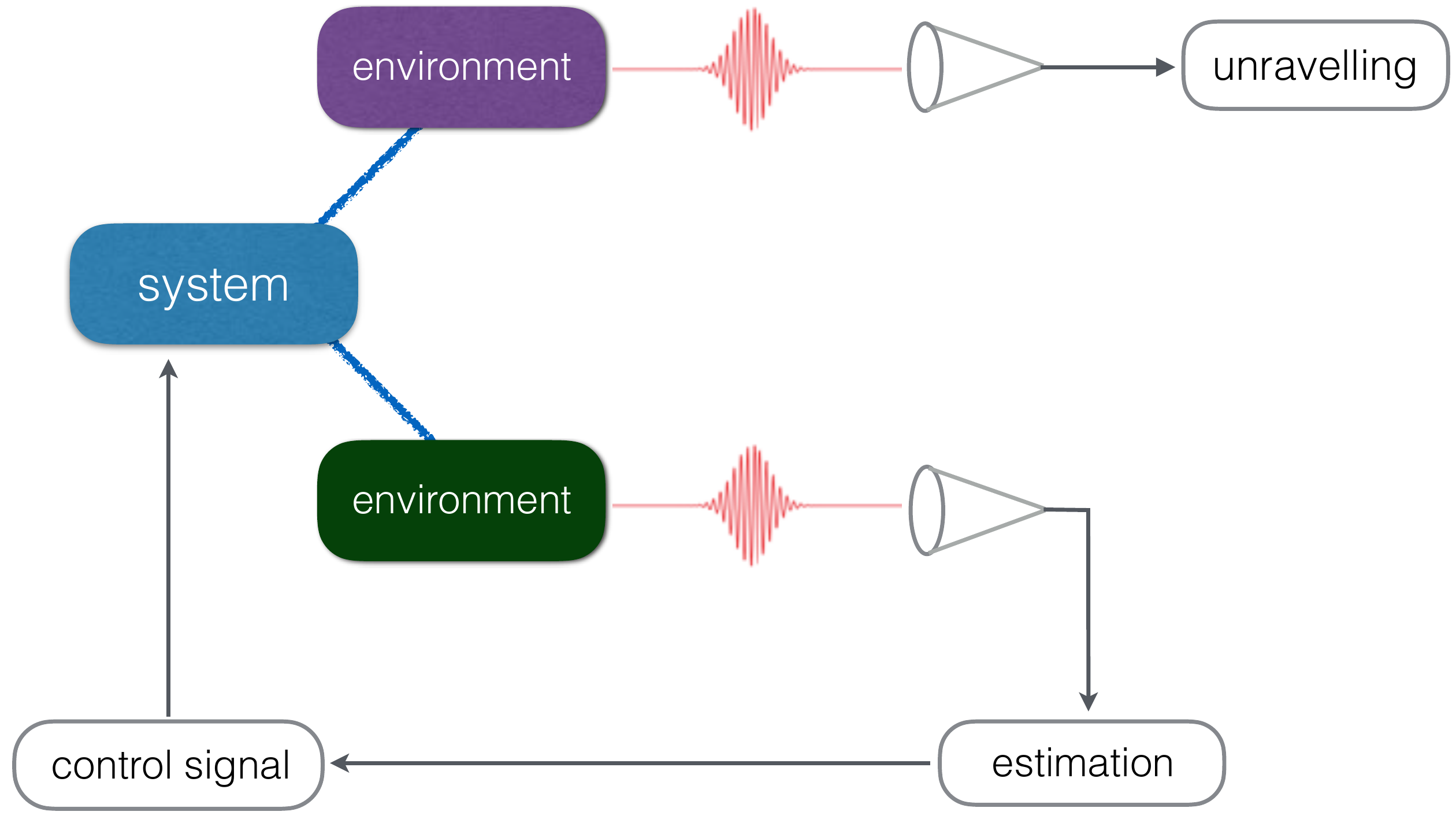}
		\label{fig_fb2}
	}
	\caption{(a) The upper panel illustrates the basic idea of measurement-based feedback control. Measurements performed on the environment provide information for estimating the system state, which is used to generate the control signal for manipulating the system dynamics by feeding it back. Note that both the estimation and signal processing steps are dealing with classical information, and are thus coloured in grey; the interaction between the system and the environment typically can generate entanglement, and is marked by a fuzzy blue line. (b) In the lower panel, the system is simultaneously coupled to two independent environments, both of which are monitored, but only one of which (bottom) is used for generating a feedback signal. See the main text for discussion.}
\end{figure}

\begin{proof}
	
	The sufficiency direction is obviously true, since any replacement channel ${\cal R}_t$ as per \blk Eq.~\eqref{channel_r} is trivially an entanglement-breaking channel, corresponding to setting  $\sigma_e^k(t) \equiv \sigma_{\rm e}(t)$ for all $k$ in Eq.~\eqref{channel_e}. 
	
	However, the necessary direction does not hold. As a first and very simple counterexample,  consider two qubits coupled through the Hamiltonian $\hat H = (\hat \sigma_z / 2) \otimes | 1 \ra\la 1 |$, where $\{ |0\ra, | 1 \ra \}$ is a basis for the bath qubit and $\hat \sigma_z$ is the Pauli $z$ operator for the system qubit. Let the initial joint state be $\rhose(0) =  \smallfrac{1}{2}( | +\ra_{\rm s}\la + |) \otimes \hat I_{\rm e}$, with $|+\rangle:=(|0\rangle+|1\rangle)/\sqrt{2}$. We set $t_0 = 0$ here for convenience, and take $\hat H_{\rm s}=\hat H_{\rm e}=0$ for simplicity. It is easy to show that the joint evolution is then described by the unitary operator
	\begin{align}
		\hat U(t) = \hat I_{\rm s} \otimes | 0 \ra \la 0 | + e^{-i \hat \sigma_z t /2}  \otimes | 1 \ra \la 1|  \, .
	\end{align}
	It immediately follows that $\rhose(t) = \smallfrac{1}{2} (\rhos \otimes | 0 \ra \la 0 |  + e^{-i \hat \sigma_z t /2} \rhos e^{ i \hat \sigma_z t /2} \otimes |1 \ra\la 1 | )$, where $\rhos = | + \ra\la + |$.
	NQIB clearly holds for this example since no entanglement is ever established between the system and the bath. Now at time $t_1 = \pi$, the reduced system state is the maximally mixed state, i.e., $\rhos(t_1) = \frac{1}{2} \hat I_{\rm s}$. But at a later time $t_2 = 2\pi$, the system state goes back to $\rhos$. That is, the system dynamics from time $t_1$ to $t_2$ is entropy-decreasing. Now assume that one performs a replacement operation at time $t_1$. Since the resulting map $\tce_{t_1}^{t_2}$ is unital, it can never be an entropy-decreasing process. That means NIB fails for this example.
	
	A second and more general counterexample, corresponding to a situation that readily arises in practice, is a measurement-based quantum feedback control scheme. As shown in Fig.~\ref{fig_fb1}, information about the system state is indirectly gained from measuring the environment~(recall Section~\ref{sec_quantum_unravelling} on unravelling), and is used for designing a control signal for further manipulating the system state. Let us assume that, in the absence of feedback, PFI holds. That is, the measurement over the bath can be done on the past part  such that it will break the entanglement between the system and the bath without affecting the system dynamics in future times (see also Theorem~\ref{PFItoQU} below). Therefore, NQIB always holds, even with the feedback operative, as the feedback control signal is classical in nature. 
	On the other hand, this control signal, which pilots the system state, contains classical information about the system from earlier times. In particular, the control operation at time $t$ is often taken to be a function of the estimated system state ${\varrho}_{\rm s}(t)$. But the latter is a functional of the measurement records from time $t_0$ to $t$, which will denote as ${\bm r}_{[t_0,t]}$ (see Ref.~\cite{WM09} for details). Now the environment or bath is everything (relevant to the system evolution) which is external to the system. 
	In this case, that includes the record ${\bm r}_{[t_0,t]}$, and hence the estimate ${\varrho}_{\rm s}(t)$, and hence the control signal. Thus, if one applies a replacement channel on the bath before time $t$, that will replace a control which is correlated with the system state by one that is uncorrelated with it, leading to quite different evolution after time $t$. That is, NIB fails. 
\end{proof}

We note that the second counterexample above, showing that NQIB and NIB are not equivalent, arises from a strong connection between NQIB and the quantum unravellings discussed in Section~\ref{sec_quantum_unravelling}. This connection, and the corresponding counterexample, will be further expounded  in the next section.

\subsubsection{Unravelling connections} 
\label{sub-unravellings-related}

In this section, we show how two concepts of quantum Markovianity, namely PFI and NQIB,  are related with the idea of quantum unravelling. To begin with, we show that quantum unravelling is a direct consequence of PFI.

\indent

\begin{theo}[label = PFItoQU]{}{}
	Past--future independence (PFI) in Definition~\ref{def_PFI} is a sufficient but not a necessary condition for many pure unravellings (MPU)  in Definition~\ref{dfmpu}.
\end{theo}

To prove this theorem we will need the following lemma, proved in~\ref{app-lemma}.

\begin{lemma}
	\label{lemma-PFI}
	Given a unitary $\hat U$ acting on the Hilbert space ${\mathbb H}_{\rm ab} = {\mathbb H}_{\rm a} \otimes {\mathbb H}_{\rm b}$, a pure state $\hat \pi_{\rm a}$  on ${\mathfrak{B}}({\mathbb H}_{\rm a})$ and a (possibly mixed) state $\rho_b$ on ${\mathfrak{B}}({\mathbb H}_{\rm b})$, such that $\rho'_{\rm a} = \tr_{\rm b} [ \hat U ( \hat \pi_{\rm a} \otimes \rho_{\rm b}) \hat U^\dagger ]$ is mixed, then either 
	\begin{enumerate}
		\item the state $\rho_{\rm b}$ is related to  $\rho'_{\rm a}$ by an isometry (and so is also mixed),  or
		\item there exists a pure-state decomposition $\rho_{\rm b} = \sum_j \wp_{\rm b}^j \hat \pi_{\rm b}^j$ of $\rho_{\rm b}$ such that $\hat U(\hat \pi_{\rm a} \otimes \hat \pi_{\rm b}^j) \hat U^\dagger$ is a  (pure)  entangled state for at least one $\hat \pi_{\rm b}^j$ with $\wp_{\rm b}^j\neq 0$. 
	\end{enumerate}
\end{lemma}

Note that the first possibility in the lemma can only hold if the dimension of ${\mathbb H}_{\rm a}$ is at least as large as the dimension of ${\mathbb H}_{\rm b}$. Theorem~\ref{PFItoQU} may now be proved as follows.

\begin{proof}
	We begin with a proof of the sufficient direction. Given an OQS described by a PFI model, let the initial system state be pure, i.e., $\hat  \pi(t_0) =  |\psi(t_0)\ra\la \psi(t_0)|$. Following the definitions of PU and MPU, we consider  measurements performed on the bath at times $t_0, t_1, t_2, \dots, t_n$. Recalling that the initial bath state is uncorrelated across the division of the bath Hilbert space, i.e., across $ {\mathbb H}_{\rm e} =  {\mathbb H}_{\rm e_0^1} \otimes {\mathbb H}_{\rm e_1^2} \otimes \dots \otimes {\mathbb H}_{\rm e_{n-1}^n}$, an initial measurement on $\rhoe(t_0)$ can be chosen to prepare conditionally (upon result ${\rm r}_0 \equiv i$, with probability $\wp_i$) a purified environment state of the form
	\begin{align}
		| \phi_i \ra = | \phi_i({\rm e}_0^1) \ra \otimes | \phi_i ({\rm e}_1^2) \ra \ldots \otimes | \phi_i ({\rm e}_{n-1}^n) \ra \otimes | \phi_i ({\rm f}_{n})\ra 
	\end{align}
	where $| \phi_i ({\rm e}_{j-1}^j) \ra$ means the purified bath state on ${\mathbb H}_{{\rm e}_{j-1}^{j}}$.  Note that each component updates by its own unitary unless interacting with the system.  Recall that this initial measurement does not change the average bath state: ${\cal M}_0 \rhoe(t_0) = \sum_i {\cal M}_0^i \rhoe(t_0) = \sum_i \wp_i | \phi_i \ra\la \phi_i | = \rhoe(t_0)$.
	
	Next, we perform a rank-1 measurement (see \srf{sec_quantum_unravelling}), 
	$\hat P_{\rm r_1}^{1}$, on the past part (${\mathbb H}_{\rm p_1}$) at time $t_1$. Note that, as per the notations we introduced in Section~\ref{sec_quantum_unravelling}, the bath measurement at time $t_1$ can be written as ${\cal M}_{\rm r_1}^1 = {\cal P}_{\rm r_1}^1 \otimes {\cal I}_{\rm f_1}$. This measurement will leave the system in a pure state, which can be written as
	\begin{align}
		\label{pf:pfitpu1}
		\hat \pi_{ {\bf r}_1 } (t_1)  = \tre \left[ ( {\cal P}_{\rm r_1}^{1} \otimes  {\cal I}_{\rm f_1} ) \cu_{t_0}^{t_1} {\cal M}^i_0  [ \, |\psi(t_0)\ra\la \psi(t_0) | \otimes  \rhoe(t_0) \, ]   \right] / \wp_{\bf r_1} \ ,
	\end{align}
	where $ \wp_{\bf r_1} $ is the probability of getting the record ${\bf r_1} = {\rm r_0}, {\rm r_1 }$. Note that the measurement on the past part will not affect the future part and its interaction with the system (recall the discussion of PFI in Section~\ref{sec_PFI}). In fact, after time $t_1$ the bath state on ${\mathfrak B}( {\mathbb H}_{\rm p_1})$ is uncorrelated with the system due to the  measurement $\hat P^1_{\rm r_1}$, and the system continues to interact with the bath state on ${\mathbb H}( \rm e_1^2 ) $ up to time $t_2$, when we can perform another rank-1 measurement $\hat P_{\rm r_2}^2$ on $\mathbb{H}_{\rm e_1^2}$ for generating another pure system state $\hat \pi_{\bf r_2}(t_2)$\footnote{Similarly, the bath measurement at time $t_2$ is ${\cal M}_{\rm r_2}^2 = {\cal I}_{\rm p_1} \otimes {\cal P}_{\rm r_2}^2 \otimes {\cal I}_{\rm f_2}$.}.
	By repeatedly doing so, one can have a set of pure system states,~$\{ \wp_{{\bf r}_j}, \hat \pi_{ {\bf r}_j} (t_j) \}_{ {\bf r}_j}$, conditional on the measurement outcomes ${\bf r}_j$ (recall the notations for unravelling in Section~\ref{sec_quantum_unravelling}).

	Now we show that one can reconstruct the dynamical map by performing the ensemble averaging over ~$\{ \wp_{{\bf r}_j}, \hat \pi_{ {\bf r}_j} (t_j) \}_{ {\bf r}_j}$. Note that for any time $t_j$, the rank-1 measurement $\hat P^j_{{\rm r}_j}$ on ${\mathbb H}_{{\rm e}_{j-1}^{j}}$ can be written as $\hat P^j_{ {\rm r}_j} =  | \varphi_j \ra \la {\rm r}_j|$, where $| \varphi_j \ra $ denotes the post-measurement bath state on ${\mathbb H}_{{\rm e}_{j-1}^{j}}$.
	The  system state at time $t_j$ then can be explicitly written as $\hat \pi_{{\bf r}_j}(t_j) = | \psi_{\bf r_j}(t_j) \ra_{\rm s} \la \psi_{\bf r_j} (t_j) | $, where 
	\begin{align}
		| \psi_{\bf r_j}(t_j) \ra_{\rm s}  =   \la {\rm r}_j | \hat U_{t_{j-1}}^{t_j} ({\rm s}, {\rm e}_{j-1}^j) | \phi_i({\rm e}_{j-1}^j, t_{j-1}) \ra   \ldots \la {\rm r}_1 | \hat U_{t_0}^{t_1} ({\rm s, e_0^1}) | \phi_i({\rm e}_0^1) \ra  | \psi(t_0) \ra_{\rm s}  / \sqrt{  \wp_{{\bf r}_j} } \ ,
	\end{align}
	where  $ | \phi_i ({\rm e}_{j-1}^j, t_{j-1}) \ra = \hat U_{t_0}^{t_{j-1}} ( {\rm e}_{j-1}^j ) | \phi_i ({\rm e}_{j-1}^j) \ra$. These factorization relations directly follow from the discussion before the proof of Theorem~\ref{PFI-GQRF}. By performing the ensemble average, one would have
	\begin{align}
		\label{pfi-mpu-1}
		{\rm E} [  \hat \pi_{{\bf r}_j} ] & = \sum_{{\bf r_j}}  \wp_{{\bf r}_j} \hat \pi_{{\bf r}_j}  \\
		& = \sum_{{\bf r_j} }  \la {\bf r}_j | \hat U_{t_0}^{t_j} ({\rm s}, {\rm e}_0^j) \hat \pi(t_0) \otimes |  \phi_i({\rm e}_0^j) \ra \la  \phi_i({\rm e}_0^j) |  \hat U_{t_0}^{t_j \dagger} ( {\rm s}, {\rm e}_0^j ) | {\bf r}_j \ra  \\
		& = \tr_{\rm e}[ \hat U_{t_0}^{t_j} \hat \pi(t_0) \otimes  \rhoe(t_0) \hat U_{t_0}^{t_j \dagger}  ] = \ce_{t_0}^{t_j} \hat \pi(t_0) \label{pfi-mpu-end} \ .
	\end{align}
	The last line, Eq.~\eqref{pfi-mpu-end}, follows from the discussion in the proof of Theorem~\ref{PFI-GQRF}. 
	
	We have thus shown that there exists at least one PU for PFI. We now show that there are in fact infinitely many, for any non-trivial PFI dynamics. That is, for any dynamics that correspond to an OQS, rather than being unitary for the system. This entails that there must be at least one time interval for which $${\rm Tr}_{{\rm e}_j^{j+1}}\left\{  {\cal U}_{t_{j}}^{t_{j+1}}({\rm s},{\rm e}_{j}^{j+1}) [\hat \pi_{{\bf r}_{j}} \otimes \rho_{{\rm e}_j^{j+1}}(t_j) ]\right\}$$ is a mixed state 
	for some $\hat \pi_{{\bf r}_{j}}$ with $\wp_{{\bf r}_{j}}\neq 0$. Thus, according to Lemma~\ref{lemma-PFI}, there are two options.
	
	The first option is that the system state $\hat \pi_{{\bf r}_{j}}$ evolves to become a mixed state, isometric to  $\rho_{{\rm e}_j^{j+1}}(t_j)$, and hence the latter state must be mixed.   
	But at the initial step ${\cal M}_0$ can realize any pure-state ensemble $\{ \wp_{i|j}, | \phi_j ({\rm e}_{j}^{j+1}) \ra \}$ such that $\sum_i \wp_{i|j} | \phi_i ({\rm e}_{j}^{j+1}) \ra \la \phi_i ({\rm e}_{j}^{j+1}) | = \rho_{{\rm e}_{j}^{j+1}}(t_0)$. Each different ensemble leads to a different set of possible conditioned states at the system, $\ket{\psi_{{\bf r}_{j+1}}}$, where in this case the only part of ${\bf r}_{j+1}$ that is relevant is ${\rm r}_0 = i$. Since there are infinitely many different pure state ensembles representing the mixed state $\rho_{{\rm e}_{j}^{j+1}}(t_j)$, it follows that there are infinitely many different PUs, by Definition~\ref{dfmpu}. 
	
	The second option is that, for some choice of initial measurement ${\cal M}_0$,  there exists some $| \phi_i ({\rm e}_{j}^{j+1}\ra$ with $\wp_i > 0$ such that $\hat U_{t_{j}}^{t_{j+1}}({\rm s},{\rm e}_{j}^{j+1}) [| \phi_i ({\rm e}_{j}^{j+1},t_j) \ra \otimes \ket{\psi_{{\bf r}_{j}}} ]$ is a pure entangled state. Therefore, by choosing different measurement basis for the environment, i.e., $\{| r_{j+1} \ra\} $,  one can project the system into any pure-state ensemble that represents the (by assumption) mixed state for the system at time $t_{j+1}$. This is the idea of pure-state steering proposed by  \sch\ in Ref.~\cite{Sch35}. Once again it follows that there are infinitely many different PUs, by Definition~\ref{dfmpu}.
	
	Therefore we conclude that PFI implies MPU. However it is worth remarking that the arguments in both the first and second options above assume that there is no limitation on the measurements that can be performed on the environment. In particular, we assume that there are no super-selection rules that limit what observables can be measured or what preparations are possible~\cite{BRS07}. For example, a classical environment could  be modelled by a quantum environment with a super-selection rule that forbids measurement in anything other than a preferred, classical basis, which is also the diagonal basis in which the environment is prepared.
	
	The converse result, that MPU implies PFI, does not hold. Consider a model as in Fig.~\ref{fig_fb2}, in which the system is coupled to two independent environments or baths (1 and 2), with each coupling satisfying PFI. Assume, for simplicity, that the initial environment states are pure, and that they are both monitored perfectly, so that the conditioned state of the system is always pure. Now suppose that the signal from monitoring bath 2 is used to estimate the system state, and that this estimate is used to control the system via a feedback loop. As we discussed in the proof of Theorem~\ref{NIBtoNQIB}, the estimation will, in general,  involve measurement records from the past, as well as the immediately obtained record. Thus the system dynamics as modified by the feedback will not satisfy PFI in general. Now for a fixed average system dynamics, the unravelling of bath 2 must be fixed. However the unravelling of bath 1 is still arbitrary. Thus there will still be many pure unravellings possible for this non-PFI system. This proves that MPU does not imply PFI. 
\end{proof}

Consistent with intuition, and also as  suggested by their names, MPU is strictly a special case of PU, as stated in the following theorem.

\indent

\begin{theo}[label = MPUtoPU]{}{}
	The existence of many pure unravellings (MPU)  as per Definition~\ref{dfmpu} is  a sufficient but not a necessary condition for the existence of a  pure unravelling as per Definition~\ref{dfpu}.
\end{theo}

\begin{proof}
	The sufficient direction holds trivially. To show the necessary direction does not hold, consider the model we used in the proof of Theorem~\ref{NIBtoNQIB}, but in its general case, where an initial pure system state is coupled to an environment with initial state of the form $\rhose(t_0) = | \psi \ra\la \psi | \otimes \sum \wp_j | j \ra\la  j |$, through the interaction Hamiltonian $\hat H_{\rm int} = \sum_j \hat H_{\rm s}^j \otimes | j\ra\la j |$, where $\{|j\ra\}$ are the energy eigenstates of a non-degenerate bath Hamiltonian $\hat H_{\rm e}$. 
	The combined state at time $t > t_0$ is then $\rhose(t) = \sum_j \wp_j e^{-i (\hat H_s +  \hat H^j_{\rm s})t} | \psi \ra\la \psi | e^{ i (\hat H_{\rm s} + \hat H^j_{\rm s})t } \otimes | j \ra \la j |$.
	It is obvious from this expression that a pure unravelling scheme for the system can be realized by applying  projective measurements $\{|j\ra\la j |\}$ on the environment, and that this is the unique pure unravelling. That is, in this case we could have considered a classical environment, with measurement in any basis other than $\{|j\ra\la j |\}$ forbidden by a super-selection rule. 
\end{proof}

Another natural consequence from PU is given by the following theorem.

\indent
\begin{theo}[label = NQIBtoPU]{}{}
	The existence of a pure unravelling as per Definition~\ref{dfpu} is a sufficient condition for no quantum information backflow (NQIB) in Definition~\ref{def_NQIB}.
\end{theo}

\begin{proof}
	
	A pure unravelling as per Definition~\ref{dfpu} requires making a  measurement on the environment at each time $t_j$ that leaves the system in a pure state. Hence, the joint state immediately after each measurement must be a tensor product, and so break any
	entanglement between the system and the bath~\cite{HSR03}. Furthermore, Eq.~\eqref{eq:dfPU} for pure unravellings requires that these entanglement-breaking  measurements do not affect the system state on average, which is in agreement with Eq.~\eqref{eq:dfNQIB} for NQIB. That is, there is no quantum information backflow.
	%
\end{proof}

Whether the converse holds, i.e., whether NQIB implies the existence of a pure unravelling, is an open question, and is briefly discussed in~\ref{app-conj}.

\subsection{The dynamical map approach} 
\label{sub:the_dynamical_approach}

We have shown the relations between Markovian concepts defined through the system-environment approach, as illustrated within the dotted border in Fig~\ref{fighie}. In this section, we will  investigate the hierarchical relations between those concepts outside the frame, corresponding to the dynamical map approach.  As many of the theorems in this section are well-known results, the reader will be guided to the original proofs or other reviews for details. 

\subsubsection{Semigroups and GKS-Lindblad equations}
\label{sec-semi-ME}

Dynamical semigroups are closely related to GKS-Lindblad master equations, as stated in the following theorem.

\indent

\begin{theo}[label = SemitoME]{}{}
	A dynamical semigroup as in Definition~\ref{df:Dynamical semigroups} is a necessary but not a sufficient condition for the existence of a strict GKS-Lindblad master equation as in Definition~\ref{df:Strict GKS-Lindblad master equation}.
\end{theo}

The necessary direction holds trivially. Given a strict GKS-Lindblad master equation as in Eq.~\eqref{strictLindblad}, it admits a dynamical map for the system, $\ce_{t_0}^t = {\rm e}^{{\cal L}(t - t_0)}$, which obviously satisfies the semigroup property. The sufficient direction fails because Definition~\ref{df:Dynamical semigroups} for dynamical semigroups  includes the case of discrete-time dynamics, where there is no master equation. It can also fail for continuous-time dynamics, with a simple example being given by the dynamical semigroup defined by $\cs_0:={\cal I}_s$ and $\cs_t X:=\sum_n |n\rangle\langle n|\,\langle n|X|n\rangle$ for $t> 0$, where $\{ |n\rangle\}$ is some orthonormal basis for $\hs$. However, it is worth emphasizing that, for continuous-time evolutions, the sufficient direction does hold  whenever the semigroup dynamical map $\cs_{t}$ is norm continuous, as per the well known result in Ref.~\cite{Lin76}.

\subsubsection{Divisibility and master equations}
\label{sec_hie_divisibility_ME}

Divisibility is a generalization of the semigroup property. Similarly to the above, there is a theorem relating it to a more general master equation. 

\indent

\begin{theo}[label = DivtoME]{}{}
	Divisibility in Definition~\ref{dfdivisibility} is a necessary but not a sufficient condition for a time-dependent Lindblad equation as in Definition~\ref{df:time-dependent Lindblad equation}.
\end{theo}

In Ref.~\cite{RHP14} this is also called the GKSL theorem, since the derivation is based on the pioneering works by Gorini, Kossakowski and Sudarshan~in Ref.~\cite{GKS76}, and Lindblad in Ref.~\cite{Lin76},  which studied the time-independent case in detail. For proofs of the time-dependent case see Refs~\cite{RH12,CK12,MJL14}.
The converse direction does not hold, similarly to Theorem~\ref{SemitoME}, since divisibility includes the case of discrete time dynamics, for which there is no differential equation.

It is obvious that the strict GKS-Lindblad master equation  is just a special case of the time-dependent GKS-Lindblad master equation. Thus we have the following trivial theorem.

\indent

\begin{theo}[label = SemitoDiv]{}{}
	A strict GKS-Lindblad master equation as in Definition~\ref{df:Strict GKS-Lindblad master equation} is a sufficient but not a necessary condition for a time-dependent GKS-Lindblad master equation as in Definition~\ref{df:time-dependent Lindblad equation}.
\end{theo}

The time-dependent GKS-Lindblad equation, which is widely used for many different physical systems, is in general hard to solve analytically. This  fact motivates the development of the
%
MCWF simulation method, which shows high efficiency in computation via the use of state vectors rather than density matrices (see discussion in Section~\ref{sec_MCWF}). In Ref.~\cite{MC96}, the authors show that quantum evolutions described by Lindblad-type master equations can always be simulated by MCWF method, and the method can be naturally generalized to discrete-time evolutions using quantum jumps. By Theorem~\ref{DivtoME}, it follows that divisible processes can always be simulated by MCWF method. 



However, MCWF simulation is not restricted to models satisfying divisibility, but also can be applied to systems explicitly identified as non-Markovian, 
as discussed in \srf{sec_MCWF}. In this sense, it is not surprising to see almost all concepts in Fig~\ref{fighie} lead to MCWF simulation. We again emphasize that the pure states generated in a typical non-Markovian MCWF simulation have no physical meaning, 
at least within the standard interpretation of quantum mechanics~\cite{GamWis03,GamWis04};  they need only serve a numerical purpose.

To summarize, we have the following theorem. 
\indent

\begin{theo}[label = MEtoSSE]{}{}
	Divisibility in Definition~\ref{dfdivisibility} is a sufficient but not a necessary condition for Monte Carlo wave function~(MCWF) simulation as in Description~\ref{def-MCWFS}.
\end{theo}
\newpage

\subsubsection{Divisibility and system state distinguishability}
\label{sec:DSSD}
\indent

\begin{theo}[label = DivitoTr]{}{}
	Divisibility in Definition~\ref{dfdivisibility} is a sufficient but not a necessary condition for decreasing system state distinguishability in Definition~\ref{dfdisting}.
\end{theo} 

This theorem is discussed and proved in detail in a recent review~\cite{RHP14}. The sufficient direction follows immediately from Eq.~(\ref{divisible}) and the fact that trace distance is non-increasing under any CPTP map. An analogous result holds if trace-distance is replaced by any other measure of system-state distinguishability that is non-increasing under such maps~\cite{RHP14,BLP16}. 
%
A counterexample for the necessary direction is provided by the `eternally  non-Markovian' qubit master equation  in Ref.~\cite{MJL14}:
\begin{align}
	\frac{{\rm d} \rhos(t)}{{\rm d} t}= \frac{1}{2} \sum_{j=1}^3 \gamma_{j}(t) [ \hat \sigma_j \rhos(t) \hat \sigma_j - \rhos(t) ] \ ,
\end{align}
where $\hat \sigma_j$ are Pauli operators and
\begin{align}
	\gamma_1(t) = \gamma_2 (t) = 1, \gamma_3 (t) = -\tanh t \ .
\end{align}
The associated dynamical map is non-divisible, but the trace distance between any two system states is non-increasing for this model at all times $t>t_0$.

We note that Chru\'sci\'nski {\it et al.}~\cite{CDA11} have shown that the property of decreasing system distinguishability is equivalent to the existence of a positive map ${\cal P}^t_{t'}$ on $\hs$, for any two times $t>t'>t_0$, such that  
\begin{align}
	\ce^t_{t_0} = {\cal P}^t_{t'}\  \ce^{t'}_{t_0}\  .
\end{align}
This property, referred to as $P$-divisibility \cite{VSL11,RHP14,BCD16}, is clearly weaker than the property of divisibility in Eq.~(\ref{divisible}). However, recalling that a map ${\cal Q}$ on $\fbs$ is completely positive if and only if ${\cal Q}\otimes {\cal I}_A$ is positive on ${\mathfrak{ B}}(\hs\otimes \mathbb{H}_A)$, for any ancilla $A$, it follows immediately that requiring decreasing distinguishability to hold for {\it joint} states of a system and any ancilla decoupled from the environment is equivalent to requiring divisibility~\cite{RHP14}.  Buscemi and Datta have recently obtained a similar result, based on the distinguishability of ensembles rather than just pairs of system states \cite{BD16}. \blk

\subsection{Bridging the two approaches}
\label{sec:bridge}

We have clarified the relations within the two approaches in previous sections. Now we complete the last elements of the hierarchy in Fig.~\ref{fighie} by bridging these two classes of Markovian concepts.

\subsubsection{Composability and dynamical semigroups}

The dynamical semigroup is connected, under certain conditions, to composability by the following theorem.

\indent

\begin{theo}[label = ComtoSemi]{}{}
	Suppose that an OQS dynamics $\{{\cal U}_{t_0}^t\}$ and $\rhoe(t_0)$ has a total Hamiltonian $\hat H_{\rm tot}$ that is time-independent, and has a bath state that in some frame is also time-independent. 
	Then the dynamics of the OQS is composable as per Definition~\ref{df_composability} if and only if 
	the family of its dynamical maps forms a semigroup as in Definition~\ref{df:Dynamical semigroups}.
\end{theo}

\begin{proof}
	First, we show there are some important properties of the generalized dynamical map following from the assumptions made in the theorem.
	Note that since $\trhoe(t) \equiv \trhoe $ for all times, it must be the same as the initial environment state, i.e., $\trhoe(t) = \rhoe(t_0)$. Recalling the definition of the generalized dynamical map in Eq.~\eqref{general_dynamical_map}, i.e., 
	\begin{align}
		\label{eq-pr-linc1}
		\tce_{t_1}^{t_2} \hat X := \tre \left[ {\cu}_{t_1}^{t_2} \left( \hat X \otimes \trhoe(t_1) \right)  \right]  = \tre \left[ {\cu}_{t_1}^{t_2} \left( \hat X \otimes \rhoe(t_0) \right)  \right] \, ,
	\end{align}
	for $\hat X \in \fbs$ and $t_2 \geq t_1 \geq t_0$, it follows from the definition of the dynamical map in Eq.~\eqref{def_dynamical_map} that
	\begin{align}
		\label{eq-pr-linc2}
		\tce_{t_0}^{t_2} \hat X = \ce_{t_0}^{t_2} \hat X \, .
	\end{align}
	Therefore, noting that the total Hamiltonian is time-independent, the dynamical map for any time interval can in general be written as
	\begin{align}
		\label{dynamical-map-semi}
		\tce_{t_1} ^ {t_2} \hat X = \tre [ e^{-i \hat H_{\rm tot} (t_2 - t_1)} (\hat X \otimes \trhoe) e^{ i \hat H_{\rm tot} (t_2 - t_1)} ]  \, 
	\end{align}	
	for all $\hat X \in \fbs$. It is easy to see that $\tce_{t_1}^{t_2}$ only depends on time difference, i.e., $\tce_{t_1}^{t_2} = : {\cal S}_{t_2 -t _1}$, and ${\cal S}_0=I$.
	
	Now, assuming that composability in Eq.~\eqref{com_2t} holds, Eqs.~\eqref{eq-pr-linc1} and~\eqref{eq-pr-linc2} immediately imply that 
	$$\cs_{t_2 - t_0} = \cs_{t_2 - t_1} \cs_{t_1 - t_0} \, .$$ 
	Letting $\tau_1 = t_1 - t_0, \, \tau_2 = t_2 - t_1$, it follows that $\cs_{\tau_1 + \tau_2} = \cs_{\tau_1}\cs_{\tau_2}$.
	That is, $\{ \cs_{\tau} :\tau \geq 0\}$ forms a  dynamical semigroup. 
	The necessary direction holds by the same argument. In particular, given the generalized dynamical map in the form of Eq.~\eqref{dynamical-map-semi}, as per the conditions of the Theorem, it immediately follows from the semigroup property that the dynamical map must be composable. 
\end{proof}

Note that the assumptions made in the statement of the theorem imply that we are considering the case of continuous time.  Thus, if one made some further continuity conditions on the dynamical map, one could show that the dynamical semigroup in the above theorem is equivalent to the existence of a strict GKS-Lindblad master equation (see discussions under Theorem~\ref{SemitoME}).

\subsubsection{Environment interventions and divisibility}
\label{sec_eid}

From Definitions~\ref{dfnib} and~\ref{dfdivisibility}, it is clear that NIB bears a formal resemblance to divisibility. However, NIB provides a physical interpretation for the construction of the CPTP map ${\cal Q}_{t_1}^{t_2}$~(see Section~\ref{sec_NIB}), while divisibility emphasizes the existence of $\cq_{t_1}^{t_2}$ (see Section~\ref{sec_divisibility}). The following theorem illustrates the relation between these two concepts.

\indent

\begin{theo}[label = NIBtoDiv]{}{}
	No information backflow (NIB) in Definition~\ref{dfnib} is a sufficient but not a necessary condition for divisibility in Definition~\ref{dfdivisibility}.
\end{theo}
\begin{proof}
	The sufficient direction is trivial by noting that divisibility is a direct consequence from the definition of NIB [see Eq.~\eqref{nib_2t}]. 
	
	The necessary direction does not hold. As a counterexample, we consider a two-atom model (TAM). 
	Given two identical two-level atoms, one (atom 1) is treated as the system and the other (atom 2) as the environment. They are coupled through the interaction Hamiltonian (we ignore $\hat H_{\rm s}$ and $\hat H_{\rm e}$ for simplicity) :
	\begin{align}
		\label{hamiltonia_tam}
		\hat H_{\rm se}(t) = g(t) ( \hat \sigma_{\rm s}^-\hat \sigma_{\rm e}^{+} + \hat \sigma_{\rm s}^{+}\hat \sigma_{\rm e}^-  )  \ , \  g(t) = \frac{1}{\sqrt{e^{2t}-1}} \ ,
	\end{align}
	where $\hat \sigma^{-}$~($\hat \sigma^{+}$) is the lowering (raising) operator, and the time-dependent function $g(t)$ represents the coupling strength, with $t_0 = 0$. Assuming the initial combined state is $\rho_{\rm se}(0) = \rho_s(0) \otimes | 0\ra\la 0| $, it can be shown that the system evolution is described by the strict GKS-Lindblad master equation 
	\begin{align}
		\label{ceta}
		\frac{{\rm d} \rhos(t)}{{\rm d} t} = 2 {\cal D} [\hat \sigma_{\rm s}^-] \rhos(t) \ .
	\end{align}
	(The derivation here is left as an exercise to the reader.)  
	The positive coefficient in the master equation means that the dynamics of the TAM is divisible (recall Theorem~\ref{DivtoME}).
	
	Now decouple these  two atoms at time $t_1$ by performing a replacement  channel ${\cal R}_{t_1}$ such that
	\begin{align}
		\left(  {\cal I}_{\rm s} \otimes {\cal R}_{t_1} \right) \rhose(t_1) = \rhos(t_1) \otimes \sigma_{\rm e}(t_1)  \, .
	\end{align}
	where $\rhos(t_1) = \tre [\rhose(t_1)]$ and $\sigma_{\rm e}(t_1)$ is an arbitrary bath state.  Following the same approach to deriving Eq.~\eqref{ceta}, it is easy to verify that the only way to get a ME of the form of Eq.~\eqref{ceta} is to choose $\sigma_{\rm e}(t_1)$ in Eq.~\eqref{action_r} as  $|0\ra\la 0|$. But this leads to a ME with a time-dependent coefficient after time $t_1$: 
	\begin{align}
		\label{eq-me-TAM}
		\frac{{\rm d} \rhos(t)}{{\rm d} t} =  2\gamma(t) {\cal D}[\sigma_{\rm s}^- ]  \rhos(t) \ , \quad  t > t_1 \ ,
	\end{align}
	where 
	$$\gamma(t) =  \frac{ g(t)g(t_1)- g^2(t) } {  g(t)g(t_1) - 1 } \, .$$ 
	Note that this $\gamma(t)$ will become negative for sufficiently large $t$, and may even always be negative if $t_1$ is sufficiently large. In either case, it is clear
	that the system is affected by such a replacement  operation. That is, the TAM does not satisfy  NIB in Definition~\ref{dfnib}. 
	This completes the proof of Theorem~\ref{NIBtoDiv}.
\end{proof}

As will be discussed in Section~\ref{sec-cl-CK}, the classical analogue of divisibility, classical divisibility, is only a necessary condition for classical Markovianity. As suggested by Theorem~\ref{NIBtoDiv}, in the quantum regime divisibility is similarly a rather weak concept of Markovianity. For example, as the reader may already have noticed, the master equation~(\ref{ceta}) for the TAM is exactly the same as for the spontaneous decay of a two-level atom coupled to a thermal bath~\cite{CH09}. As another example, the AFL model admits the same master equation as that of a pure dephasing process (See~\ref{app-AFL}). This observation, that different physical models can have the same master equation description, shows that detailed information about the total dynamics is lost when employing dynamical map approach, where such information is crucial for characterizing many concepts related to quantum Markovianity. 

\subsubsection{Quantum white noise, master equations and pure unravellings}
\label{sec-QWN-ME-PU}

To complete the last elements of the hierarchy in Fig.~\ref{fighie}, we now consider the relations between stochastic descriptions and quantum unravellings, as per the following two theorems.

\indent

\begin{theo}[label = QWNtoME]{}{}
	Quantum white noise (QWN) in Definition~\ref{def_QWN} is a sufficient but not necessary condition for a time dependent GKS-Lindblad equation as in Definition~\ref{df:time-dependent Lindblad equation}.
\end{theo}

\begin{proof}
	To prove the sufficient direction, one can formally write a stochastic differential equation for the system from  Eq.~\eqref{Utpdt}, 
	\begin{align}
		\label{st-system}
		{\rm d} \rhos(t)  = \tre[ {\rm d} \hat U_{t_0}^{'t} \rhose(t_0) \hat U_{t_0}^{'t \dagger}  +  \hat U_{t_0}^{'t} \rhose(t_0) {\rm d} \hat U_{t_0}^{'t \dagger}  +  {\rm d} \hat U_{t_0}^{'t} \rhose(t_0) {\rm d} \hat U_{t_0}^{'t \dagger}  ] \ .
	\end{align}
	By choosing the initial total state as $\rhose(t_0) = \rhos(t_0) \otimes | \bm \beta \ra\la\bm \beta |$ where ${\rm d}\hat{\bm B}(t) | \bm \beta \ra = {\bm \beta}(t) {\rm d}t $ (see discussion in Section~\ref{sec_quantum_white_noise}), and using the It\^{o} rules as per Eqs.~\eqref{wqnBL}-\eqref{coherentinput}, Eq.~\eqref{st-system} leads to the master equation 
	\begin{align}
		\frac{{\rm d} \rhos(t)}{{\rm d} t} = -i[ \hat H + (\hat{\bm L}^\dagger\underline{\hat S}{\bm \beta(t)} - {\bm \beta}^\dagger(t)\underline{\hat S}^\dagger \hat{\bm L} ) /  (2i) , \rhos(t) ] +  {\cal D}[ {\hat{\bm L}} + \underline{\hat S}{\bm \beta}(t) ] \rhos(t) \, 
	\end{align}
	for the system. Here we have broadened the definition in Eq.~(\ref{superD}) to allow ${\cal D}$ to take a vector argument, by defining ${\cal D}[\hat{\bm L}] = \sum_{j=1}^K {\cal D}[{\hat L_j}]$.
	Detailed discussions are given in Refs.~\cite{GC09,CKS17}.
	
	The necessary direction does not hold in general. For example, the stochastic terms in Eq.~\eqref{Utpdt} are defined in terms of field operators. But there is no stochastic evolution equation in this sense for the TAM discussed in the proof of Theorem~\ref{NIBtoDiv}, as the combined system merely consists of two qubits.
\end{proof}

Recalling the discussion in Sections~\ref{sec_quantum_unravelling} and~\ref{sec_MCWF}, both PU and MCWF provide a stochastic description of system dynamics. The relation between them is given by the last theorem of this section.

\indent

\begin{theo}[label = PutoSSE]{}{}
	Pure unravelling (PU) in Definition~\ref{dfpu} is a sufficient but not necessary condition for Monte Carlo wave function (MCWF) simulation in Description~\ref{def-MCWFS}.
\end{theo}

It is easy to understand the sufficient direction, since by PU one can generate ensembles of pure system states,  from which the statistical information about the system can be recovered---a procedure that MCWF simulation aims for. To see that the necessary direction does not hold, note that MCWF simulation can be performed for an OQS coupled to a (nonclassical) bath with finite correlation time, for which a PU does not exist as proved in Refs.~\cite{GamWis03,HJ08}.

\section{An analogous classical hierarchy} 
\label{sec_acsp}

\begin{figure}[htbp]
	\centering
	\includegraphics[width=1.\linewidth]{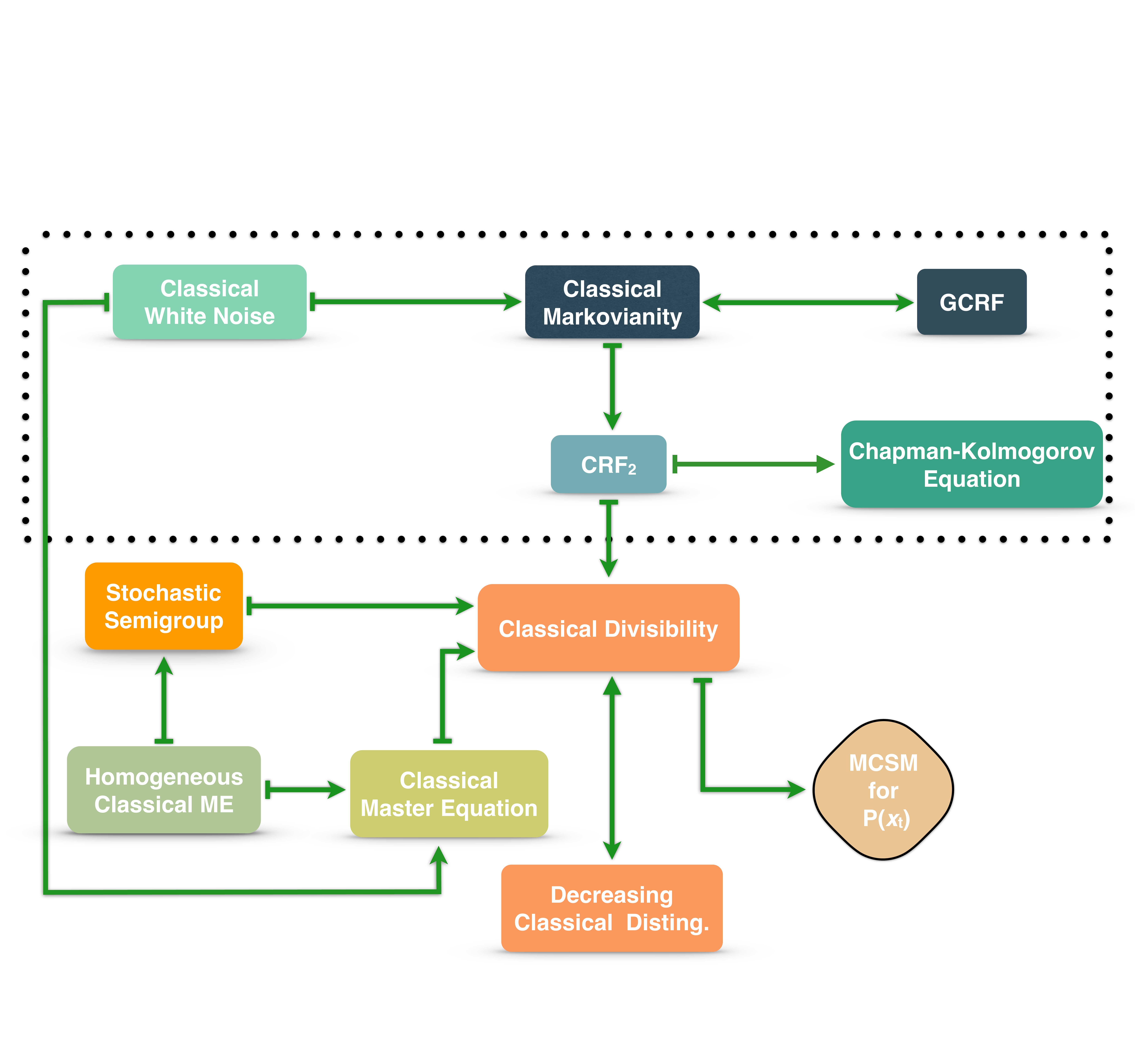}
	\caption{Classical analogue to the quantum hierarchy of non-Markovianity. We use the following abbreviations:
	GCRF for general classical regression formula; CRF$_2$ for the two-time classical regression formula;
	ME for Master Equation; and MCSM for Monte Carlo simulation method. The concepts within the dotted frame require specification of the {\em stochastic process} at multiple times, while those outside are based merely on the properties of transition matrices. Note that MCSM, like MCWF in the quantum case, appears in a differently styled box because it is not a formally-defined Markovianity-related concept (and therefore has a black border line as in the quantum hierarchy). Arrows indicate the same logical relation as those in the quantum hierarchy figure, Fig.~\ref{fighie}. The colours match analogous quantum concepts in Fig.~\ref{fighie}, where applicable. See text for detailed explanations.} 
	\label{fig_cl}
\end{figure}

In the previous sections we have formulated  a hierarchy of possible definitions of quantum Markovianity, as per Fig.~\ref{fighie}. The necessity (and complexity) of such a hierarchy arises because there is no direct quantum analogue of the definition of classical Markovianity (CM) in Eq.~(\ref{def_classical_Markovianity}). As discussed in Section~\ref{intro}, quantum properties  at different times cannot be represented by a joint probability distribution, and nor does  the quantum state --- an obvious quantum generalization of a probability distribution --- exist across multiple times. However, a number of the quantum concepts we cover are either physically or mathematically analogous to particular aspects of classical Markovianity. In this section we will explore the classical analogue of our quantum hierarchy, as summarized in Fig.~\ref{fig_cl}.

In the quantum case, we made a fundamental distinction between two classes of concepts: those that required the environment to formulate them (via $\rhoe$ and ${\cal U}$), and those that could be formulated purely in terms of the map $\ce$ relating the system state $\rhos$ at different times. In the classical case the term `environment' never enters the mathematicians' lexicon when defining Markovianity or related concepts. However, the quantum--classical difference here is not as great as it seems. There are also two classes of concepts related to classical Markovianity. Those of the first class, including CM itself, as well as classical regression formulas, the Chapman-Kolmogorov equation and classical white noise, involve consideration of the {\em stochastic process}, whether formulated in terms of a multi-time probability distribution, or as a correlation function for the process across multiple times. Those of the second, weaker, class --- including the semigroup property, classical divisibility, and its differential form -- can be formulated in terms of a {\em transition matrix} $T$ relating the single-time probability distribution at different times.  

Classical concepts in the {\em stochastic process} class are analogous to quantum processes that involve the environment. Indeed, in some instances, the definitions of concepts are almost identical. The quantum regression formula, involving correlations of system operators at multiple times, looks almost identical to the classical regression formula, involving correlations of system variables at multiple times. The former is an `environment' concept in that to consider, in the same formula, system observables at different times requires use of the \hei\ picture, in which system operators at the initial time become `contaminated' by environment operators which can be thought of as ``quantum noise'', as discussed in Section~\ref{sec_oqs:HP}. In the classical case the system variables are similarly `contaminated' by classical noise. This classical noise comprises random variables that are distinct from the system variables under consideration. 

If, as we do in the quantum case (by the assumption of unitary evolution), we postulate that all physical processes are fundamentally reversible, then the origin for classical noise in a classical stochastic process must lie in something physical which is external to the system of interest. That is, it must lie in the environment. It is noteworthy that the mathematical study of stochastic processes began with physicists in the early 20th century~\cite{SB68}, with the elucidation of Brownian motion by Sutherland~\cite{WS05}, Einstein~\cite{AE05}, and Langevin~\cite{LP08}. For these physicists, the noise variable appearing in their classical equations was explicitly understood to be of environmental origin: the liquid molecules battering the particle under study. 

For mathematicians, the physical origins of the noise in their stochastic processes are irrelevant, and we will not make explicit mention of the environment in our definitions below. If we were to introduce the environment explicitly then we could introduce further analogues to quantum concepts of Markovianity, such as the factorization approximation, the failure of dynamical decoupling, past--future independence,  no information backflow, and composability. Moreover these could be defined solely in terms of the stochastic process itself, as mathematicians do, by formulations along the lines of ``there exists some initial probability distribution over a set of environmental variables, and some logically reversible dynamics on these variables and the system variables, which generate the stochastic process for the system, such that \ldots''. As conceptually interesting as such a discussion may be, in the interests of brevity we will consider only the type of classical concepts that are generally considered by mathematicians. This leads to a classical hierarchy (Fig.~\ref{fig_cl}) much simpler than the corresponding quantum one (Fig.~\ref{fighie}).

\subsection{Concepts deriving from the stochastic process}
\label{sec-cl-process}

Mathematically, a classical stochastic process can be represented by a 1-parameter family of random variables, often denoted as $\{\bm X (t): t \geq t_0 \}$. Formally, a random variable can be defined as a function which maps the {\em sample space} to some set, e.g., the real numbers~\cite{GS02}. Here $\bm X(t)$ is a vector of random variables, encoding the system configuration. We use the term `configuration', rather than `state', to avoid any suggestion that the reader should  regard $\bm X(t)$ as being analogous to the quantum state $\rho(t)$. The configuration \blk takes the value $\bm x_t$ at time $t$ with probability ${\rm Pr}( \bm X(t) = \bm x_t)$. For simplicity, we employ the following notations:
\begin{align}
	& P(\bm x_t) := {\rm Pr} ( \bm X(t) = \bm x_t ) \ , \\ 
	\label{notate}
	& P(\bm x_{t_2} | \bm x_{t_1}) := {\rm Pr} ( \bm X(t_2) = \bm x_{t_2} | \bm X(t_1) = \bm x_{t_1} ) \ ,\\
	& P(\bm x_{t_n}, \dots, \bm x_{t_0}) := {\rm Pr} ( \bm  X(t_n)
 = \bm x_{t_n}, \ldots, \bm X(t_1) = \bm x_{t_0} ) \ ,
\end{align}
for any $t_n > t_{n-1} > \ldots > t_0$. Now we proceed to the discussion of classical Markovian concepts based on the properties of the process $\{ \bm X(t): t \geq t_0 \}$.

\subsubsection{Classical Markovianity}

The standard definition of classical Markovianity in Eq.~(\ref{def_classical_Markovianity}) can be formally stated as follows:
\begin{tcolorbox}
	\begin{df}[Classical Markovianity]
		\label{def_CM}
A classical stochastic process $\{\bm X(t): t \geq t_0 \}$ satisfies classical Markovianity (CM) if and only if
\begin{equation}
	\label{cmdef}
	P(\bm x_{t_n} | \bm x_{t_{n-1}},\dots, \bm x_{t_0})=
	P(\bm x_{t_n} | \bm x_{t_{n-1}}) 
\end{equation}
for all $t_n>t_{n-1}>\dots>t_0$.
\end{df}
\end{tcolorbox}
As discussed in Section~\ref{intro}, this captures the notion of memoryless evolution in an obvious way. The same remarks as made in Section~\ref{sec_oqs:IF} about applicability to either discrete or continuous time also apply here. 

\subsubsection{Classical white noise}
\label{sec-cl-CWN}

The analogue of quantum white noise in Section~\ref{sec_quantum_white_noise} is classical white noise. This applies only for continuous-time evolution and can be defined as follows:
\begin{tcolorbox}
	\begin{df}[Classical White Noise]
\label{def_cwn}  
	A classical stochastic process $\{\bm X(t): t\geq t_0 \}$ is driven by classical white noise (CWN) if $\bm x_t$ satisfies a stochastic differential equation of the form
		\begin{align}
		\label{classical_sde}
			{\rm d}  {\bm x}_t = \bm A({\bm x}_t, t) {\rm d} t + B({\bm x}_t, t)  {\rm d} {\bm W}(t) + C({\bm x}_t, t) {\rm d} \bm N (t)\;,
		\end{align}
	where ${\rm d} {\bm W}$ and ${\rm d}{\bm N}$ are increments of the Wiener and jump processes respectively, which are both assumed to be vector-valued to allow full generality.
\end{df}
\end{tcolorbox} 

The first term on the right hand side of Eq.~\eqref{classical_sde} describes the deterministic component of the process, and $\bm A(\bm x_t , t)$ is called the drift vector~\cite{GC09}.
The remaining  two terms are stochastic terms. 

In the first stochastic term, ${\rm d}{\bm W}(t)$ is the increments of a vector Wiener process ${\bm W}(t)$, defined by
\begin{align}
	{\rm d}{\bm W}(t) = {\bm W}(t +{\rm d}t ) - {\bm W}(t) \ ,
\end{align}
where $W(t)$ consists of statistically independent components, each of the form 
\begin{align} 
\label{Wint}
W_j(t) := 
\int_{t_0}^t \xi_j(s)ds \ . 
\end{align}
Here $\xi_j(t)$ is a Gaussian white noise process, defined by the moments
\begin{align}
		{\rm E} [ \xi_j(t)] = 0 \ , \quad {\rm E} [ \xi_j (t) \xi_k (t') ]  = \delta_{jk} \delta(t-t') \ .
\end{align}
The singular correlation function here is directly related to the singular commutation relations in the case of quantum white noise, Eq.~(\ref{singcomm}), while the integral for $W_{j}$ in \erf{Wint} is analogous to that of the quantum Wiener increment ${\rm d}\hat B_{j}$ in \erf{wqnBL}. The singularity implies that $\{W_j(t), t \geq t_0 \}$ is a continuous but non-differentiable function (for a rigorous proof, see Ref.~\cite{AR90}), characteristic of a diffusion process. Thus, the matrix product $D(\bm x_t,t) := B(\bm x_t , t) B^\top(\bm x_t, t)$ is known as the diffusion matrix~\cite{GC09} (see also  Section~\ref{sec-cl-dCKE} below). 

In the second stochastic term, ${\rm d} \bm N (t)$ is the increment in a vector of counting processes ${\bm N}(t)$. That is, each element $N_j(t)$ is an integer that can only increase with time. Thus ${\rm d} N_j(t)$ takes a value of either 1 or 0, i.e., $({\rm d} N_j(t))^2 = {\rm d}N_j(t)$. The increments in the on-diagonal elements of the matrix of quantum operators $\hat{\Lambda}(t)$ defined in \erf{wqnBL} satisfy exactly the same relation. The statistics of $N_j(t)$ are determined solely by the 
single-time probability that ${\rm d}N_j(t)= 1$. In order for the expectation value of 
$N_j(t)$ to remain finite,  this probability must be of order ${\rm d}t$. Since it may also depend on the system configuration~\cite{BP81}, so we have 
 \begin{align}
 	{\rm Pr}({\rm d}N_j(t)=1)= {\rm E} [ {\rm d} N_j(t) ] = \lambda_j( \bm x_t , t) {\rm d} t  \ .
 \end{align}
Whenever ${\rm d}N_j(t)=1$, for some $j$, there is, from \erf{classical_sde}, 
 a discontinuous change in the system configuration ${\bm X}$. 
 Thus, $N_j(t)$ can be interpreted as the total number of jumps, of type $j$, 
 accumulated from time $t_0$ till time $t$, and $\lambda_j \geq 0$ can be interpreted 
 as the jump rate. 
%
Finally, the effect of each type of jump on the system configuration is determined by the 
matrix $C(\bm x_t, t)$.

From the above definitions, it is easy to verify that CWN implies $\{{\bf X}(t)\}$ is a classical Markov process according to definition~\ref{def_CM}. It is less obvious, but can be shown, that it is the most general Markov process in continuous time, subject to certain continuity conditions on $P({\bf x}_t)$. For details, see for example Ref.~\cite{RW00}.

\subsubsection{Classical regression formula}
\label{sec-cl-CRF}\label{sec:nonlinearP} 

The classical analogue of the general quantum regression formula discussed in Section~\ref{sec_correlation_function} is given by:

\begin{tcolorbox}
	\begin{df}[General Classical Regression Formula]
\label{dfgcrf}
 A classical stochastic process $\{\bm X(t): t \geq t_0 \}$ satisfies the general classical regression formula (GCRF) if and only if its multitime correlation functions can be calculated via 
\begin{align}
\label{gcrf}
	& \la  A_n ( \bm x_{t_n} ) A_{n-1}(\bm x_{t_{n-1}}) \dots\, A_0(\bm x_{t_0} )\ra  \nonumber \\
&~~~~=   \int {\rm d}  \mu ( \bm x_{t_n} ) \dots  {\rm d }  \mu ( \bm x_{t_0}  ) P(\bm x_{t_n} | \bm x_{t_{n-1}})P(\bm x_{t_{n-1}} | \bm x_{t_{n-2}})\dots P(\bm x_{t_1} | \bm x_{t_0}) P( \bm x_{t_0} ) \, A_n( \bm x_{t_n})\dots\, A_0 ( \bm x_{t_0} ) \ ,
\end{align}
for all time sequences $t_n > t_{n-1}>\dots>t_1>t_0$ and initial densities $P(\bm x_{t_0})$. Here the $A_j(\bullet)$ are arbitrary functions, and ${\rm d}\mu ( \bm x)$ is the measure on configuration space, such that $\int {\rm d}  \mu(\bm x)P({\bm x})=1$. 
\end{df}
\end{tcolorbox}

Thus, the average of a product of quantities calculated at different times may be evaluated by replacing the joint probability density $P(\bm x_{t_n} \dots,\bm x_{t_1} , \bm x_{t_0})$ by $P(\bm x_{t_n} | \bm x_{t_{n-1}}) \dots P(\bm x_{t_1} | \bm x_{t_0}) P( \bm x_{t_0})$. Note that, unlike the quantum case in Definition~\ref{def_GQRF}, the order of the terms in the product on the left-hand side is irrelevant because all terms commute.\footnote{One could alternatively consider defining GCRF using the classical propagator $G(t'|t)$ of the process~\cite{HTGT78,HT82} in place of $P(\bm x_{t'}|\bm x_t)$, analogously to the alternative definition of GQRF mentioned in footnote~\ref{footgqrf}. However, the physical interpretation and motivation for such a definition is unclear, given that the propagator can take negative values in general.}

One can show that GCRF is in fact equivalent to classical Markovianity, as indicated in Fig.~\ref{fig_cl}. First, since $A_0,A_1,\dots,A_n$ are arbitrary functions, GCRF is equivalent to the factorization property 
\begin{equation} \label{factor}
P(\bm x_{t_n} \dots,\bm x_{t_1} | \bm x_{t_0}) = P(\bm x_{t_n} | \bm x_{t_{n-1}}, \dots, \bm x_{t_1}, \bm x_{t_0}) \dots P(\bm x_{t_1} | \bm x_{t_0}) 
= P(\bm x_{t_n} | \bm x_{t_{n-1}}) \dots P(\bm x_{t_1} | \bm x_{t_0}) \ , 
\end{equation}
for any set of times $t_n > \dots > t_1 > t_0$, where the first equality follows from Bayes rule. Clearly, this property is implied by CM in Eq.~(\ref{cmdef}).  Conversely, note for $n=2$ that the second equality simplifies to
\beq \label{crf2}
P(\bm x_{t_2}|\bm x_{t_1}, \bm x_{t_0})=P(\bm x_{t_2}|\bm x_{t_1}) \ 
\eeq 
for all $t_2>t_1>0$, corresponding to Eq.~(\ref{cmdef}) for $n=2$. Recursively considering $n=3,4,\dots$ yields CM for arbitrary $n$.

It is also of interest to consider the $n$th-order classical regression formula, CRF$_n$,  corresponding to a given value of $n$ in either of Eqs.~(\ref{gcrf}) and~(\ref{factor}).  Note that CRF$_1$ corresponds to the trivial identity $ P(\bm x_{t_1} | \bm x_{t_0})= P(\bm x_{t_1} | \bm x_{t_0})$, and hence always holds.  However,  CRF$_2$, which is defined as follows, corresponds to the nontrivial factorization property 
$P(\bm x_{t_2},\bm x_{t_1}|\bm x_{t_0}) = P(\bm x_{t_2} | \bm x_{t_1}) P(\bm x_{t_1} | \bm x_{t_0})$, and is the classical analogue of QRF in Eqs.~(\ref{qrf1})-(\ref{qrf2}): 
\begin{tcolorbox}
	\begin{df}[Classical Regression Formula]
\label{df_crf}
 A classical stochastic process $\{\bm X(t): t \geq t_0 \}$ satisfies the classical regression formula (CRF$_2$) if and only if its two-time correlation functions can be calculated via 
\begin{align}
\label{crf}
	& \la  A_2 ( \bm x_{t_2} ) A_{1}(\bm x_{t_1}) A_0(\bm x_{t_0} )\ra 
	 \nonumber \\
&~~~~=   \int {\rm d}  \mu ( \bm x_{t_2} )  {\rm d }  \mu ( \bm x_{t_1}  ) {\rm d }  \mu ( \bm x_{t_0}  ) P(\bm x_{t_2} | \bm x_{t_{1}})P(\bm x_{t_{1}} | \bm x_{t_{0}}) P(\bm x_{t_0}) \, A_2( \bm x_{t_2})\, A_1( \bm x_{t_1} ) \, A_0 ( \bm x_{t_0} ) \ ,
\end{align}
 for all time sequences $t_2>t_1>t_0$ and initial densities $P(\bm x_{t_0})$.
\end{df}
\end{tcolorbox}
More generally, as is partially indicated in Fig.~\ref{fig_cl}, these formulas form a strict hierarchy. To be specific, if CRF$_n$ holds for some classical stochastic process, then ${\rm CRF}_{n-1}$ also holds. This is trivially true as one can simply choose $A_n(\bm x_{t_n})=1$ in Eq.~(\ref{gcrf}). However, the reverse implication is not valid in general, and a family of explicit counterexamples is given in~\ref{sec:hCRF}.

Finally, it may be noted that for a general classical stochastic process, the two-time conditional probabilities $P(\bm x_{t_2}|\bm x_{t_1})$ in the above definitions for GCRF and CRF$_2$ will typically depend nonlinearly on the initial probability density $P( \bm x_{t_0})$~\cite{HTGT78,HT82}, via
\beq \label{hanggi}
P(\bm x_{t_2}|\bm x_{t_1})= \frac{\int {\rm d }  \mu ( \bm x_{t_0}  ) P(\bm x_{t_2}, \bm x_{t_1}| \bm x_{t_0})  P( \bm x_{t_0}) }{ \int {\rm d }  \mu ( \bm x_{t_0}  ) P( \bm x_{t_1}| \bm x_{t_0}) P( \bm x_{t_0})} = \frac{\int {\rm d }  \mu ( \bm x_{t_0}  ) P(\bm x_{t_2}| \bm x_{t_1}, \bm x_{t_0}) P(\bm x_{t_1}| \bm x_{t_0}) P( \bm x_{t_0}) }{ \int {\rm d }  \mu ( \bm x_{t_0}  ) P( \bm x_{t_1}| \bm x_{t_0}) P( \bm x_{t_0})} \ .
\eeq
However, if GCRF is satisfied, or even merely CRF$_2$, then any such dependence factorizes out via Eq.~(\ref{crf2}), implying all memory of the initial configuration is removed. The assumption that the stochasticity  in the evolution for $t>t_0$ is independent of $P( \bm x_{t_0})$ is analogous to the assumption of initial system-bath factorization, \erf{initialfactorization}, which was fundamental to our entire quantum analysis.

\subsubsection{Chapman-Kolmogorov equation}
\label{sec_cke}

The Chapman-Kolmogorov equation is a composition property of the two-time conditional probability density:
\begin{tcolorbox}
	\begin{df}[Chapman-Kolmogorov Equation] 
		The two-time conditional probability density of a classical stochastic process $\{\bm X(t): t \geq t_0 \}$ satisfies the Chapman-Kolmogorov equation (CKE) if and only if 
		\begin{align} \label{cke}
		P(\bm x_{t_3}| \bm x _{t_1}) = \int {\rm d}{\mu(\bm x_{t_2})} P( \bm x_{t_3} | \bm x_{t_2} )
		P( \bm x_{t_2}| \bm x_{t_1} ) \ , 
		\end{align}
		for all times $t_3\geq t_2\geq t_1\geq t_0$ and initial probability densities $P(\bm x_{t_0})$.
\end{df}
\end{tcolorbox}

As is indicated in Fig.~\ref{fig_cl}, CKE is not only weaker than CM \cite{FW59}, but is weaker than the classical regression formula CRF$_n$ for any $n\geq 2$. 
To show this, first recall that CRF$_2$ is equivalent to taking $n=2$ in Eq.~(\ref{factor}). Integrating over $\bm x_{t_1}$  then gives 
\begin{equation} \label{cke0}
P(\bm x_{t_2} | \bm x_{t_0}) = \int {\rm d}{\mu(\bm x_{t_1})} P( \bm x_{t_2} | \bm x_{t_1} )
P( \bm x_{t_1} | \bm x_{t_{0}} ) \ ,
\end{equation}
which, applied recursively, yields
\[ P (\bm x_{t_3} | \bm x_{t_0}) = \int {\rm d}{\mu(\bm x_{t_2})} P ( \bm x_{t_3} | \bm x_{t_2} )
P( \bm x_{t_2} | \bm x_{t_{0}} ) = \int {\rm d}{\mu(\bm x_{t_1})} {\rm d}{\mu(\bm x_{t_2})} P( \bm x_{t_3} | \bm x_{t_2} )
P( \bm x_{t_2} | \bm x_{t_{1}} ) P( \bm x_{t_1} | \bm x_{t_{0}} )\ . \]
Subtracting this result from Eq.~(\ref{cke0}) with $t_2$ replaced by $t_3$, then multiplying by $P(\bm x_{t_0})$ and integrating over $\bm x_{t_0}$, leads to
\[ \int {\rm d}{\mu(\bm x_{t_1})} P( \bm x_{t_1})\left[ P( \bm x_{t_3} | \bm x_{t_1} ) -\int {\rm d}{\mu(\bm x_{t_2})} P( \bm x_{t_3} | \bm x_{t_2} ) P( \bm x_{t_2} | \bm x_{t_{1}} )
\right] =0 \ 
\]
for all possible densities $P( \bm x_{t_1})$, implying Eq.~(\ref{cke}) as desired. Finally, a counterexample by Feller \cite{FW59} shows that the converse does not hold in general (see also the discussion of semi-Markov processes in Refs.~\cite{FW50,VSL11}).

\subsection{Concepts deriving from the transition matrix}
\label{sec_transition_matrices}

The one-time probability density $P(\bm x_t)$ of a classical statistical process is related to the initial probability density via $P(\bm x_t) = \int {\rm d}{\mu(\bm x_{t_0})} P(\bm x_t|\bm x_{t_0}) P(\bm x_{t_0})$.  Thus, the classical conditional probability density $ P(\bm x_t|\bm x_{t_0})$, also called  the transition matrix, plays a role to analogous to the quantum dynamical map $\ce^t_{t_0}$ for open quantum systems.	Note that, unlike general two-time probability densities in Eq.~(\ref{hanggi}), the transition matrix is always independent of the initial probability density $P(\bm x_{t_0})$~\cite{HTGT78,HT82}. To emphasise this, we will write it in the form 	 
\begin{align}
\label{def_transition_matrices}
	T(\bm x_{t}, \bm x_{t_0}) := P( \bm x_{t} | \bm x_{t_0} )  \  ,
\end{align}
for any $t \geq t_0$, with no assumption that the process is Markovian. For a discrete-valued stochastic process, $T$ is indeed a matrix, depending on $t$,  with elements $T(\bm x_{t}, \bm x_{t_0})$. We will also use the matrix terminology for continuous-valued processes. 

The following subsections are devoted to the discussion of notable properties of transition matrices that hold for Markovian processes, and which are analogous to  concepts of quantum Markovianity defined in section~\ref{sub_Mathematician} in terms of the quantum dynamical map $\ce$.

\subsubsection{Classical divisibility}
\label{sec-cl-CK} 

Recall from Section~\ref{sec_divisibility} that quantum divisibility corresponds to being able to describe the evolution of the state of an open quantum system, between two arbitrary times, by a quantum information channel. Hence, it is natural to define classical divisibility as corresponding to being able to describe the evolution of the probability density $P(\bm x_t)$, between two arbitrary times, by a classical information channel. Any such classical channel, mapping $P(\bm x_{t_1})$ to $P(\bm x_{t_2})$,  is defined by a {\em stochastic matrix} $S(\bm x_{t_2}, \bm x_{t_1})$, with nonnegative elements and satisfying $\int {\rm d}\mu (\bm x_{t_2}) S(\bm x_{t_2}, \bm x_{t_1}) = 1$~\cite{GS02}. Clearly, the transition matrix $T(\bm x_{t}, \bm x_{t_0})$ is an example of a stochastic matrix.  We now formally define:
\begin{tcolorbox}
	\begin{df}[Classical Divisibility]
	The transition matrix $T(\bm x_{t} , \bm x_{t_0})$ of a classical stochastic process $\{\bm X(t): t \geq t_0 \}$ satisfies classical divisibility (CD) if and only if for all times $t_2 \geq t_1$ there exists a stochastic matrix 
	$S(\bm x_{t_2}, \bm x_{t_1})$ such that
	\begin{align}
	\label{cdivisible}
	T(\bm x_{t_2} , \bm x_{t_0})  = \int {\rm d}{\mu(\bm x_{t_1})} \, S(\bm x_{t_2}, \bm x_{t_1})\, T(\bm x_{t_1} , \bm x_{t_0})  \ .
	\end{align}
\end{df}
\end{tcolorbox}

The integral in Eq.~(\ref{cdivisible}) may be regarded as a generalized matrix product. As is indicated in Fig.~\ref{fig_cl}, classical divisibility is not only weaker than CM \cite{FW59}, but is weaker than the classical regression formula CRF$_n$ for any $n\geq 2$. In particular, it follows immediately  from Eq.~(\ref{cke0}) that classical divisibility is satisfied, with $S(\bm x_{t_2}, \bm x_{t_1})=P(\bm x_{t_2}| \bm x_{t_1})$, whenever CRF$_2$ holds.

\subsubsection{Classical master equation}
\label{sec-cl-dCKE}

In the analysis of some classical stochastic processes, it is both possible and convenient to use a differential (in time) form of  divisibility. Various terminologies are used in the literature, but here we refer to it as the classical master equation. 
\begin{tcolorbox}
	\begin{df}[Classical master equation]
		The transition matrix $T(\bm x_{t} , \bm x_{0})$  of a time-continuous classical stochastic process $\{\bm X(t): t \geq t_0 \}$ admits a classical master equation (CME) if it  satisfies
		\begin{align} 
		\label{dCKE}
			\frac{\partial T(\bm x_t ,\bm x_{t_0})}{\partial t} = & - \nabla \cdot [\bm A(\bm x_t) T(\bm x_t ,\bm x_{t_0}) ] + \frac{1}{2}\tr[\nabla \nabla^\top D(\bm x_t) T(\bm x_t ,\bm x_{t_0}) ] \nonumber \\ & +\int  {\rm d}\mu(\bm y_{t}) \left[ J(\bm x_t | \bm y_t)T(\bm y_t ,\bm x_{t_0}) - J(\bm y_t | \bm x_t)T(\bm x_t ,\bm x_{t_0})\right]
		\end{align}
		for all times $t\geq  t_0$, where $\bm A$ and $D\geq0$ are (differentiable) vector and matrix functions respectively,  and $J \geq 0$ is a non-negative function.
	\end{df}
\end{tcolorbox}

The existence of a classical master equation (CME) implies classical divisibility holds in the sense that, solving Eq.~\eqref{dCKE} with the appropriate initial condition gives a solution satisfying  classical divisibility~\cite{GS04}.  Thus, the relation between classical divisibility and CME in the classical hierarchy is analogous to that between divisibility and time-dependent GKS-Lindblad equation in the quantum hierarchy (see Theorem~\ref{DivtoME}). This suggests one may take  CME  as the classical analogy of the time-dependent GKS-Lindblad equation. Recall that the existence of the latter can be derived from the open quantum system dynamics being a QWN process (Section~\ref{sec_quantum_white_noise}). 
Similarly, the CME can be derived from the classical stochastic process being 
described by CWN (Section~\ref{sec-cl-CWN}), as discussed below. 
Note, however, that even though QWN  has analogues of the jump and diffusion terms in CWN, the GKS-Lindblad master equation has only one type of irreversible term, unlike the CME which inherits its jump and diffusion terms directly from CWN. Indeed, in the MCWF method (Section~\ref{sec_MCWF}), any Lindblad term can be simulated by either saltatory or diffusive evolution for $\ket{\psi(t)}$.

For the details of the derivation of the CME from CWN, see Ref.~\cite{GC09}. The parameters of the two descriptions are related as follows. 
The drift vector $\bm A(\bm x_t, t)$ is the same in both descriptions. 
The positive semidefinite and symmetric diffusion matrix $D$ is given by 
$D(\bm x_t, t) = B(\bm x_t, t) B^\top (\bm x_t,  t)$. Finally, the jump function in the DCKE is given by $J(\bm x_t | \bm y_t ) = \sum_j \lambda_j (\bm x_t, t) \delta ( \bm x _t- \bm y_t - \bm c_j(\bm y_t) )$, where $\bm c_j$ is the $j$-th column vector of matrix $C$ in Eq.~\eqref{classical_sde}. Note that the sum here is over the number of jump channels (the dimensionality of the vector ${\rm d}{\bm N}$). For a discrete configuration space, the number of jump channels can be taken to be countable. In this case jumping  is the only possible dynamics, so $\bm A=\bm 0$ and $B=0$, and one must 
 replace the Dirac-$\delta$ functions above by Kronecker-$\delta$ functions. 
 In this situation the $J(\bf x|\bm y)$ are the elements of the so-called jump rate matrix~\cite{GC09}.  For a continuous configuration space, the number of jump channels may be uncountable, allowing for $J$ 
to be a smooth function. In this case it is not uncommon to consider models 
with $J=0$, for which Eq.~\eqref{dCKE} reduces to the well known Fokker-Planck equation. Note that the term ``master equation'' is often restricted to the opposite case, where ${\bf A}$ and $D$ are identically zero~\cite{GC09}, contrary to our more general 
meaning which is inspired by the analogue with the quantum master equation.   

\subsubsection{Stochastic semigroups}
\label{sec-cl-semigroup}

In this subsection, we consider a special type of stochastic process, which is homogeneous in time. First, recalling the notation in Eq.~(\ref{notate}), we define the stochastic matrix $S_\delta(\bm x,\bm y)$ for any time displacement $\delta\geq0$ by
\begin{align} \label{deftdelta}
S_{\delta} (\bm x, \bm y) :=  {\rm Pr} ( \bm X(t_0+\delta) = \bm x | \bm X(t_0) = \bm y )  \ . 
\end{align}
Note that $S_0=I$, where $I$ denotes the `identity' matrix with $I(\bm x, \bm y)= \delta(\bm x-\bm y)$ for continuous state spaces and $I(\bm x, \bm y)= \delta_{\bm x \bm y}$ for discrete configuration spaces. Note also that 
\beq \label{st}
S_{\delta} (\bm x_{t_0+\delta}, \bm x_{t_0})=T(\bm x_{t_0+\delta}, \bm x_{t_0}) \ ,
\eeq  
i.e., $S_\delta$ is determined by the transition matrix $T$ of the stochastic process.
	
It is natural to define the product of any two such stochastic matrices $S_r$ and $S_s$ via
\begin{align}
(S_{r}    \circ S{_s}) (\bm x, \bm y) 
:=  \int {\rm d}\mu ( \bm  z) \,S_{r} ( \bm x, \bm z ) S_{s} ( \bm z, \bm y ) \ .
\end{align}
The classical analogue of a quantum dynamical semigroup (see Definition~\ref{df:Dynamical semigroups}) is then given by: 
 \begin{tcolorbox} 
	\begin{df}[Stochastic semigroup]
		\label{def_cl_Semigroup}
		The transition matrix $T(\bm x_{t} , \bm x_{t_0})$ of a classical stochastic process  $\{\bm X(t) : t\geq t_0 \}$  generates a stochastic semigroup if and only if for any $r, s\geq 0, \, S_\delta = T(\bm x_{t_0+\delta} , \bm x_{t_0})$  satisfies
		\begin{align}
		\label{cl_semigroup_con}
		S_{r+s}  = S_{r} \circ S_{s}\ .
		\end{align}
	\end{df} 
\end{tcolorbox}
It is easy to see that the concept of stochastic semigroup is strictly stronger than classical divisibility, as is depicted in Fig.~\ref{fig_cl}. In particular, Eqs.~(\ref{st}) and~(\ref{cl_semigroup_con}), for the choices $r=t_2-t_1$ and $s=t_1-t_0$, immediately yield Eq.~(\ref{cdivisible}) with $S= S_{t_2-t_1}$.

Recall that in the quantum case, under certain conditions of continuity, a  dynamical semigroup is equivalent to a time-independent GKS-Lindblad equation (see Theorem~\ref{SemitoME}). Similarly, in the classical time-continuous case a stochastic semigroup is equivalent to a homogeneous classical master equation, where the latter is defined as follows: 
\indent
\begin{tcolorbox} 
	\begin{df}[Homogeneous classical master equation]
	\label{def_cl_hdCKE}
	The transition matrix $T(\bm x_{t} , \bm x_{t_0})$ of a classical stochastic process  $\{\bm X(t) : t\geq t_0 \}$  admits a homogeneous classical master equation (HCME) if and only if Eq.~\eqref{dCKE} holds for time-independent $\bm A$, $B$ and $J$. 
	\end{df}
\end{tcolorbox}
This classical equivalence was noted in Refs.~\cite{DE69,Lin76}. For a rigorous discussion  
of these technical assumptions, we refer readers to Ref.~\cite{BSW13}.

\subsubsection{Decreasing distinguishability}
\label{sec_cl_decreasing}

The maximum probability of correctly distinguishing between any two classical probability densities $P(\bm x)$ and $P'(\bm x)$,  having respective prior probabilities $w$ and $1-w$, is $\half(1+\int {\rm d}\mu(\bm x) |wP(\bm x)-(1-w) P'(\bm x)|)$~\cite{HC69}.  Thus, the classical analogue of decreasing system distinguishability for open quantum systems (Definition~\ref{dfdisting}) is:
\begin{tcolorbox}
	\begin{df}[Decreasing classical distinguishability]
	The transition matrix $T(\bm x_{t} , \bm x_{t_0})$ of a classical stochastic process $\{\bm X(t) : t \geq t_0 \}$  exhibits decreasing classical distinguishability if and only if
	\begin{equation} 
	 \int {\rm d}\mu(\bm x_{t_2}) \left|w P(\bm x_{t_2})-(1-w)P'(\bm x_{t_2})\right| \leq 	\int {\rm d}\mu(\bm x_{t_1}) \left|w P(\bm x_{t_1})-(1-w) P'(\bm x_{t_1})\right| \ ,  
		\end{equation} 
	for all $w\in (0,1)$, initial distributions $P(\bm x_{t_0})$, $P'(\bm x_{t_0})$, and for all times $t_2>t_1\geq t_0$, where $P(\bm x_t) := \int {\rm d} \mu(\bm x_{t_0})  T(\bm x_t, \bm x_{t_0} ) P(\bm x_{t_0})$.
\end{df}
\end{tcolorbox}

It follows from Theorem~2.4 of Ref.~\cite{RHP14} that decreasing classical distinguishability is equivalent to classical divisibility~(\ref{cdivisible}).  Thus, the strict quantum hierarchy between divisibility and decreasing system distinguishability (Fig.~\ref{fighie}) collapses for the corresponding classical concepts (Fig.~\ref{fig_cl}). This difference is due to 
the distinction in quantum mechanics between completely positive and merely positive maps (see \srf{sec:DSSD}), a distinction which does not exist in classical processes.

\subsubsection{Monte Carlo  simulations}
\label{sec-cl-MCSM}
An alternative way to find the probability $P(\bm x_t)$ for a classical stochastic process is through numerical simulations. Mathematically, the idea is to reconstruct $P(\bm x_t)$ from \emph{sampling}. This is useful when an analytical solution cannot be found, especially for large systems where an exact numerical solution is impractical. Sampling has become increasingly used because of the development of efficient algorithms and increasing capacity in computation.
A sufficiently large ensemble of stochastic paths allows one to further draw statistical inferences of the system which are the same as that from $P(\bm x_t)$. Any sample-based numerical simulation method, in this regard, is usually referred to as a Monte Carlo simulation method~\cite{NS49}. Using the notations developed for MCWF in Section~\ref{sec_MCWF}, we describe  Monte Carlo simulation methods (MCSMs) as follows:

\begin{tcolorbox}
\begin{ds}[\bf Monte Carlo Simulation]
\label{ds:MCSM}
	A classical stochastic process $\{\bm X(t): t \geq t_0 \}$ can be simulated by a Monte Carlo simulation method (MCSM) if, for any system configuration $\bm X(t_0) = \bm x_0$ and  for any finite sequence of times $t_n > \dots > t_1 > t_0$, it is possible to numerically generate, in parallel, an ensemble of $M$ system trajectories ${\Upsilon(\bm x_0, {\rm  \bf r}_t)}$, each of the form
	\begin{align*}
		{\Upsilon(\bm x_0, {\rm \bf r}_n)} \equiv \left[ \bm x_0 \rightarrow \wp_{{\rm r}_1}(t_1) \; \bm x_1 \rightarrow \dots \rightarrow \wp_{{\rm \bf r}_j}(t_j) \; \bm x_j \rightarrow \dots \rightarrow \wp_{{\rm \bf r}_n}(t_n) \; \bm x_{{\rm \bf r}_n} \right] \ ,
	\end{align*}
	such that, at each time $t_j$ in the sequence,
	\begin{align}
		\lim_{M \to \infty} {\rm E}_M \left[ f (\bm x_j) \right] = \int {\rm d}\mu(\bm x_j)  P(\bm x_j) f(\bm x_j) \ , 
	\end{align}
	where  ${\rm E}_{ M} [\bullet]$ denotes an equally-weighted average over the numerically-generated trajectories; $f(\bm \bullet)$ is an arbitrary function;
	and $P(\bm x_{j})$ is the true probability distribution for $\bm X(t_j) = \bm x_j$. 
	Moreover, the method of generation cannot rely on explicit knowledge of the $P(\bm x_{j})$. 
\end{ds}
\end{tcolorbox}

It should be emphasized that not every numerical simulation method of classical stochastic processes, which may be known under the name of a Monte Carlo method, is formulated as above, and it is certainly not our intention to claim so. We thus call it  a description, rather than a definition. In particular, following our characterization of the MCWF method for the  quantum case, we have restricted to equally-weighted ensembles, but weighted ensembles can of course be used, with weights calculated along with the samples~\cite{WL97}. 
Also note that, again following the MCWF case, we have required only that MCSM reproduce the correct 1-time statistics as described by $P(\bm x_{j})$, for all $t_j$. That is, 
we do not require it to reproduce any multi-time correlation functions. 

It is easy to see that any stochastic differential equation as per Eq.~\eqref{classical_sde} can be simulated using MCSM. In particular, the increments, ${\rm d} \bm W(t)$, can be  sampled from Gaussian distributions, and the jumps from an adapted binomial distribution~\cite{RD16}. This can be used to generate a record of the random variable $\{\bm X(t) = \bm x_t , t \geq t_0 \}$, which, similar to MCWF, is called a trajectory  (or `realization') of $\bm X(t)$~\cite{GS02}. This direct connection from the classical white noise stochastic differential equation to the MCSM is an important difference from the quantum case. In the latter case, the quantum white noise differential equation, Eq.~\eqref{HPEqn}, describes 
operators on the joint system--environment Hilbert space. It has no direct correspondence to stochastically evolving wavefunctions in the system Hilbert space as used in MCWF.

Note that MCSM can be applied to both discrete and continuous-time processes. As long as the conditional probability distribution $P(\bm x_t|\bm x_{t'})$ can be sampled for a fixed $\bm x_{t'}$, it is possible to stochastically update the simulated system configuration from  $\bm x_{t'}$ to $\bm x_{t}$. In this sense, any Markovian process can be simulated by MCSM, as shown in Fig.~\ref{fig_cl}. Thus, MCSM can be implied by any other concept that we have discussed in this section, 
at least if we assume that the evolution after $t_0$ is independent of $P(\bm x_{t_0})$; see \srf{sec:nonlinearP}.  In this regard, it plays a similar role to that of MCWF for the quantum hierarchy (see Section~\ref{sec_MCWF}).


\section{Concluding remarks}
\label{sec_conclusion}

In this report, we have rigorously defined many Markov-related concepts of OQSs, within a very general framework. These concepts naturally fall into two distinct classes: those requiring knowledge of the system-environment interaction and those relying solely on the properties of the system dynamical map. On this basis, we then derived the complex hierarchical relations between those concepts, as summarized in Fig.~\ref{fighie}. We do not propose that any of our concepts should be identified as quantum Markovianity. Rather, we expect  the hierarchy figure to bring some clarity to the long-argued issue of what constitutes quantum Markovianity or non-Markovianity. Developing new criteria, that are similarly rigorous, to distinguish amongst classes of OQSs that are non-Markovian by all our definitions is a remaining challenge. 

We hope that readers have gained, if nothing else, an understanding that `quantum Markovianity' is a term which is highly dependent on the context. This is also the main reason why we have not devoted much discussion to non-Markovian `measures', despite the fact that they are widely used in the literature. It now should be clear, from our hierarchy Figure~\ref{fighie}, that the variability in constructing those measures originates  in part  from the fact that there are many different perspectives that one could take to define quantum Markovianity. For example, many possible measures of non-Markovianity based on the concepts of divisibility and decreasing system-state distinguishability are reviewed in Refs.~\cite{RHP14,MJL14,BLP16}, while a measure based on the quantum regression formula has been recently proposed in Refs.~\cite{GAV14,ALT15}.  Our hierarchy of concepts in Figure~\ref{fighie} is of direct value in this regard: if a given measure detects non-Markovianity relative to a particular concept (e.g., divisibility), then it also immediately detects non-Markovianity relative to any concepts further up the hierarchy (e.g, PFI, QRF and NIB).


To finish, we emphasize that we make no claims about the completeness of this work. There is no reason to think that our hierarchy has included all quantum Markov-related concepts of interest;   
further exploration of dynamical decoupling is one area where we know there is current activity~\cite{GV}. Also, as we have made plain, there are still some remaining questions in our hierarchy. For some we have conjectured answers, while others are completely open questions. For the reader's convenience, these are listed and discussed in~\ref{app-conj}.


\section*{Acknowledgements}
\addcontentsline{toc}{section}{Acknowledgements}%

On the connection between dynamical decoupling and quantum Markovianity, we acknowledge motivating discussions with Daniel Burgarth and Lorenza Viola. We are grateful to an anonymous referee for a number of valuable suggestions, including the need for a clear distinction between two-time conditional probabilities and stochastic matrices for classical processes. We also thank Josh Combes, Hsi-Sheng Goan, John Gough, Joe Hope, Kavan Modi, \'Angel Rivas, Wolfgang Schleich and Weimin Zhang for useful discussions and comments.
L. L. acknowledges support by a Griffith University International Postgraduate Research Scholarship.
This work was supported by the ARC Centre of Excellence for Quantum Computation and Communication Technology (CQC2T), project numbers CE110001027 and CE170100012.

\appendix

\section{Conjectures and open questions}
\label{app-conj}

\subsection*{From GQRF to PFI}

We are tempted to conclude that a system fulfilling GQRF might satisfy PFI for time-discrete evolutions, when PFI can be strictly fulfilled (recall the discussion in Section~\ref{sec_PFI}). We leave the necessary direction of Theorem~\ref{PFI-GQRF} open.

\subsection*{From Composability to QRF}
We note that that QRF is {\it prima facie}  more general than composability, and therefore conjecture that the reverse direction of Theorem~\ref{QRFtoCom} does not hold. However, a concrete counterexample is still missing and the reverse direction (from composability to QRF) is left open in Fig.~\ref{fighie}.

\subsection*{From NQIB to PU}

It is unclear whether NQIB can imply PU. Note that NQIB is defined for three times (recall Definition~\ref{def_NQIB}), while PU is defined by performing measurements on the bath at multiple times. To our knowledge, NQIB cannot be generalized to multiple times as was done for composability and NIB. Thus it is not obvious that NQIB should imply PU. However, we have not found a counterexample and thus leave this direction (in Theorem~\ref{NQIBtoPU}) open in Figure~\ref{fighie}.

\subsection*{From NIB to composability}

According to Definition~\ref{df_composability} of composability and Definition~\ref{dfnib} of NIB, the latter is a `stronger' version of composability, since it only requires the existence of an environment state which satisfies Eq.~\eqref{nib}, while composability requires that such an environment state must be unitarily equivalent to the initial one. We thus conjecture that NIB is only a necessary condition for composability. However, we have not found a counterexample to support this conjecture.

\subsection*{GQRF and MPU}

For those concepts in Fig.~\ref{fighie} that are not directly linked by any kind of arrows, one can still sort out the relations between them based on the established hierarchy. However, there does exist an exception: the relation between GQRF and MPU is not clear. It is known that MPU does not imply GQRF; recall the counterexample in the proof of Theorem~\ref{PFItoQU}. However, it is not known whether GQRF can imply MPU and we here leave it as an open question.

\section{FA, PFI and QRF}
\label{app-fa}
Here we supply the promised details in the proofs of Theorems~\ref{Fa-QRF} and~\ref{PFI-GQRF}, for deriving QRF from FA and PFI respectively.
\begin{itemize}
	\item[I.] FA and QRF (Theorem~\ref{Fa-QRF}):
	\begin{align}
		\la \check A(t_1) \check B(t_2)  \ra & = \trse \left[  (\cu_{t_0}^{t_1 \dagger} \hat A) (\cu_{t_0}^{t_2 \dagger} \hat B ) \rhose(t_0)   \right] \nonumber \\
		& = \trse \left[  \hat B \hat U_{t_1}^{t_2} \rhose(t_1) \hat A  \hat U_{t_1}^{t_2 \dagger}  \right]  \nn \\
		& \overset{\rm FA}= \trse \left[  \hat B \hat U_{t_1}^{t_2} \rhos(t_1) \hat A \otimes \trhoe(t_1) \hat U_{t_1}^{t_2 \dagger}  \right]  \label{fa-in-qrf2} \\
		& = \trs \left[ \hat B \tce_{t_1}^{t_2} [ \rhose(t_1) \hat A ] \right]  \label{fa-to-qrf2} \ ,
	\end{align}
	which is the same as Eq.~\eqref{qrf2}.
	\item[II.] PFI and QRF (Theorem~\ref{PFI-GQRF}):
	\begin{align}
		\me{\check A(t_1) \check B(t_2)} 
		& = \trse \left[  (\cu_{t_0}^{t_1 \dagger} \hat A) (\cu_{t_0}^{t_2 \dagger} \hat B ) \rhose(t_0)   \right] \nn \\
		&   = \trse \left[ \hb {\cal U}_{t_1}^{t_2} \left( \rhose(t_1) \ha \right)\right]  \nonumber \\ 
		& \overset{\rm PFI}= \trse \left[ \hat B \bigg( {\cal U}_{t_1}^{t_2}({\rm p_1}) \otimes {\cal U}_{t_1}^{t_2}({\rm s,e_1^2})  \otimes {\cal U}_{t_1}^{t_2}({\rm f_2}) \bigg) \bigg( \rho_{\rm sp_1}(t_1) \ha \otimes  \rho_{\rm e_1^2}(t_1) \otimes \rho_{\rm f_2}(t_1) \bigg)   \right] \nn \\
		& = \trs\bigg[\hat B\, \tr_{\rm e_1^2f_2} \left[  \bigg( {\cal U}_{t_1}^{t_2}({\rm s,e_1^2})\otimes {\cal U}_{t_1}^{t_2}({\rm f_2}) \bigg) \left(\rhos(t_1) \ha \otimes \rho_{\rm e_1^2}(t_1) \otimes \rho_{\rm f_2}(t_1)\right)  \right] \bigg]  \nn \\
		& = \trs \left[ \hat B \tce_{t_1}^{t_2} [ \rhos(t_1) \hat A  ]  \right] \ ,
	\end{align}
	which is the same as  Eq.~\eqref{qrf2}.
\end{itemize}

\section{The AFL model and its relations to FA, QRF and GQRF}
\label{app-AFL}

In this Appendix, we discuss the AFL model\footnote{This model was first proposed in Ref.~\cite{AFL82}. It was independently introduced recently in Ref.~\cite{AHFB14}, where the authors call it the shallow pocket model. In this report, we follow the latter formulation, which takes the environment to be a quantum superposition, rather than the former, which takes it to be a classical mixture.}, and show it satisfies QRF but fails FA and GQRF, as claimed in Theorems~\ref{Fa-QRF} and~\ref{GQRF-QRF} respectively. Consider a qubit coupled to a single environment field mode (or particle), through the time-independent Hamiltonian
\begin{align}
	\hat{H} = \frac{g}{2} \hat{\sigma}_z \otimes \hat{x} \ ,
\end{align}
where $g$ is a real number describing the coupling strength, $\hat{\sigma}_z$ is the Pauli $z$ matrix of  the system and  $\hat{x}$ is the position operator of the field.
Let $t_0 = 0$ for simplicity. The initial combined state is assumed to be $\rhose(0) = \rhos(0) \otimes | \psi \ra\la \psi |$, in which the initial field state is pure with wave function:
\begin{equation}
\label{eq-AFL-wf}
\psi(x) = \la x | \psi \ra = \sqrt{\frac{\gamma}{\pi}} \frac{1}{x + i \gamma} \ ,
\end{equation}
where $\gamma$ is a positive real number. Note that the interaction Hamiltonian is unbounded. Despite the environment comprising only a single mode, by tracing over the field one finds that the system obeys a dephasing-like master equation~\cite{AHFB14}:
\begin{equation}
\label{ME_AFL}
\frac{{\rm d} \rhos(t)}{{\rm d} t} =  \frac{g \gamma}{2} {\mathcal D}[\hat \sigma_z] \  \rhos(t)  \  .
\end{equation}
Note that if the initial system state is pure, it will becomes mixed according to Eq.~\eqref{ME_AFL}. Since the total state $\rhose(t)$  is pure, it must be entangled, which means the AFL models must fail FA.

We now show the AFL model strictly satisfies QRF in Definition~\ref{def_QRF}. Let $c_{kj} = \la k | \rhos(0) | j \ra$, where $| k \ra ,| j \ra$ (and similar below) are the eigenstates of $\hat \sigma_z$ with eigenvalues $\lambda_j = (-1)^j$. Setting $\gamma = g/2 = 1$ for convenience, the combined system-environment state can be written as
\begin{align}
	\rhose(t) & = \hat U_{0}^t  \left[ \rhose(0) \right] \hat U_{0}^{t \dagger}  \nonumber \\
	& = \exp \left(-i \hat \sigma_z \hat{x} t \right)  \left( \sum_{k,j} c_{kj} | k\ra\la j | \otimes | \psi \ra\la \psi | \right) \exp \left( i \hat \sigma_z \hat{x} t \right) \nn \\
	& = \sum_{k,j} \iint {\rm d}x\,{\rm d}x'\,  \exp \big( -i\lambda_k x \, t \big) c_{kj} \psi(x)\psi^{\ast} (x') \left(| k\ra\la j | \otimes | x \ra\la x'| \right) \exp \left( i \lambda_j x't \right)  \ , \label{eq-AFL1}
\end{align}
By tracing over the field, the reduced system state reads:
\begin{align}
	\rhos(t) & = \ce_{t_0}^t \rhos(t_0 ) \nn \\ 
	& = \sum_{k,j} \int {\rm d}x \, \exp \bigl( -i (\lambda_k-  \lambda_j) x \, t \bigr) | \psi (x)|^2 c_{kj} | k \ra\la j | \ , \\
	& = \sum_{k,j} \chi [ f(\lambda_k-  \lambda_j, t) ] c_{kj} | k \ra\la j | \label{AFL2} \ ,
\end{align}
where we define  $ f( \delta ,  s  )  = \exp \bigl( -i \delta \, x \,  s  \bigr)$, and $\chi [ \bullet ] = \int \bullet \  | \psi (x)|^2 \, {\rm d} x$.
Now, given two system operators $\hat A$ and $\hat B$, one has
\begin{align}
	\label{eq-app2-qrf1}
	\la \check A(t_2) \check B(t_1) \ra = \trse \left[  \hat U_{t_1}^{t_2\dagger} \hat A 
	\hat U_{t_1}^{t_2} \hat B \, \hat U_{0}^{t_1} \rhose(0) \hat U_{0}^{t_1\dagger} \right]
\end{align}
Using the notation introduced in Eq.~\eqref{AFL2}, this can be rewritten as
\begin{align}
	\label{eq-app-con-exa}
	\la \check A(t_2) \check B(t_1) \ra =  \trs \left[ \sum_{k,m,j} \chi [ f(\lambda_k-  \lambda_j,  t_1  ) f(\lambda_m -  \lambda_j, \Delta t_2 ) ] c_{kj}B_{mj} \hat A  |m \ra\la j | \right] \ ,
\end{align}
For QRF to hold for this correlation function, one requires that
\begin{align}
	\label{eq-app-con-qrf}
	\la \check A(t_2)  \check B(t_1) \ra & =  \trs [ \hat A \tce_{t_1}^{t_2} \rhos(t_1) \hat B ] \nonumber \\
	& =\trs \left[ \sum_{k,m,j} \chi [ f(\lambda_k-  \lambda_j,  t_1  )] \chi[ f(\lambda_m -  \lambda_j, \Delta t_2 ) ] c_{kj}B_{mj} \hat A  |m \ra\la j | \right] \ .
\end{align}
It follows from Eqs.~\eqref{eq-app-con-exa} and~\eqref{eq-app-con-qrf}, recalling that $\hat A$ and $\hat B$ are arbitrary system operators, that a sufficient and necessary condition for QRF to hold is
\begin{align}
	\label{AFLcon}
	\chi [ f(\lambda_k-  \lambda_j,  t_1  )]\  \chi[ f(\lambda_m -  \lambda_j, \Delta t_2 ) ] =
	\chi [ f(\lambda_k-  \lambda_j,  t_1  ) f(\lambda_m -  \lambda_j, \Delta t_2 ) ]\ .
\end{align}
But from  Eq.~\eqref{eq-AFL-wf}, $| \psi(x)|^2 = 1/\pi(1 + x^2)$, and one thus has  
\begin{align}
	\chi \left[ f(\lambda_k-  \lambda_j, \Delta t) \right] = \exp ( -|\lambda_k-  \lambda_j| \Delta t)	\, .
\end{align}
Noting that $\lambda_{k,j,m} \in \{ -1, 1\}$, it follows that Eq.~\eqref{AFLcon} is strictly satisfied, as required.  One can similarly show that $\la \check A(t_1) \check B(t_2) \ra $ also satisfies QRF.

However, the AFL model does not satisfy GQRF. As a counterexample, one can calculate $\la \check A(t_2) \check B(t_3) \check D(t_1) \ra$. The sufficient and necessary condition for GQRF to hold for $\la  \check A(t_2) \check B(t_3) \check D(t_1) \ra$ is
\begin{align}
	& \chi [ f(\lambda_k-  \lambda_j,  t_1  )]\  \chi[ f(\lambda_k -  \lambda_n, \Delta t_2 ) ] \  \chi[ f(\lambda_l -  \lambda_n, \Delta t_3 ) ] \nonumber \\ = &
	\chi [ f(\lambda_k-  \lambda_j,  t_1  ) f(\lambda_k -  \lambda_n, \Delta t_2 )  f(\lambda_l -  \lambda_n, \Delta t_3 )  ]\ \ .
\end{align}
This does not hold, for example, for the choice $\lambda_k = \lambda_n =1, \lambda_j = \lambda_l= -1$ and $ t_1  = \Delta t_2= \Delta t_3$.

\section{Proof of Lemma 1}
\label{app-lemma}

The proof of Lemma~\ref{lemma-PFI} in section~\ref{sub-unravellings-related} is given here.  We first repeat the lemma for convenience.

\paragraph{\bf Lemma 1} 
{\it 
	%
	Given a unitary $\hat U$ acting on the Hilbert space ${\mathbb H}_{\rm ab} = {\mathbb H}_{\rm a} \otimes {\mathbb H}_{\rm b}$, a pure state $\hat \pi_{\rm a}$  on ${\mathfrak{B}}({\mathbb H}_{\rm a})$ and a (possibly mixed) state $\rho_b$ on ${\mathfrak{B}}({\mathbb H}_{\rm b})$, such that $\rho'_{\rm a} = \tr_{\rm b} [ \hat U ( \hat \pi_{\rm a} \otimes \rho_{\rm b}) \hat U^\dagger ]$ is mixed, then either 
	\begin{enumerate}
		\item  the state $\rho_{\rm b}$ is  related to  $\rho'_{\rm a}$ by an isometry (and so is also mixed),  or
		
		\item there exists a pure-state decomposition $\rho_{\rm b} = \sum_j \wp_{\rm b}^j \hat \pi_{\rm b}^j$ of $\rho_{\rm b}$ such that  $\hat U(\hat \pi_{\rm a} \otimes \hat \pi_{\rm b}^j) \hat U^\dagger$ is a pure entangled state for at least one $\hat \pi_{\rm b}^j$ with $\wp_{\rm b}^j\neq 0$.  
	\end{enumerate}
}

\begin{proof}
	We proceed by showing that if property~2 of the lemma does not hold, then property~1 must hold.
	
	First, let $\tilde{\mathbb H}_{\rm b}\subseteq \mathbb H_{\rm b}$ denote the support of $\rho_{\rm b}$, i.e., the Hilbert space spanned by those eigenvectors of $\rho_{\rm b}$ corresponding to nonzero eigenvalues. If property~2 does not hold it then follows that $\hat U(\hat \pi_{\rm a} \otimes \hat \pi_{\rm b}) \hat U^\dagger$ is unentangled for every $\hat\pi_{\rm b}=|\psi\rangle\langle \psi|$ on the support of $\rho_{\rm b}$, i.e., for all $|\psi\rangle\in \tilde{\mathbb H}_{\rm b}$. Hence, writing $\hat \pi_{\rm a}=|\phi\rangle\langle \phi|$, we must have
	\begin{align} \label{uabapp}
		\hat U | \phi\rangle\otimes |\psi\rangle = |\phi'\rangle\otimes |\psi'\rangle
	\end{align}
	for all $|\psi\rangle\in \tilde{\mathbb H}_{\rm b}$. The Hilbert space  $\tilde{\mathbb H}^U_{\rm b}:= \{ \hat U | \phi\rangle\otimes |\psi\rangle : |\psi\rangle\in \tilde{\mathbb H}_{\rm b} \}$, is therefore isomorphic to $\tilde{\mathbb H}_{\rm b}$ and consists solely of factorisable states in ${\mathbb H}_{\rm a} \otimes {\mathbb H}_{\rm b}$.
	
	Second, note that the Hilbert space $\tilde{\mathbb H}^U_{\rm b}$ must be closed under superposition. However, for some fixed $|\phi'_0\rangle\otimes |\psi'_0\rangle \in \tilde{\mathbb H}^U_{\rm b}$, its superposition with any other factorisable state $|\phi'\rangle\otimes |\psi'\rangle$ is factorisable if and only if $|\phi'\rangle=f |\phi'_0\rangle$ or $|\psi'\rangle=g |\psi'_0\rangle$ for some multipliers $f, g$ (this is the crucial property required for the proof).  Hence, every state $|\phi'\rangle\otimes |\psi'\rangle \in \tilde{\mathbb H}^U_{\rm b}$ is either of the form $|\phi'_0\rangle\otimes |\psi'\rangle$ or $|\phi'\rangle\otimes |\psi'_0\rangle$, for two fixed states $|\phi'_0\rangle\in {\mathbb H}_{\rm a}, |\psi'_0\rangle \in {\mathbb H}_{\rm b}$.
	
	Third, one might think that ${\mathbb H}_{\rm b}$ could contain a state of each form,  i.e, two states $|\phi'_0\rangle\otimes |\psi'\rangle$ and $|\phi'\rangle\otimes |\psi'_0\rangle \in \tilde{\mathbb H}^U_{\rm b}$. But we know from the preceding paragraph that $|\phi'\rangle=f |\phi'_0\rangle$ or $|\psi'\rangle=g |\psi'_0\rangle$ for some multipliers $f, g$. This means that one of the two states must be equal to the fixed state $|\phi'_0\rangle\otimes |\psi'_0\rangle$, up to some multiplying factor.  It follows that in fact ${\mathbb H}_{\rm b}$ can only contain normalized states of just {\it one} form. That is,  recalling such states are of the form $\hat U | \phi\rangle\otimes |\psi\rangle$ as per Eq.~(\ref{uabapp}), we have
	\begin{align} 
		\label{possu}
		\hat U | \phi\rangle\otimes |\psi\rangle=|\phi'_0\rangle\otimes |\psi'\rangle ~~\forall~|\psi\rangle\in \tilde{\mathbb H}_{\rm b},\qquad {\rm or}\qquad \hat U | \phi\rangle\otimes |\psi\rangle =|\phi'\rangle\otimes |\psi'_0\rangle ~~\forall~|\psi\rangle\in \tilde{\mathbb H}_{\rm b},
	\end{align}
	for some fixed states $|\phi'_0\rangle\in {\mathbb H}_{\rm a}, |\psi'_0\rangle \in {\mathbb H}_{\rm b}$.
	
	Finally, note that the first possibility in Eq.~(\ref{possu}) implies immediately that 	$\rho'_{\rm a} = \tr_{\rm b} [ \hat U ( \hat \pi_{\rm a} \otimes \rho_{\rm b}) \hat U^\dagger ]=|\phi'_0\rangle\langle \phi'_0 |$, i.e., that $\rho'_{\rm a}$ is pure, which contradicts the assumption  in the statement of the lemma. Hence, the second possibility in Eq.~(\ref{possu}) must hold. Further, since $\hat U$ must preserve inner products, then any orthonormal set of states $\{ |\psi_j\rangle \}$ in $\tilde{\mathbb H}_{\rm b}$ must be mapped to a corresponding orthonormal set of states $\{ |\phi'_j\rangle \}$ in ${\mathbb H}_{\rm a}$. It follows immediately, for an orthogonal decomposition $\rho_{\rm b}=\sum_j p_j |\psi_j\rangle\langle \psi_j|$, that
	\begin{align}
		\rho'_{\rm a} = \tr_{\rm b} [ \hat U ( \hat \pi_{\rm a} \otimes \rho_{\rm b}) \hat U^\dagger ] = \sum_j p_j |\phi'_j\rangle\langle \phi'_j| .
	\end{align}
	Hence, recalling $\rho'_{\rm a}$ is mixed, as per the statement of the lemma,  it follows immediately that $\rho_{\rm b}$ is mixed, and related to $\rho'_{\rm a}$ by an isometry, as per property~1 of the lemma.
	
\end{proof}

\section{Hierarchy of classical regression formulas}
\label{sec:hCRF}

The $n$th order classical regression formula, denoted by CRF$_n$, is equivalent to (see Section~\ref{sec-cl-CRF})
\beq
P(\bm x_{t_n},\dots,\bm x_{t_2},\bm x_{t_1}|\bm x_{t_0}) = P(\bm x_{t_n}|\bm x_{t_{n-1}})\dots P(\bm x_{t_2}|\bm x_{t_1}) \ P(\bm x_{t_1}|\bm x_{t_0}) .
\eeq
Here we show that the sequence $\{{\rm CRF}_n\}$ forms a strict hierarchy: in particular, while CRF$_n$ clearly implies CRF$_{n-1}$ (choose $t_{n-1}=t_n$), the converse does not hold.  

Feller has given an example of a non-Markovian classical stochastic process which satisfies the Chapman-Kolmogorov equation~(\ref{cke}) \cite{FW59}. Feller's example further satisfies CRF$_2$ but not CRF$_3$. However, it does not readily generalize to higher CRF$_n$. We therefore require a different example, that generalizes to all values of $n$. We will only consider the case of discrete classical stochastic processes and times.  

First, consider a set of 3 random variables $\{x_1,x_2,x_3\}$, each taking values in $\{\pm 1\}$, and having joint probability distribution 
\beq \label{p0}
p_0(x_1,x_2,x_3):=\frac{1}{8}(1 + x_1x_2x_3)\  .
\eeq 
Thus these variables are statistically correlated.  Further, summing over one or more variables gives
\beq \label{counter}
p_0(x_j) = \frac{1}{2}\ ,\qquad p_0(x_j,x_k)=\frac{1}{4}\ , \qquad j<k.
\eeq
We now extend these random variables to an infinite set $\{x_n|n=0,1,2,\dots \}$, where $n$ is identified with a fixed discrete time $t_n$ and $x_n=\pm1$.  First, each block $B_m:=(x_{3m+1},x_{3m+2},x_{3m+3})$ for $m\geq 0$ is defined to have associated joint probability distribution $p_0(x_{3m+1},x_{2m+2},x_{3m+3})$. Second, $x_0$ is permitted to have some arbitrary probability distribution $q(x_0)$. Third, each block is statistically independent of $x_0$ and all other blocks (thus, for example, $p(x_0,x_1,x_3,x_5)=q(x_0)p_0(x_1,x_3)p_0(x_5)=q(x_0)/8$).  It follows immediately that all sets of two or fewer variables are uncorrelated, with joint probabilities as per Eq.~(\ref{counter}) for $j>0$.  Further, $p(x_m|x_0) = p(x_m)=\frac{1}{2}$ and $p(x_m|x_n) = p(x_m,x_n)/p(x_n)=\half$ for all $m,n>0$, yielding
\beq
p(x_m,x_n|x_0) = p(x_m,x_n)= \frac{1}{4} = p(x_m|x_n)\ p(x_n|x_0) 
\eeq
for all $m>n>0$. Thus, CRF$_2$ is valid.  However, it is clear from Eqs.~(\ref{p0}) and (\ref{counter}) that CRF$_3$ fails.  Note that, by iterative insertion of further statistically independent blocks between any two consecutive times, one can construct further examples where the time difference between consecutive variables becomes arbitrarily small.

The above counterexample is easily generalized to CRF$_n$, and arbitrary block sizes, as follows.  First, for integers $M\geq N\geq2$ let $x_1,x_2,\dots,x_M$ be $M$ random variables with $x_n=\pm1$, with a joint probability distribution of the form
\beq
p_{MN}(x_1,x_2,\dots,x_M) := 2^{-M}\ \binom{M}{N}^{-1} \sum_{1\leq j_1<j_2<\dots<j_N\leq M}(1 + \alpha_{j_1j_2\dots j_N}\  x_{j_1}\dots x_{j_N})\ ,
\eeq
with coefficients $0<|\alpha_{j_1j_2\dots j_N}|\leq 1$.  Thus, $p_0$ above corresponds to the case $M=N=4$ and $\alpha_{1234}=1$.  It is straightforward to check that any $N$ of these variables are correlated, with marginal distribution
\beq
p_{MN}(x_{j_1},\dots, x_{j_N}) = 2^{-N}\left[ 1+ \binom{M}{N}^{-1}\alpha_{j_1j_2\dots j_N}\ x_{j_1}\dots x_{j_N} \right].
\eeq
Extending these random variables to an infinite set as before, with blocks $B_m=(x_{mM+1},\dots,x_{mM+M})$ of length $M$ having joint distribution $p_{MN}(x_{mM+1},\dots,x_{mM+M})$ and being statistically independent of each other and $x_0$, one finds that CRF$_{n}$ is satisfied for all $n<N$, but fails for $n=N$. We note this general example can be extended to a stationary process, similarly to Feller's example \cite{FW59}, by averaging over $M$ consecutive time-shifts. 


\phantomsection
\addcontentsline{toc}{section}{References}
\section*{References} 

\bibliographystyle{elsarticle-num}

\bibliography{pr}

\end{document}